%% file: manuscript.tex
\documentclass[iop]{emulateapj}
\usepackage{times, graphicx, setspace, subfigure, latexsym, amssymb, amsmath, fancyhdr, epsf}
\usepackage{epstopdf}
\usepackage{dcolumn}
  \newcolumntype{d}[1]{D{.}{.}{#1}}    
\usepackage[backref,breaklinks,colorlinks,citecolor=blue]{hyperref}
\usepackage{natbib}
\usepackage{multirow}
\usepackage{color}
\usepackage{rotating}
\usepackage{subfigure}
\shorttitle{The NANOGrav Nine-Year Data Set: Dispersion Measure Variations}
\shortauthors{Jones et al.}

\begin{document}

\title{The NANOGrav Nine-Year Data Set: Measurement and Interpretation of Variations in Dispersion Measures}

\author{M. L. Jones\altaffilmark{1,2,20}, M. A. McLaughlin\altaffilmark{1,2}, 
M. T. Lam\altaffilmark{1,2}, 
J. M. Cordes\altaffilmark{3}, 
L. Levin\altaffilmark{4}, 
S. Chatterjee\altaffilmark{3},
Z.\,Arzoumanian\altaffilmark{5}, 
K.\,Crowter\altaffilmark{6}, 
P.\,B.\,Demorest\altaffilmark{7}, 
T.\,Dolch\altaffilmark{8}, 
J.\,A\,Ellis\altaffilmark{9,21},  
R.\,D.\,Ferdman\altaffilmark{10,11}, 
E.\,Fonseca\altaffilmark{10},
M.\,E.\,Gonzalez\altaffilmark{6,12},
G.\,Jones\altaffilmark{13},
T.\,J.\,W.\,Lazio\altaffilmark{9},
D.\,J.\,Nice\altaffilmark{14},
T.\,T.\,Pennucci\altaffilmark{13},
S.\,M.\,Ransom\altaffilmark{15},
D.\,R.\,Stinebring\altaffilmark{16},
I.\,H.\,Stairs\altaffilmark{6},
K.\,Stovall\altaffilmark{7,17},
J.\,K.\,Swiggum\altaffilmark{18},
W.\,W.\,Zhu\altaffilmark{19}}

\altaffiltext{1}{Department of Physics and Astronomy, West Virginia University, Morgantown, WV 26506, USA}
\altaffiltext{2}{Center for Gravitational Waves and Cosmology, West Virginia University, Chestnut Ridge Research Building, Morgantown, WV 26505}
\altaffiltext{3}{Department of Astronomy and Cornell Center for Astrophysics and Planetary Science, Cornell University, Ithaca, NY 14853, USA}
\altaffiltext{4}{Jodrell Bank Centre for Astrophysics, Alan Turing Building, School of Physics and Astronomy, The University of Manchester,
Oxford Road, Manchester, M13 9PL, UK}
\altaffiltext{5}{Center for Research and Exploration in Space Science and Technology and X-Ray Astrophysics Laboratory, NASA Goddard Space Flight Center, Code 662, Greenbelt, MD 20771, USA}
\altaffiltext{6}{Department of Physics and Astronomy, University of British Columbia, 6224 Agricultural Road, Vancouver, BC V6T 1Z1, Canada}
\altaffiltext{7}{National Radio Astronomy Observatory, P.O.~Box 0, Socorro, NM, 87801, USA}
\footnotetext[8]{Department of Physics, Hillsdale College, 33 E. College Street, Hillsdale, MI 49242, USA}
\footnotetext[9]{Jet Propulsion Laboratory, California Institute of Technology, 4800 Oak Grove Dr. Pasadena CA, 91109, USA}
\footnotetext[10]{Department of Physics, McGill University, 3600 rue Universite, Montreal, QC H3A 2T8, Canada}
\footnotetext[11]{Department of Physics, Univ. of East Anglia, Norwich NR4 7TJ, UK}
\footnotetext[12]{Department of Nuclear Medicine, Vancouver Coastal Health Authority, Vancouver, BC V5Z 1M9, Canada}
\footnotetext[13]{Department of Physics, Columbia University, 550 W. 120th St. New York, NY 10027, USA}
\footnotetext[14]{Department of Physics, Lafayette College, Easton, PA 18042, USA}
\footnotetext[15]{National Radio Astronomy Observatory, 520 Edgemont Road, Charlottesville, VA 22903, USA}
\footnotetext[16]{Department of Physics \& Astronomy, Oberlin College, Oberlin, OH 44074, USA}
\footnotetext[17]{Department of Physics and Astronomy, University of New Mexico, Albuquerque, NM, 87131, USA}
\footnotetext[18]{Center for Gravitation, Cosmology and Astrophysics, Department of Physics, University of Wisconsin-Milwaukee, P.O. Box 413, Milwaukee, WI 53201, USA}
\footnotetext[19]{Max-Planck-Institut f\"{u}r Radioastronomie, Auf dem H\"{u}gel 69, D-53121, Bonn, Germany}
\footnotetext[20]{STEM Mountains of Excellence Fellow}
\footnotetext[21]{Einstein Fellow}


\begin{abstract}
We analyze dispersion measure (DM) variations of 37 millisecond pulsars in the 9-year NANOGrav data release and constrain the sources of these variations. Variations are significant for nearly all pulsars, with characteristic timescales comparable to or even shorter than the average spacing between observations. Five pulsars have periodic annual variations, 14 pulsars have monotonically increasing or decreasing trends, and 13 pulsars show both effects. Several pulsars show correlations between DM excesses and lines of sight that pass close to the Sun.  Mapping of the DM variations as a function of the pulsar trajectory can identify localized ISM features and, in one case, an upper limit to the size of the dispersing region of 13.2 AU. Finally, five pulsars show very nearly quadratic structure functions, which could be indicative of an underlying Kolmogorov medium. Four pulsars show roughly Kolmogorov structure functions and another four show structure functions less steep than Kolmogorov. One pulsar has too large an uncertainty to allow comparisons. We discuss explanations for apparent departures from a Kolmogorov-like spectrum, and show that the presence of other trends in the data is the most likely cause.
\end{abstract}

\hypersetup{linkcolor=blue}
\section{Introduction}
The principal goal of the North American Nanohertz Observatory for Gravitational Waves \citep[NANOGrav;][]{mcl13} is to detect gravitational waves in the nanohertz regime of the gravitational wave spectrum using a pulsar timing array (PTA). Sensitivity improves as more millisecond pulsars (MSPs) are added to the PTA, and therefore it is essential to have as many well-timed MSPs as possible \citep{sie13,vig16}.
For every MSP, we must construct an accurate timing model that accounts for all known effects on the pulsar times-of-arrival (TOAs) over decade timescales \citep{jen05,cor10}. 
One of the parameters that must be fit in the timing model is the dispersion measure \citep[DM;][]{lor12}.
As the pulsar signal travels through the interstellar medium (ISM), it encounters ionized plasma and electron density variations along the way. The DM is the integrated column density of free electrons along the line of sight to a pulsar

\begin{equation}
	\rm{DM} = \int_{0}^{\it{d}} \it{n_e(l) dl}~,
	\label{eq:def}
\end{equation}
where $n_{e}$ is the free electron density along a line of sight $l$ and $d$ is the pulsar distance. When the pulsar signal propagates through the ISM, interactions with these free electrons cause dispersion that is characterized by a frequency dependent time delay

\begin{equation}
	\Delta t \simeq 4.15 \times 10^{6}~\rm{ms} \times \rm{DM} \left( \frac{1}{\nu_1^{2}}-\frac{1}{\nu_2^{2} } \right) ~,
	\label{eq:delay}
\end{equation}
where $\nu_1$ and $\nu_2$ are two different frequencies in MHz and DM is in pc cm$^{-3}$.
Observing at least two frequencies is therefore necessary to solve for the DM for a measured time delay.
This time delay can be significant when compared to the pulsar period, and therefore the DM must be fit when creating a timing model and corrected for at each epoch \citep[e.g.][]{dem13,arz15}. 

Inhomogeneities in the ISM or solar wind and differences in the relative velocity of the pulsar and the Earth, among other processes, can change the free electron density along the line of sight \citep[LOS;][]{lam16}. The result is a DM that varies with time, changing on timescales of hours to years. 
In this paper we discuss the variations seen in the NANOGrav 9-year data release \citep{arz15}, constrain the possible sources of these variations, and use these constraints to characterize the ISM along the LOS.

In \S 2, we discuss the data used for this analysis. In \S 3, we discuss the significance and trends seen in the DM time series. In \S 4, we perform a structure function analysis on select MSPs and put the results in the context of a Kolmogorov spectrum.  In \S 5, we discuss the results of these analyses and in \S 6 we present conclusions.

\section{Data}

Our analysis uses data from the NANOGrav 9-year data set \citep{arz15}. Pulsars were included in the data set based on the anticipated stability of their timing, their TOA precision, and their detection over a wide frequency range. Of the 37 MSPs included in the data release, 17 were reported on in \cite{dem13}. Observations took place roughly once a month between 2004 and 2013 with observing time spans of individual pulsars ranging from 0.6 to 9.0 years. 
Those MSPs with declinations between 0 and 39$^{\circ}$ were observed with the 305-m William E. Gordon Telescope at the Arecibo Observatory, and the rest were observed with the 100-m Robert C. Byrd Green Bank Telescope (GBT) of the National Radio Astronomy Observatory (NRAO); PSRs J1713+0747 and B1937+21 were observed with both. 
Every MSP was observed at multiple frequencies to account for frequency-dependent dispersion effects. Dual frequency observations occurred within $\sim$1 hour at Arecibo and within several days at the GBT. The typical length of an observation was $\sim$25 minutes.
A more detailed and thorough description of these observations can be found in \cite{arz15}. 
 
For each pulsar, the DM was measured at nearly every observing epoch and recorded using the DMX parameter as part of the TEMPO software package\footnote[1]{TEMPO software package: http://tempo.sourceforge.net}, where DMX is the difference between the fiducial DM and the DM at each epoch. A single DM value was held constant for a 14 day window; the possible errors in DM$(t)$ estimation using this method are discussed in \cite{lam15}. 
Data from early single-receiver observations were omitted for PSRs J1741+1351, J1853+1303, J1910+1256, J1944+0907, and B1953+29 as it was not possible to independently measure DM and other timing properties.
We plot  DMs vs time (i.e., DM$(t)$) for all of the pulsars in Figures \ref{fig:trendremoval1} through \ref{fig:trendremoval5}. Values from the 9-year data release used in this analysis can be seen in Table \ref{tab:DMV}.

\section{Determining Significance and Trends in the Variations}
\label{sec:variations}

Many pulsars exhibit time variability of DM; it has been a long known effect. The first detection of temporal variations was for the Crab pulsar \citep{ran71}. These variations were later determined to be most likely due to variations in the surrounding nebula \citep{isa77}. The Vela pulsar exhibits a decreasing time-dependent DM attributed to the pulsar motion through the enveloping supernova remnant \citep{ham85}.

We first determine whether the DM variations we see are significant or if they are consistent (within errors) with a constant DM value. We calculate the reduced $\chi^2$ for each pulsar as 
\begin{equation}
\chi_r^2 = \frac{1}{N_{\rm{DOF}}}\sum \frac{(\rm{DM}\it{(t)}-\overline{\rm{DM}})^{\rm{2}}}{\sigma(t)^2}~,
\label{chi}
\end{equation}
where $N_{\rm{DOF}}$ is the number of degrees of freedom, $\overline{\rm{DM}}$ is the average DM value for the data span for a pulsar, DM$(t)$ is the DM at a particular time $t$, and $\sigma(t)$ is the error associated with each DM(t) value.
All but two of the pulsars (PSRs J1923+2515 and J2214+3000) have $\chi^2_r$ $\geq$ 1. 
Of these, 15 pulsars show moderate variations (1$~\leq \chi^2_r \leq$ 10), and 20 pulsars have significant variations ($\chi^2_r~\geq~10$).
We therefore conclude that the DMs are intrinsically variable for all of the MSPs in our sample with the possible exceptions of J1923+2515 and J2214+3000, which both show visible variation at a low level despite the statistical test. Both of these pulsars have short data sets (2.2 and 2.1 years, respectively).

\input{Table1}

\label{tab:DMV}

\subsection{DM Trends}
\label{sec:Trends}
\input{Table2}

\input{Table3}

DMs can vary in many ways, with components that appear linear, periodic, or random. We do not attempt to model random variations that may be due to noise or a purely stochastic medium here. 
Sources of linear and periodic trends include a changing distance between the Earth and the pulsar, a wedge with linear density changes in the ISM or the orbital motion of the Earth, among others; the possible geometries from which these trends arise are explained in detail in \cite{lam16}. Both linear and periodic trends have been seen in Parkes Pulsar Timing Array (PPTA) data \citep{you07,kei13,pet13,rea16}. 
\cite{pet13} determine the significance of a linear trend by calculating the error of a fit to the slope; linear trends were deemed significant if the errors are less than 35\% of the slope value and highly significant if the errors on the slope measurement are less than 20\%. This method is not applicable for this data set as a large number of pulsars exhibit sinusoidal trends without linear trends.

In order to determine the scale and structure of the variations, the options being linear, periodic, both, or variations consistent with white noise, we applied a non-linear least squares fit to the data using three functions, 

\begin{equation}
	\begin{aligned}
	& \overline{\rm{DM}}\it{(t)} = a_0 t+a_1, \\
	& \overline{\rm{DM}}\it{(t)} = b_0\cos(b_1t+b_2)+b_3, \\
	& \overline{\rm{DM}}\it{(t)} = c_0\cos(c_1t+c_2)+c_3t+c_4, 
	\end{aligned}
\label{eq:trends}
\end{equation}
with the $\chi_r^2$ calculated for the time series after each of these fits was individually subtracted off. For each fit, $N_{\rm{DOF}} \approx N-N_p$, where $N$ is the number of DM measurements and $N_p$ is the number of free parameters being fit in that function ($N_p=1$ when $\overline{\rm{DM}}\it{(t)}$ is a constant value). The three $\chi_r^2$ values were then compared to the original value; the result producing the lowest $\chi_r^2$ was assigned as the trend.
The results of these fits can be seen in Table \ref{tab:Trends}.  
There is a known complication when estimating the number of degrees of freedom for a non-linear model \citep{and10}. The $\chi_r^2$ is only used as a metric to compare the fits of models we know the be incomplete; as stated earlier, the ISM is more complicated than a purely linear tend plus annual component. The fitting routine incorporates a non-linear least squares fit which is locally linearizing around the minimum $\chi_r^2$. 
Later on, we describe $\chi_r^2$ surfaces in the full parameter space and analyze the degree of covariance between fit parameters, finding it agrees with this fitting routine.

The periodic term in the function was fit using an initial guess of 365 days.
Due to a change in sign of $d\rm{DM}/\it{dt}$ or the appearance/disappearance of a trend partway through the data span, PSRs J0613--0200, J1600--3053, J1640+2224, J1918--0642, and B1937+21 are not well characterized by a single fit. These MSPs were fit using piecewise functions, using the $\chi_r^2$ of the fit to identify the applicable MJD range for each fit. The results of this partial fitting can be seen in Table \ref{tab:PartialTrends}. 
The fits can be seen in Figures \ref{fig:trendremoval1} through \ref{fig:trendremoval5}. The $\chi_r^2$ values listed in Table \ref{tab:PartialTrends} are for the individual fit regions; these differ from the values shown in Figures \ref{fig:trendremoval1} through \ref{fig:trendremoval5} because those values incorporate both fits as well as any regions excluded from the fit.

\input{Figures1to5}
\label{fig:trendremoval}

We applied a Lomb-Scargle periodogram analysis to corroborate the best-fit periods and possibly identify other periodicities (also seen in Table \ref{tab:Trends}). This analysis is able to detect periodicities in unevenly sampled data \citep{sca82} for which a false alarm probability (FAP) may be calculated. The FAP is the likelihood that these periods would occur as a result of random white noise. We ignored periods found by the periodogram that coincided with either the length of the dataset or the observing cadence. The resolution of the analysis is equal to the cadence of the observations. Any linear trend in the DM variations will mask the periodic effect, and therefore was removed from those identified to have linear effects before applying the periodogram analysis. 

\subsection{DM Variation Timescale}
The DM value can vary on timescales of years, days, or even hours. The DMX parameter fits the DM variations by holding it constant for a specified time window. Therefore it is important to know on what timescale this DM is accurate. 
The time $\delta t$ for DM to change by $\sigma_{\rm{DM}}$, the rms DM in the DM time series, gives us a rough estimate for how long a single DM estimation is valid. For a linear trend
\begin{equation}
	\frac{\sigma_{\rm{DM}}}{\delta t} = \frac{d\rm{DM}}{dt} = a_0~,
\end{equation}
where $a_0$ is the slope of the linear trend, seen also in Equation \ref{eq:trends}. The time associated with a periodic trend
\begin{equation}
\frac{\sigma_{\rm{DM}}}{\delta t} = \frac{d\rm{DM}}{dt} \approx A\omega~,
\end{equation}
where $A$ and $\omega$ are the amplitude and angular frequency of the periodic trend respectively. The variation time for a DM time series showing both trends combines the $d\rm{DM}/dt$ components from both the periodic and linear components
\begin{equation}
	\frac{\sigma_{\rm{DM}}}{\delta t} \approx \frac{d\rm{DM}}{dt} + A \omega~.
\end{equation}
The $\delta t$ values for the MSPs showing trends are seen in Tables \ref{tab:Trends} and \ref{tab:PartialTrends}. This $\delta t$ can inform on what timescale our DM measurement is constant and the importance of observing at epochs with spacing smaller than this timescale.

\subsection{Solar-Angle Correlation}

\begin{figure*}
	\begin{center}			
		\includegraphics[width=0.57\textwidth]{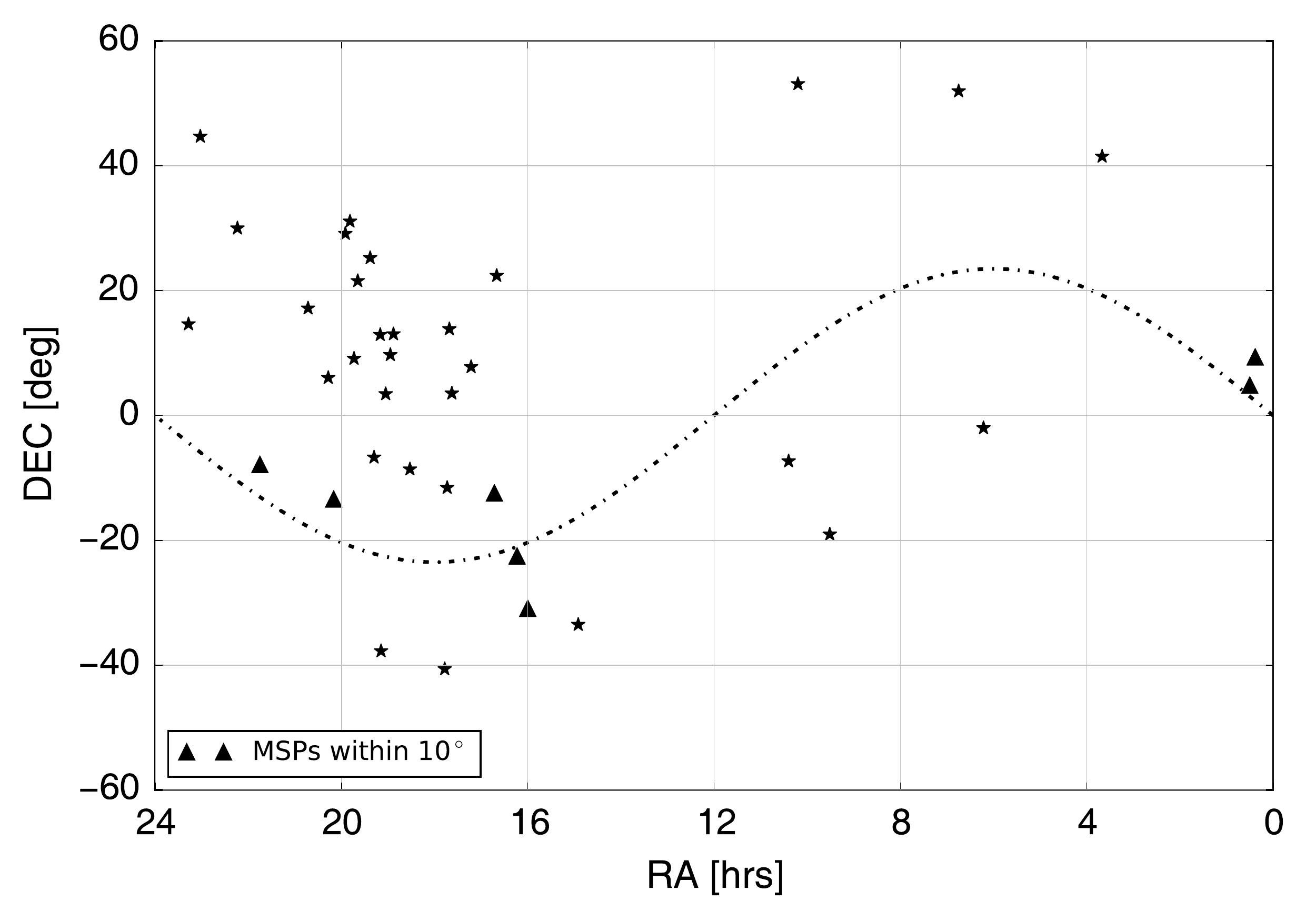}
	\caption{MSP positions with respect to the ecliptic, shown by the dashed line. Sources that lie within $\sim10^{\circ}$ of the ecliptic are signified by triangles. A number of pulsars pass near enough to the Sun for the signal to potentially encounter the solar ionosphere, resulting in a variation in the number of free electrons over an annual cycle.}
	\label{fig:pos}		
	\end{center}
\end{figure*}

Pulsars that lie close to the ecliptic (within approximately 10$^{\circ}$) will have their LOS pass near the Sun during Earth's orbit. This proximity can cause a sinusoidal trend in DM variations due to the large variation in $n_e$ along the LOS from the solar wind.

We examine the pulsar positions with respect to the ecliptic to determine for which MSPs this effect could be significant.
As can be seen in Figure \ref{fig:pos}, PSRs J0023+0923, J0030+0451, J1614--2230, and J2010--1323 reside close (within approximately 6.3$^{\circ}$, 1.5$^{\circ}$, 6.8$^{\circ}$, and 6.5$^{\circ}$ respectively on closest approach in the data set) to the ecliptic. The DM as a function of solar angle can be seen in Figure \ref{fig:sunpos}.
PSRs J0023+0923 and J2010--1323 show a slight peak in DM at the smallest pulsar-Sun angles. 
PSRs J0030+0451 and J1614--2230 show significant peaks at the minimal solar angle. It should be noted that the two highest DM points for J0030+0451 were omitted from \cite{arz15} as outliers but were included for this analysis. 

\input{Figure7}
\label{fig:sunpos}

\subsection{Pulsar Trajectories}

We have plotted the pulsar trajectories through the ISM as seen from Earth, color coded to signify the DM value at each epoch (Figures \ref{fig:trajectory1} through \ref{fig:trajectory4}). For this, we assumed that all of the free electrons along the line of sight are sitting in a stationary phase screen located halfway between the Earth and the pulsar. 
The trajectories are the projected motions of the pulsar as seen on this phase screen.
Using proper motion and distance estimates with errors from the NE2001 model \citep{cor02}, the transverse velocity can be calculated and used to track the pulsar's trajectory in the sky. Proper motions were taken from the data release (seen in Table \ref{tab:DMV}). 
These trajectory maps can be useful in isolating features in the ISM as well as visualizing trends in the DM time series.

\input{Figures8to11}
\label{fig:trajectory}

\section{Structure Functions}

\input{Figure12}
\input{Figure13}
\def\mean#1{\left< #1 \right>}

Turbulence in the ISM is typically described as having a Kolmogorov power spectrum, meaning we expect to find larger variations over longer timescales. 
The power spectrum used to derive the structure function has the form
\begin{equation}
	P(q) \propto q^{-\beta}~,~q_{\rm{outer}} \le q \le q_{\rm{inner}}
\end{equation}
where $q$ is the  reciprocal of the size scale, and $\beta$ is the power spectrum exponent.
A Kolmogorov medium corresponds to a $\beta$ value of 11/3, while the highest value expected for turbulence in the ISM (for an inner scale shorter than 10$^9$m) is $\beta = 4$ \citep{ric90}. 
The outer scale is described as the size at which the ISM ceases to be homogeneous, and the inner scale is the point at which dissipation occurs in the material along the line of sight. 

The DM structure function (SF) is an effective analytical tool for characterizing interstellar turbulence over various time and size scales \citep{ric90, cor90, you07, kei13, fon14, lam16}. We compute SFs by binning the change in time across all epochs into equally log-spaced bins after calculating
\begin{equation}
D_{\textrm{DM}}(\tau) = \mean{[\textrm{DM}(t+\tau) - \textrm{DM}(t)]^2}~,
\end{equation}
where $\tau$ is the time lag in days \citep{cor98}.  
The diffractive timescale $\Delta \tau_D$, the scale during which the diffraction intensity varies as a result of irregularities in the ionized plasma along the line of sight, is used to anchor the SF
\begin{equation}
	D_{\rm{DM}}(\tau) = (A_{\nu} \nu)^2(\tau/\Delta \tau_{\rm{D}})^{\beta - 2}~,
	\label{eq:tau_D}	
\end{equation}
where $A_{\nu}$ = 3.84$\times$10$^{-5}$ MHz$^{-2}$ pc cm$^{-3}$ s$^{-1}$ and $\nu$ is the observing frequency. Epoch to epoch variations of $\Delta \tau_{\rm{D}}$ are expected of order 10$\%$ or more \citep{lam16}.  
The SF is poorly estimated at large time lags, and so some functions may appear Kolmogorov at shorter timescales but fall below at longer time lags; this is an indication of an underlying Kolmogorov spectrum \citep{you07}. This is also why SF values at high time lags may have large errors on them. Quadratic SFs occur when the time lag is smaller than the timescale necessary to adequately probe the structure in a region, if any \citep{lam16}.

Several models were applied to the SF in order to constrain a range for $\beta$. Fitting only for Equation \ref{eq:tau_D} is problematic because there are other contributions to the SF, such as trends and noise, among other possibilities. However, over-fitting the SF will cause the fit to fail for a number of reasons, discussed in this section. The models are of the form
\begin{equation}
D_{\rm{DM}}(\tau) = D_{\rm{sto}}(\tau) + D_{\rm{lin}}(\tau) + D_{\rm{per}}(\tau) + D_{\rm{noise}}~,
\end{equation}
where the first term is the stochastic component, the second term is the linear component, the third term is the periodic component, and the last term is the noise component.

One method is to fit the stochastic and noise components while fixing the linear trend and periodic components to the trend values found in the time series analysis. However, we were only able to successfully fit the SF for one pulsar (PSR J1643$-$1224) using this model. In all other cases, the SF of the two trends is higher than the calculated SF; fitting for a stochastic component on top of that would only increase the chi-squared, and the fit fails. This failure is evidence that there is contamination between the stochastic and trend components. In many cases, a linear trend fit over the time series will absorb part of the stochastic component. Therefore, the ``true" linear trend may be different than the one we fit for in the time series, which will bias the component in the SF high. Therefore, we are unable to obtain a proper fit using this model.

To bypass this contamination between the trend and stochastic aspects, we tried fitting for all parameters (stochastic, noise, linear, and periodic) without using any prior information in the hopes of allowing a fit and comparing values with those previously found. Due to the covariant nature of the parameters, we found values for the stochastic, periodic amplitude, and noise components, but the amplitude of the periodic trend was not much larger than its uncertainties while the linear trend component was found to be zero. The period was then fixed to one year in order to eliminate some of the covariance; fixing the period again gave significant values for the stochastic, noise, and periodic amplitude components, and no significance for an additional linear component. This is further evidence of the high covariance between the stochastic and linear trend components. The periodic amplitudes found here agree very well with those found previously by the trend analysis. Simultaneously fitting for both the stochastic and linear components does not appear to yield significant values for both. 

The simplest model is then to only fit for stochastic and noise components while not fitting for a linear trend component. This model can be applied to all pulsars. Because the periodic component was not found to be highly covariant with the stochastic component, the periodic amplitudes were fitted as well and compared to the values from the trend analysis. In order to constrain a linear trend, we would need some prior information on the shape of the stochastic component in the time series relative to the true linear component that we do not have at this time.

\input{Table4}

Diffractive timescales, listed in Table \ref{tab:tdiff}, were calculated by creating a 2-dimensional
dynamic spectrum of each 1500-MHz observation in the data set and computing the
2D autocorrelation function of each spectrum, which in turn is summed
over time and frequency separately. A Gaussian function is fitted to the
1D frequency-summed autocorrelation function, and the scintillation timescale is defined as
the half-width at $e^{-1}$.
This is following the same procedure as described in \cite{lev16}. 

Most observations in the 9-year data set are around 30 minutes long, and for many pulsars, the scintillation timescale is much longer than this integration time. Therefore, it was only possible to measure diffractive timescales in this way for a few of the pulsars in the sample.

Lag bins are equally spaced in log space.
Errors on the SF were calculated by combining the propagated errors from the DM time series values and the errors due to uncertainty in the specific realizations in a stochastic process. To constrain those errors, we used simulations of different spectral slopes and different timespans of data, the standard deviation of the SF was calculated for 10$^4$ realizations at every time lag $\tau$. This was saved as a 3D grid of values (timespan, $\beta$, $\tau$) and then a function was used to interpolate over that grid to give the realization error of the SF value at each $\tau$. We held $\beta$ constant to the value for a Kolmogorov medium as varying the slope had a negligible effect on the derived errors.

\section{Results}
\label{sec{results}}

\subsection{Linear Trends and Annual Periodicities}
Of the 37 MSPs in the data release, 32 show the presence of DM trends, of which five MSPs show piecewise trends over different time spans. With the least squares fitting procedure, we find periods roughly consistent with an annual periodicity $\pm$ 54 days for 17 pulsars; eight of these periods were also detected by the periodogram. 
\cite{kei13} predicted that annual modulations would be seen based on the spectral analyses for PSRs J1024--0719 and J1909--3744, and that they are dominated by the steep linear trend. The authors suggest that a more significant detection could occur through combining datasets. Our trend analysis did not find an annual trend in PSR J1024--0719. We did, however, calculate a period of 366$\pm$5 days for PSR J1909--3744. 

A linear trend was found in 27 pulsars, 13 of which also exhibit annual trends. Five pulsars exhibit annual trends without a linear trend.
\cite{rea16} models the DM variations for 20 PPTA-observed MSPs, nine of which are also included in the NANOGrav 9-year data release (PSRs J0613$-$0200, J1024$-$0719, J1600$-$3053, J1643$-$1224, J1744$-$1134, B1855+09, J1909$-$3744, B1937+21, and J2145$-$0750). 
The trends assignments agree between the NANOGrav and PPTA data for PSRs 1024$-$0719, J1600$-$3053,  J1643$-$1224, J1713+0747, B1937+21, and J2145$-$0750. The PPTA data did not show a linear trend for PSR J0613$-$0200 or periodic trends for PSRs B1855+09 and J1909$-$3744.
Discrepancies could arise due to the difference in data spans, trends that continue or vary after the end of the data set, variations in methods of DM fitting, as well as differences in the trend fitting algorithm; we fit for the period whereas it is fixed at one year by \cite{rea16}. There is also the possibility that the DM measured for observed epochs do not agree between the sites due to the difference in observing frequencies used and spatial location; the two data sets may essentially be sampling a different ISM due to multi-path scattering \citep{cor16}.

\input{Table5}

The DM is highly correlated with solar angle for PSRs J0030+0451 and J1614--2230, as can be seen in Table \ref{tab:sunpos}.
There is a moderate correlation in PSRs J0023+0923 and J2010--1323.
The Lomb-Scargle periodogram detected roughly semi-annual periodicities for PSRs J0340+4130 and J0645+5158. These half-year periods arise from the solar-ionospheric effects along the LOS \citep{lam16}.

The DM varies by one error bar over a timescale between one month and one year (30 days $< \delta t<$ 365 days) for 18 MSPs. An additional 10 MSPs had a $\delta t$ approximate to or smaller than the cadence of our observations ($\sim$one month). This illustrates how quickly the DM can vary by a significant amount and the necessity of observing at $\sim$week cadences and fitting for DM at every epoch.

\subsection{Structure Functions}
We have computed SFs for MSPs whose diffractive timescales could be calculated or obtained from the literature, seen in Table \ref{tab:tdiff}. This could be done for 15 of 37 MSPs in the data release, which can be seen in Figure \ref{fig:structure}. Three MSPs (PSRs J1832--0836, B1953+29, and J2017+0603) 
whose diffractive timescales were available were omitted because there was less than two years of continuous DM measurements.
Fit power spectral values can also be seen in Table \ref{tab:tdiff}.
PSRs J0030+0451 and J2145--0750 show fairly flat SFs, which is most likely a result of a white noise dominated data set. \cite{kei13} show a SF for J2145--0750 exhibiting similar structure at shorter time lags, but the error bars are too large to allow a detailed comparison.

For J1600--3053, \cite{kei13} measure a Kolmogorov SF; \cite{you07} mention it as being quadratic at shorter time lags becoming less steep at higher lags. Our analysis shows apparent white noise domination for time lags below $\sim$100 days, beyond which the SF is bit more shallow than Kolmogorov for the majority of time lags.

PSRs J0613--0200, J1024--0719,  J1643--1224, B1855+09, and J2317+1439 show nearly quadratic power spectra. Our calculated SF for J0613--0200 agrees with \cite{you07}.
The SF for J1643--1224 resembles that in \cite{kei13}. \cite{you07} show a power law exponent between the expected values for a quadratic and Kolmogorov medium at shorter time lags, with the SF exhibiting a power spectrum below Kolomogorov at higher lags. Our analysis shows a nearly quadratic power law with a distinct turnover present at a time lag of one year before climbing again at higher lags. This could be indicative of an underlying Kolmogorov medium.

\cite{you07} do not calculate a SF for J1744--1134, but predict it would be Kolmogorov based on previous $d\rm{DM}/\it{dt}$ measurements. Our calculation compares well to \cite{kei13}, starting out dominated by white noise then ends roughly Kolmogorov at higher lags. Our analysis finds a power spectrum below Kolmogorov.

PSRs J1614--2230, J2145--0750, J1909--3744, and J1944+0907 are roughly Kolmogorov.
PSR J1909--3744 compares well to the calculation in \cite{lam16}.
PSRs J1600--3053, J1713+0747, J1744--1134, and B1937+21 have power-law indices lower than expected for a Kolmogorov medium. 
PSR J0030+0451 has too large an uncertainty on $\beta$ for a definitive comparison.

The periods found by the SF fitting analysis agree within errors with those found by the DM time series trend analysis for all pulsars except one. A periodic trend was detected in the DM time series PSR B1855+09, but was not found by the SF analysis.

Here we discuss specific pulsars of interest.

\subsubsection{PSR J1713+0747}
\cite{you07} show the SF for J1713+0747 as being less steep than a quadratic power law at higher time lags, as does \cite{kei13}. The SF calculated here looks almost white noise dominated; removing the DM event occurring around MJD 54750 (2008--2009) and re-calculating the SF still yields what looks like a white noise dominated spectrum. We do not believe this is purely white noise because of the correlated structure we see in the time series. However, if that power is evenly distributed over the range of time lags we are concerned with then the SF will appear constant with lag. 

The DM event lasts $\sim$675 days; given the transverse velocity and distance from Table \ref{tab:DMV} and assuming the structure responsible is located at the pulsar gives an upper limit size to the dispersing region of 13.2$\pm$0.4 AU.

\subsubsection{PSR B1855+09}
PSR B1855+09 shows a very linear SF that does not align with the Kolmogorov or quadratic trendlines when using the diffractive timescale listed in Table \ref{tab:tdiff}. This could be due to an incorrect or varying diffractive timescale; adding the diffractive timescale as another fitted parameter when fitting the SF gives a value of $\Delta \tau_D=$ 55 $\pm$ 1 minutes, which is more than double the calculated values. The resulting SF is nearly quadratic with some white noise dominating at small time lags. 

\subsubsection{PSR B1937+21}
There is an extensive history of SF analysis for B1937+21. \cite{kas94} found a power law exponent of $\beta$=3.874$\pm$0.011 with a little more than 8 years of data. \cite{cor90} found a similar $\beta$ value that falls between a Kolmogorov and quadratic power spectrum, which agrees with our data at lower lags. \cite{ram06} finds a lower value of $\beta$=3.66$\pm$0.04 (compared to our 3.59$\pm$0.01) from 1983 to 2004 that is consistent with a Kolmogorov medium. 
As with previous studies, \cite{kei13} also shows a steady decrease in the DM through the end of 2010, and show a similar Kolmogorov-consistent SF. However, in 2011 the DM started to continuously increase through 2013, which is not a date range that any of the previously calculated SFs covered. It is likely that this latest increase in DM is the reason for the dip at higher lags that are not present in previous datasets.

\section{Discussion}

DM measurements can inform us about the free electron density along the LOS to a pulsar. In addition, trends due to the changing LOS over time aid in investigating structure in the ISM.
Linear trends may be caused by parallel or transverse motion when the free electron density may be changing to a higher or lower than average density in a region. \cite{lam16} show that DM variations due to a changing distance between Earth and the pulsar is dominated by parallel motion and that the transverse motion is negligible, entering only as a second order consideration.
The free electron density along a particular LOS is typically assumed to be temporally invariant. Examining the scintillation parameters and flux densities of MSPs exhibiting linear DM trends can inform if this is an accurate assumption for that particular LOS.
Annual trends may be due to a variety of solar effects and their amplitude is influenced by the relative velocity of the MSP when compared to the Earth's orbital motion as well as the Sun's velocity as it moves through the Galaxy.

Five MSPs show only annual trends and 14 show only linear trends, while 13 exhibit both trends. 
More than half of the MSPs showed significant DM variation beyond our measurement error over the timescale of one month to one year. There are 10 MSPs with timescales less than 31 days, which is on par with the average cadence of our observations. Of those, four MSPs have timescales of 14 days or less, which is the size of the fitting window used for DMX in the 9-year data release. It is therefore imperative that we fit for DM at every epoch due to the scale of the variations over these timescales, as well as observe as often as possible to minimize the time between DM measurements due to the rapid variation seen in some MSPs.

For three PSRs, the SFs appear to be dominated by white noise, resulting in a flat power spectrum.
Three MSPs have very nearly quadratic power spectra, with two (PSRs J1024--0719 and J2317+1439) having a $\beta$ value within 1\% of quadratic. \cite{lam16} suggest discrete structures in the ISM as well as the changing distance will contaminate the SF resulting in a quadratic power spectrum. This steeper than Kolmogorov value could be indicative that the time lag is smaller than the crossing time for the structure probed during the time series. Higher values than consistent with a Kolmogorov medium could be attributed to present trends or systematic variations in addition to a Kolmogorov medium, particularly with the previously discussed difficulties in disentangling a linear trend component from a stochastic one. We cannot impose priors without assuming something about the contributions from the ISM that we are trying to constrain, and which may not actually be the case.
There are also more possible sources of error in the calculation of the 
$\beta$ values than we have included. We have accounted for the random and stochastic uncertainties but not systematic uncertainties, which can result from variability of white noise statistics over time from changing backends, the variation in the diffractive timescale, and the fact that the models used here could be incomplete in describing the ISM. 
Therefore, while the $\beta$ values presented here are illustrative,  their errors bars are likely under-estimated. In addition, care should be taken when using these values to make inferences about the ISM due to possible covariances and systematics present.


\vskip0.35cm
The NANOGrav project received support from the National Science Foundation (NSF) PIRE program award number 0968296 and NSF Physics Frontier Center award number 1430284. MLJ acknowledges support from West Virginia University through the STEM Mountains of Excellence Fellowship. NANOGrav research at UBC is supported by an NSERC Discovery Grant and Discovery Accelerator Supplement and the Canadian Institute for Advanced Research. JAE acknowledges support by NASA through Einstein Fellowship grant PF3-140116. Portions of this research were carried out at the Jet Propulsion Laboratory, California Institute of Technology, under a contract with the National Aeronautics and Space Administration. TTP was a student at the National Radio Astronomy Observatory (NRAO) while this project was undertaken. Data for the project were collected using the facilities of the NRAO and the Arecibo Observatory. The NRAO is a facility of the NSF operated under cooperative agreement by Associated Universities, Inc. The Arecibo Observatory is operated by SRI International under a cooperative agreement with the NSF (AST-1100968), and in alliance with the Ana G. M\'{e}ndez-Universidad Metropolitana, and the Universities Space Research Association.

{\it Author contributions.} MLJ carried out the analysis and prepared the majority of the text, figures, and tables. MAM, MTL, JMC, LL and SC helped with the development of the framework and text. ZA, KC, PBD, TD, JAE, RDF, EF, MEG, GJ, MLJ, MTL, LL, MAM, DJN, TTP, SMR, IHS, KS, JKS, and WWZ all ran observations and developed timing models for the NG9 data set; additional specific contributions are outlined in \cite{arz15}. 

\bibliography{dm}{}

\end{document}

%% file: Table1.tex
\begin{table*}
\begin{center}
\caption{Properties of NANOGrav MSPs in the 9-Year Data Release.}
\label{tab:DMV}
\begin{tabular}{lllllllllll}
\hline 
\hline
PSR & $\mu_{\lambda}$ & $\mu_{\beta}$ & $\mu_{\alpha}$ & $\mu_{\delta}$ & DM  &  d$_{\textrm{DM}}$ & d$_{\textrm{PX}}$ & $\chi_r^2$ & V$_T$ \\
& (mas~yr$^{-1}$) & (mas~yr$^{-1}$)& (mas~yr$^{-1}$) & (mas~yr$^{-1}$) & (pc~cm$^{-3}$) & (kpc) & (kpc) &  & (km~s$^{-1}$) \\
	\hline	
J0023+0923	& --13.9(2)	& --1(1) & --12.3(6) & --6.7(9)	& 14.3 	& 0.7(2) & --- 		& 4.2 & 46(13)\\    
J0030+0451	& --5.52(1) & 3.0(5) & --6.3(2) & 0.6(5) & 4.3  	&  0.3(1)& 0.30(2) 	& 11 & 8.9(7) \\
J0340+4130	& --2.4(8) 	& --4(1) &--1.3(7) & --5(1)	& 49.6 	&  1.7(4)&---		& 6.8 & 38(12) \\    
J0613--0200	& 2.12(2) 	& --10.34(4)& 1.85(2) & --10.39(4) & 38.8 	& 1.7(4) & 1.1(2) 	& 70 & 55(10)\\
J0645+5158	& 2.1(1) 	& --7.3(2) & 1.4(1) & --7.5(2)	& 18.2 	& 0.7(2) & 0.8(3) 	& 2.7 & 29(11) \\
 & & &  \\
J0931--1902	& ---		& --- & ---		& ---		& 41.5 	& 1.8(5) & --- 		& 2.2 & --- \\  
J1012+5307	& 13.9(1) 	& --21.7(3) & 2.5(2) & --25.6(2) & 9.0 	& 0.4(1) & ---		& 1.8 & 49(12) \\
J1024--0719	& --14.36(6)& --57.8(3) & --35.2(1) & --48.0(2) & 6.5 	& 0.4(1) &---		& 15 & 113(28) \\   
J1455--3330	& 8.16(7) 	& 0.5(3) &7.9(1) & --2.0(3)	&  13.6 & 0.5(1) & --- 	& 2.4 & 19(4) \\
J1600--3053	& 0.47(2) 	& --7.0(1) & --0.95(3) & --7.0(1) & 52.3 	& 1.6(4) & 3.0(8) 	& 42 & 100(27)  \\
 & & & \\
J1614--2230	& 9.46(2) 	& --31(1) & 3.8(2) & --32(1) &  34.5 & 1.3(3) & 0.65(5) 	& 20 & 100(8) \\  
J1640+2224	& 4.20(1)	& --10.73(2) & 2.09(1) & --11.33(2) & 18.5 	& 1.2(3) & --- 		& 295 & 66(16) \\
J1643--1224	& 5.56(8) 	& 5.3(5) & 6.2(1) & 4.5(5)	& 62.4  & 2.3(6) & ---		& 112 & 84(22) \\
J1713+0747	& 5.260(2) 	& --3.442(5) & 4.918(2) & --3.914(5) & 16.0 	& 0.9(2) & 1.18(4) 	& 29 & 35(1) \\
J1738+0333	& 6.6(2) 	& 6.0(4) & 6.9(2) & 5.8(4)	&  33.8 & 1.4(4) & --- 		& 5.0 & 59(17) \\ 
 & & &  \\
J1741+1351	& --8.8(1) 	& --7.6(2) & --9.1(1) & --7.2(2) & 24.2 	& 0.9(2) & ---		& 4.7 & 50(11)\\ 
J1744--1134	& 19.01(2) 	& --8.68(8) & 18.76(2) & --9.20(8) & 3.1 	& 0.4(1) & 0.41(2) 	& 17 & 41(2)\\
J1747--4036	& 0.1(8)		& --6(1) & 0(1) & --6(1)	& 153.0 & 3.3(8) & --- 		& 133 & ---& \\
J1832--0836	& --- 		& --- 	& --- 		& --- 	& 28.2 	& 1.1(3) &  --- 	& 16 & ---  \\
J1853+1303	& --1.8(2)  & --2.9(4) & --1.48(2) & --3.1(4)	& 30.6 	& 2.0(5) & ---		& 5.8 & 32(9) \\
 & & &  \\
B1855+09	& --3.27(1) & --5.10(3) & --2.651(15) & --5.45(3) &  13.3 & 1.2(3) & ---		& 1335 & 34(9) \\
J1903+0327	& --3.5(3) 	& --6.2(9) & --2.7(3) & --6.5(9) & 297.6 & 6(2)   & --- 		& 27 & 202(71) \\
J1909--3744	& --13.868(4)& --34.34(2) & --9.518(4) & --35.79(2) & 10.4 & 0.5(1) & 1.07(4) 	& 1375 & 188(7)\\  
J1910+1256	& --0.7(1) 	& --7.2(2) & 0.3(1) & --7.2(2)	&  38.1 & 2.3(6) & --- 		& 2.7 & 79(21) \\
J1918--0642	& --7.93(2) & --4.85(9) & --7.18(3) & --5.90(9) & 26.6 	& 1.2(3) & 0.9(2) 	& 171 & 40(9) \\ 
 & & &  \\
J1923+2515	& --9.5(2) 	& --12.8(5) & --6.6(2) & --14.5(5) & 18.9 	& 1.6(4) &---		& 0.9 & 121(30) \\
B1937+21	& --0.02(1) & --0.41(2) & 0.07(1)& --0.40(2) & 71.0 & 3.6(7) &---		& 1162 & 7(1) \\
J1944+0907	& 9.4(1) 	& --25.5(4)& 14.37(11) & --23.1(4) & 24.3  & 1.8(5) &  --- 	& 147 & 232(64)\\
J1949+3106	& 13(15)	&  10(13)	&	10(11)	& 13(16) & 164.1 & 3.6(9) & --- 		& 1.4 & --- \\
B1953+29	& --1.8(9) 	& --4.4(14) & --0.4(12) & --5(1) & 104.5 & 5(1) & --- 		& 6.6 & 113(24) \\
 & & &\\
J2010--1323	& 1.16(4)	& --7.3(4) & 2.71(9) & --6.9(4)	& 22.2 	& 1.0(3) & --- 		& 70 & 35(11) \\
J2017+0603	& 2.3(6) 	& --0.1(7)	& 2.2(7) & 0.5(6)	& 23.9 &  1.6(4)& ---		& 2.8 & --- \\
J2043+1711	& --8.97(7) & --8.5(1) & --5.85(7) & --10.9(1) & 20.7 	& 1.7(4) & 1.3(4) 	& 6.3 & 100(23)\\ 
J2145--0750	& --12.04(4)& --3.7(4) & --10.1(1) & --7.5(4) 	& 9.0 	& 0.6(2) & 0.8(2) 	& 24 & 48(12)\\
J2214+3000	& 17.1(5) 	& --10.5(9) & 20.0(6) & --1.7(8) & 22.6 	& 1.5(4) &  --- 	& 1.0 & 143(38) \\ 
 & & &  \\
J2302+4442	& --3.3(6) 	& --1(2)	& --2(1) & --3(2)	&  13.7 &  1.1(3)& --- 		& 1.5 & --- \\
J2317+1439	& 0.19(2)	& 3.80(7) & --1.39(3) & 3.55(6)	& 21.9 	& 0.8(2) & --- 		& 18732 & 14(4) \\
	\hline
	\hline
  \end{tabular}
  \end{center}
{\small {\bf Notes.} Columns are pulsar name, ecliptic proper motion (latitude and longitude), proper motion in RA and Dec, the DM, the DM-derived distance, the parallax-derived distance, the reduced chi-squared of the DM time series, and the transverse velocity. The ecliptic proper motions and DMs were calculated for the 9-year data release \citep{arz15}. Proper motion in RA and Dec as well as parallax distances were calculated through timing observations and discussed in \cite{mat16}. Proper motion values that are smaller than their uncertainties were not used in subsequent analysis. The DM derived distances were calculated from the NE2001 model assuming 20\% error (Cordes \& Lazio 2002). The value for $\chi_r^2$ was calculated using Equation \ref{chi}.  Dashes indicate that no significant measurement was possible. The transverse velocity V$_T$ was calculated from the proper motion and the distance estimate with the smaller error (i.e. d$_{\rm{DM}}$ or d$_{\rm{PX}}$).}
\end{table*}

%% file: Table2.tex
\begin{table*}
  \begin{center}
  \caption{Fitted Trends in the DM Time Series for MSPs in the 9-Year Release}
  \label{tab:Trends}
  \begin{tabular}{llllllllllll}
	\hline 
	\hline
	PSR &  Trend & $d\rm{DM}/\it{dt}$  &  Amplitude & Period & $\chi^2_r$ & P$_{\rm{LS}}$ & FAP & Length & $\delta t$\\
		   && (10$^{-3}$~pc~cm$^{-3}$~yr$^{-1}$) & (10$^{-4}$~pc~cm$^{-3}$) &  (days) & & (days) & (\%) & (days) & (days)\\
	\hline	
J0023+0923     & None & --- & --- & --- & --- & --- & --- & 841 & ---\\    
J0030+0451   & Periodic & --- & 1.2(3) & 373(5) & 9.2 & 371 & 6.3 & 3204 & 33\\
J0340+4130   & Linear & 0.88(9) & --- & --- & 1.7 & 241 & 6.3 & 613 & 73\\  
J0645+5158   & Periodic & --- & 0.9(3) & 377(29) & 2.6 & 199 & 0.54 & 881 & 78\\
J0931--1902   & None & --- & ---&---  &--- & --- & --- & 235 & ---  \\  
J1012+5307   & Linear & 0.11(2) & ---& --- & 0.94 & --- & --- & 3368 & 2286\\
J1024--0719  & Linear & 0.39(2) & --- & --- & 2.7 & ---& --- & 1467 & 148 \\  
J1455--3330  & Linear & 0.15(2) & ---& --- & 1.0 & 361 & 7.3 & 3368 & 904 \\
J1614--2230  & Periodic & ---  & 3.1(5) & 370(9) & 11 & 370 & 0.17 & 1860 & 14\\  
J1643--1224 & Both & --1.02(3) & 8(1) & 387(4) & 10 & 387 & 0.01 & 3293 & 104\\
J1713+0747   & None & --- & ---& --- & --- & ---& --- & 3199 & ---\\ 
J1738+0333   & Linear & --0.8(2) & ---& --- & 1.4 & ---& --- & 1456 & 213\\ 
J1741+1351   & Linear & --0.12(4) & ---& --- & 1.9 & ---& --- & 1224 & 287\\ 
J1744--1134   & Both & --0.069(7) & 0.4(2) & 383(16) & 8.0 & ---& --- & 3369 & 66 \\
J1747--4036  & Linear & --7.3(4) & --- & --- & 10.0 & 459 & 6.7 & 608 & 16\\ 
J1832--0836   & None  & --- & --- & --- & ---& ---& --- & 231 & ---\\ 
J1853+1303     & Linear & 0.12(9) & --- & --- & 5.2 & ---& --- & 1468 & 361\\
B1855+09     & Both & 0.382(7) & 0.5(3) & 364(11) & 15.7 & ---& --- & 3240 & 27 \\ 
J1903+0327   & Both & --3.0(4) & 31(6) & 375(11) & 12 & 371 & 1.0 & 1456 & 99\\
J1909--3744  & Both & --0.239(4) & 0.7(1) & 366(5) & 28 & 366 & 0.25 & 3306 & 9\\  
J1910+1256   & Linear & 0.51(6) & --- & --- & 0.90 & 404 & 2.9 & 2574 & 443\\
J1923+2515    & None &--- & ---& ---&--- & ---& --- & 803 & ---\\ 
J1944+0907    & Linear & 1.3(2) & --- & --- & 44 & ---& --- & 1467 & 43\\
J1949+3106    & Periodic & --- & 10(3) & 391(37) & 1.0 & ---& --- & 455 & 51 \\ 
B1953+29      & Both & --1.3(3) & 3(2) & 356(72)& 2.1 & --- & ---& 1967 & 136\\
J2010--1323    & Both & 0.38(2) & 2.2(4) & 372(9) & 14 & 372 & 0.56 & 1490 & 16\\
J2017+0603    & Both & 0.23(7) & 2.3(5) & 440(37) & 0.94 & ---& ---& 609 & 38\\
J2043+1711   &  Both & --0.12(4) & 1.0(4) & 390(38) & 3.7 & ---& --- & 834 & 189\\
J2145--0750  & Linear & 0.08(2) & --- & --- & 18 & ---& ---& 3318 & 568\\
J2214+3000    & Periodic & --- & 4(1) & 319(25) & 0.83 & --- & --- & 755 & 17 \\
J2302+4442    & Linear & --0.6(2) & --- & --- & 1.5 & --- & --- & 613 & 202 \\
J2317+1439    & Both & --0.550(9) & 0.9(3) & 311(6) & 321 & ---& --- & 3243 & 5\\ 
	\hline
	\hline
  \end{tabular}
  \end{center}
  {\small {\bf Notes.} Results of fitting periodic and linear trends to the DM variations, where 1$\sigma$ uncertainty in the last significant digit is expressed in parentheses. Columns are the detected trend, the slope, the amplitude of the periodic fit, the period of the fit, the reduced chi-squared after the fitting, the period found by the Lomb-Scargle periodogram, the false alarm probability for that period, the length of the data span for that pulsar, and the average time it takes the DM time series to vary by $1\sigma$. Pulsars were not listed here if no significant trends were found. Values and corresponding errors were found by implementing a non-linear least squares Marquardt-Levenberg algorithm. Variations were determined to exhibit a trend if the post-fit $\chi^2_r$ was lower than the pre-fit $\chi^2_r$ value. The period found by the Lomb-Scargle periodogram analysis, P$_{\rm{LS}}$, was omitted if it corresponded to the length of the dataset or the cadence of observations. The resolution of the Lomb-Scargle analysis (and hence error on P$_{\rm{LS}}$) is equal to the cadence of the observations for that particular pulsar. Those MSPs that are identified as having a linear trend had that trend subtracted off prior to applying the periodogram analysis.} 
\end{table*}

%% file: Table3.tex
\begin{table*}
  \begin{center}
  \caption{Pulsars with Piecewise Trends in the DM Variations from the 9-Year Release}
  \label{tab:PartialTrends}
  \begin{tabular}{llllllllllll}
	\hline 
	\hline
	PSR &  Trend & $d\rm{DM}/\it{dt}$  &  Amplitude & Period & $\chi^2_r$  & Length & Start & End & $\delta t$\\
		   && (10$^{-3}$~pc~cm$^{-3}$~yr$^{-1}$) & (10$^{-4}$~pc~cm$^{-3}$) &  (days) &   & (days) & (MJD) & (MJD) & (days) \\
	\hline	

J0613--0200  & Both & 0.066(7) & 1.8(1) & 358(4) & 0.75  & 3137 & 53448 & 54970 & 18\\
			 & Both & 0.161(7) & 1.2(2) & 352(5) & 3.5 & ---  & 54970 & 56380 & 23\\
J1600--3053  & Linear & --0.73(4) & --- & --- & 2.8  & 2184 & 54400 & 55300 & 40\\
			 & None & --- & --- & --- & --- & --- &  --- & --- & ---\\
J1640+2224   & Linear & 0.145(3) & --- & --- & 7.0  & 3254 & 53344 & 55850 & 78\\
			 & None  & --- & --- & --- & --- & ---  & --- & --- & ---\\ 
J1918--0642  & Both & --0.49(1) &  1.2(4) & 385(11) & 4.3 & 3293 & 53292 & 56000 & 24\\ 
			 & Both & 0.23(3) & 1.2(3) & 541(47) & 2.9  & 3293 & 56000 & 56585 & 31\\
B1937+21     & Both & --0.34(3) & 3.2(4) & 395(11) & 28  & 3327 & 53267 & 54550 & 5\\
			 & Both & 0.050(3) & 3.7(2) & 469(14) & 10 & 3327 & 55970 & 56594 & 5\\

	\hline
	\hline
  \end{tabular}
  \end{center}
  {\small {\bf Notes.} For notes on columns, see Table \ref{tab:Trends} The last column gives the final start and end dates used in the fit. The Lomb-Scargle periodogram found a period for PSR J0613$-$0200 of 371 days with a FAP of 5.4\%.} 
\end{table*}

%% file: Figures1to5.tex
\begin{figure*}
	\begin{center}
	
		\includegraphics[width=0.49\textwidth]{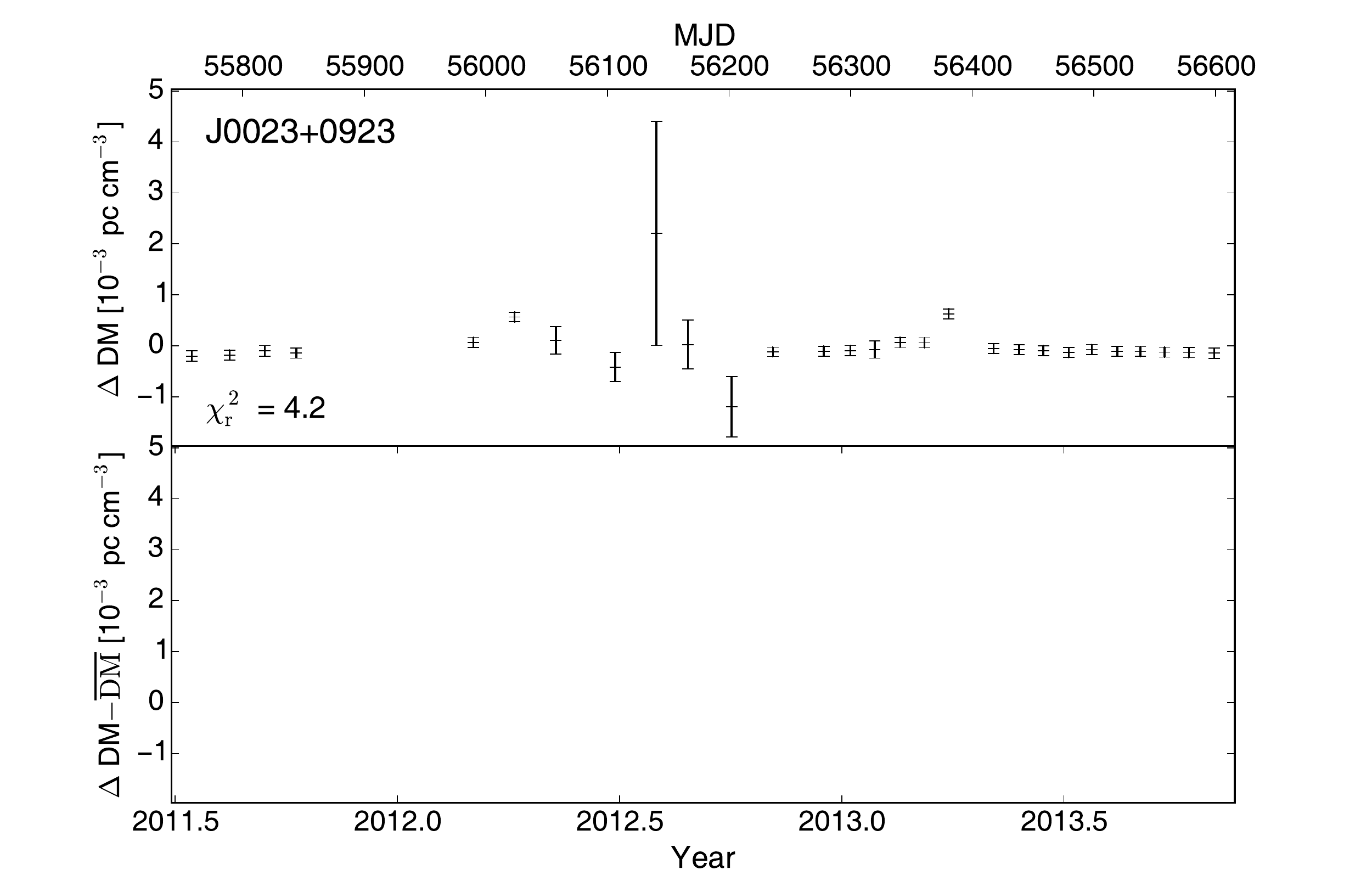}
		\includegraphics[width=0.49\textwidth]{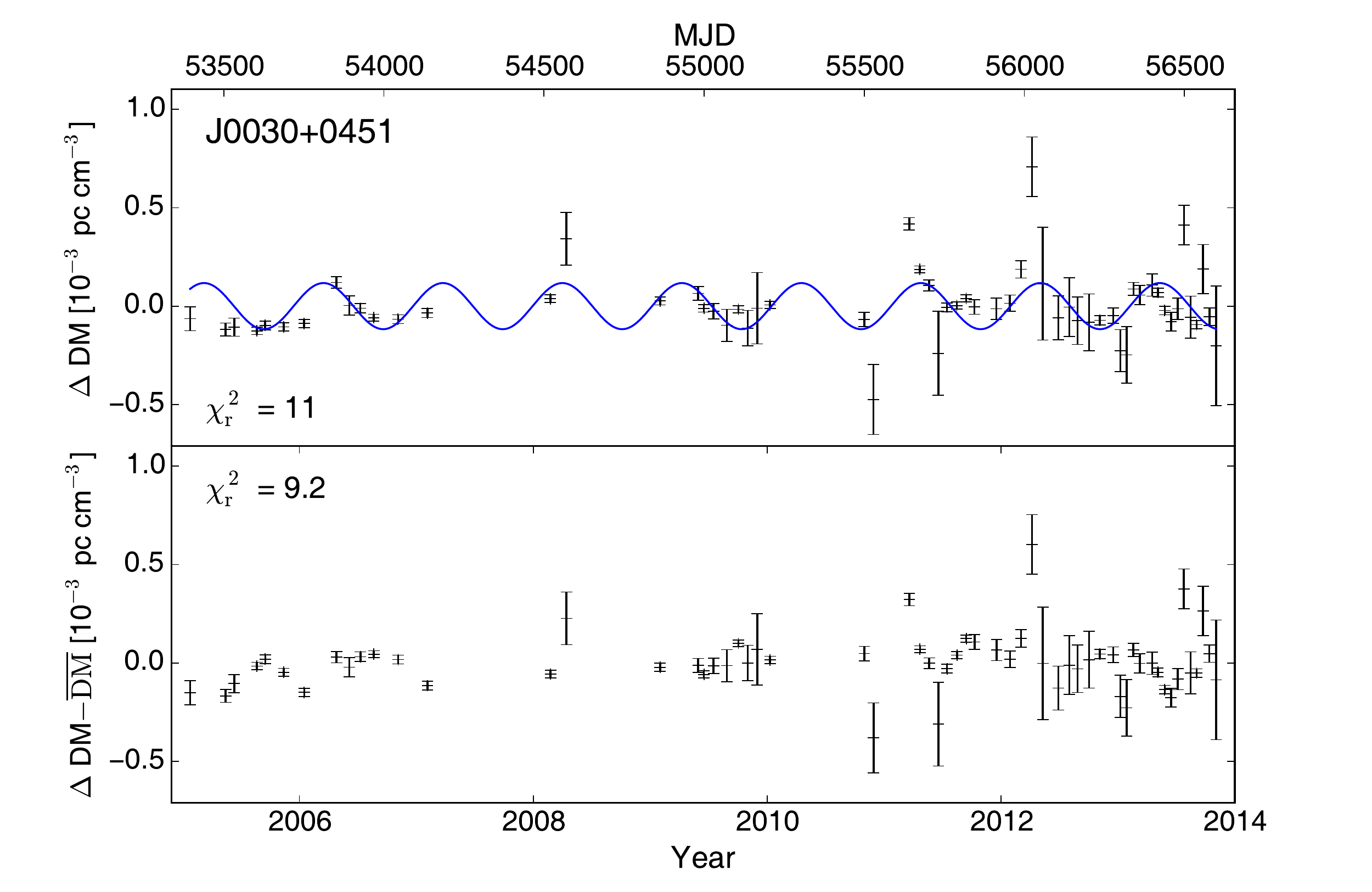}
		\includegraphics[width=0.49\textwidth]{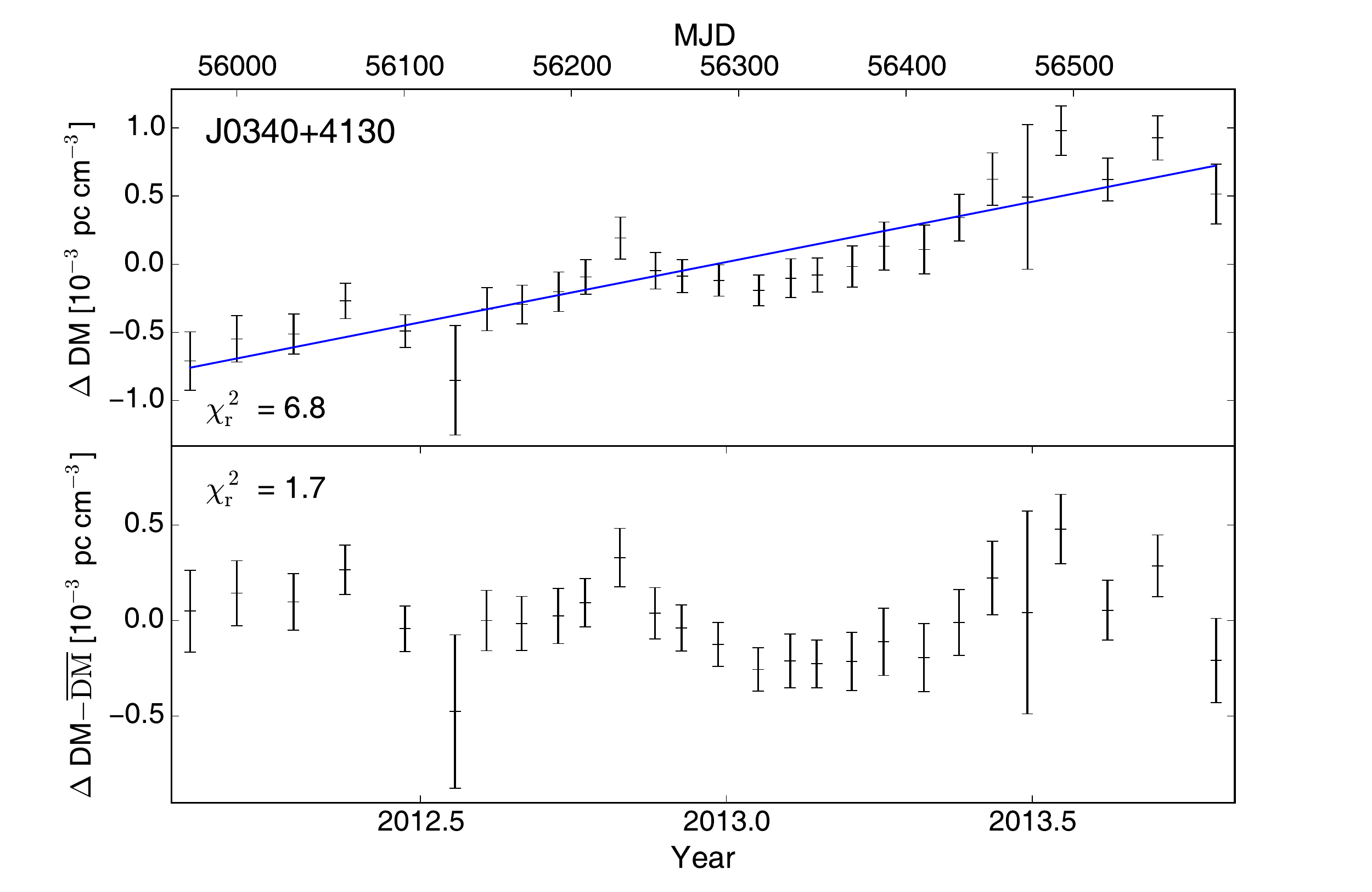}
		\includegraphics[width=0.49\textwidth]{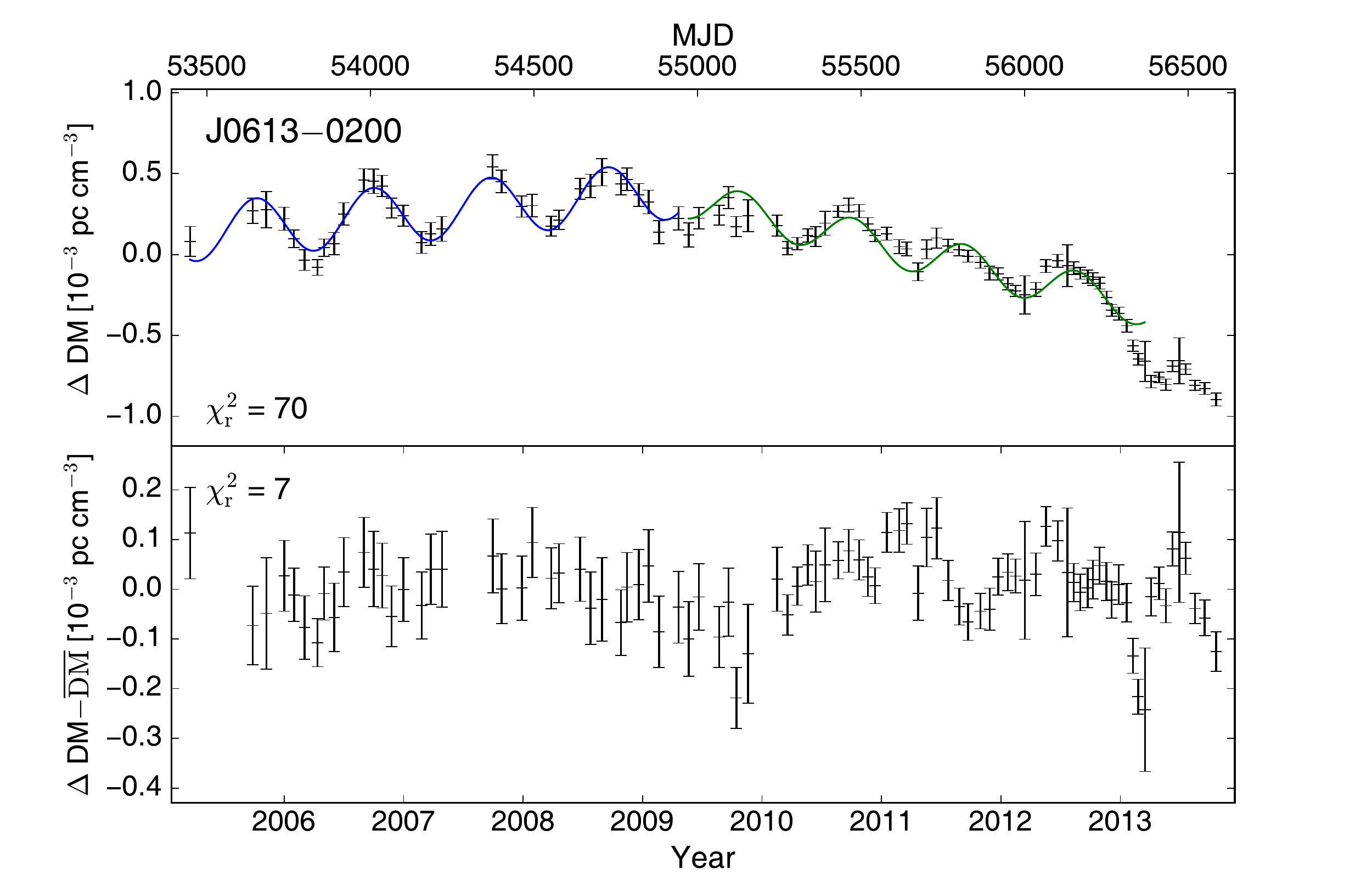}
		\includegraphics[width=0.49\textwidth]{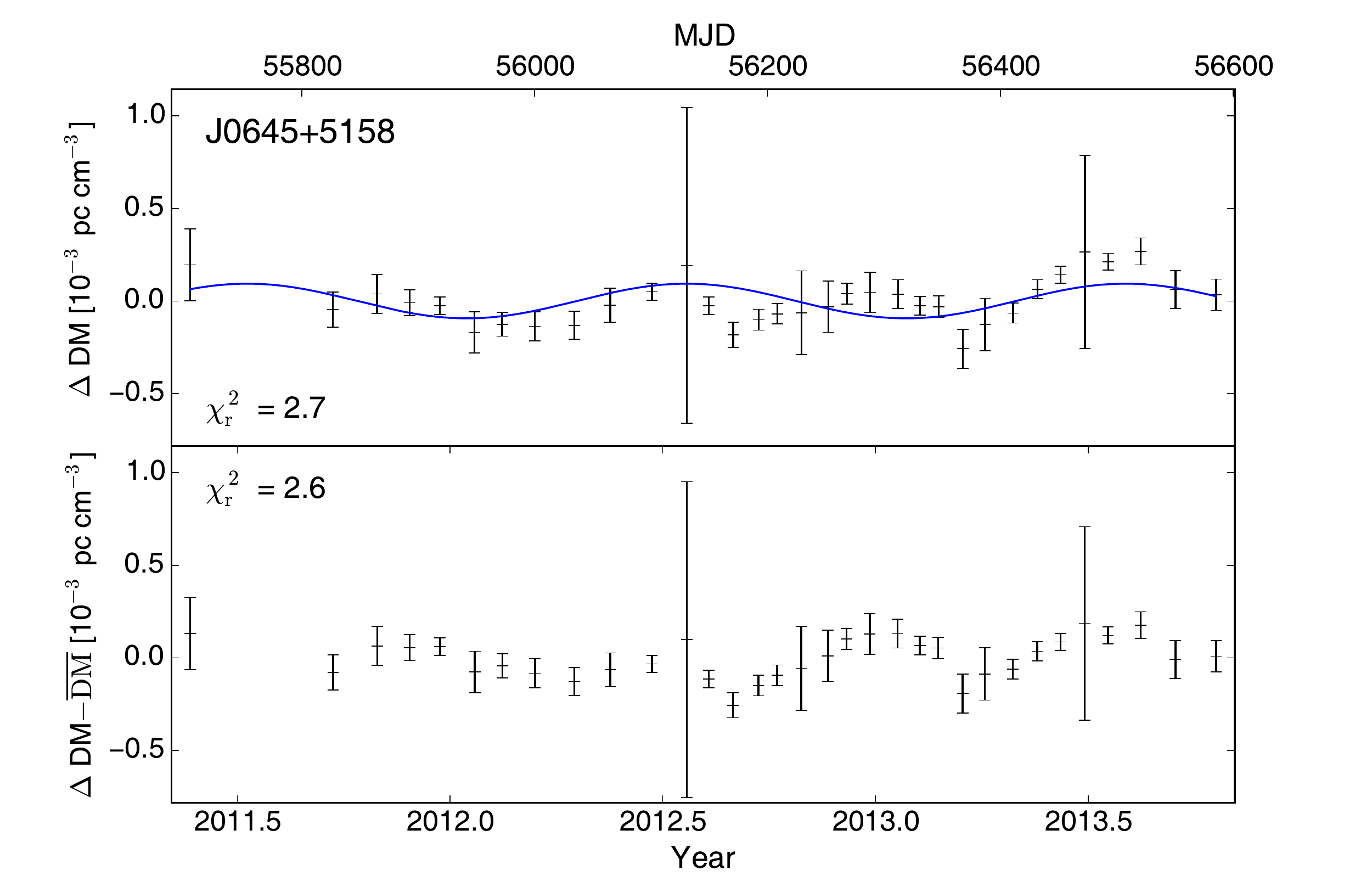}	
		\includegraphics[width=0.49\textwidth]{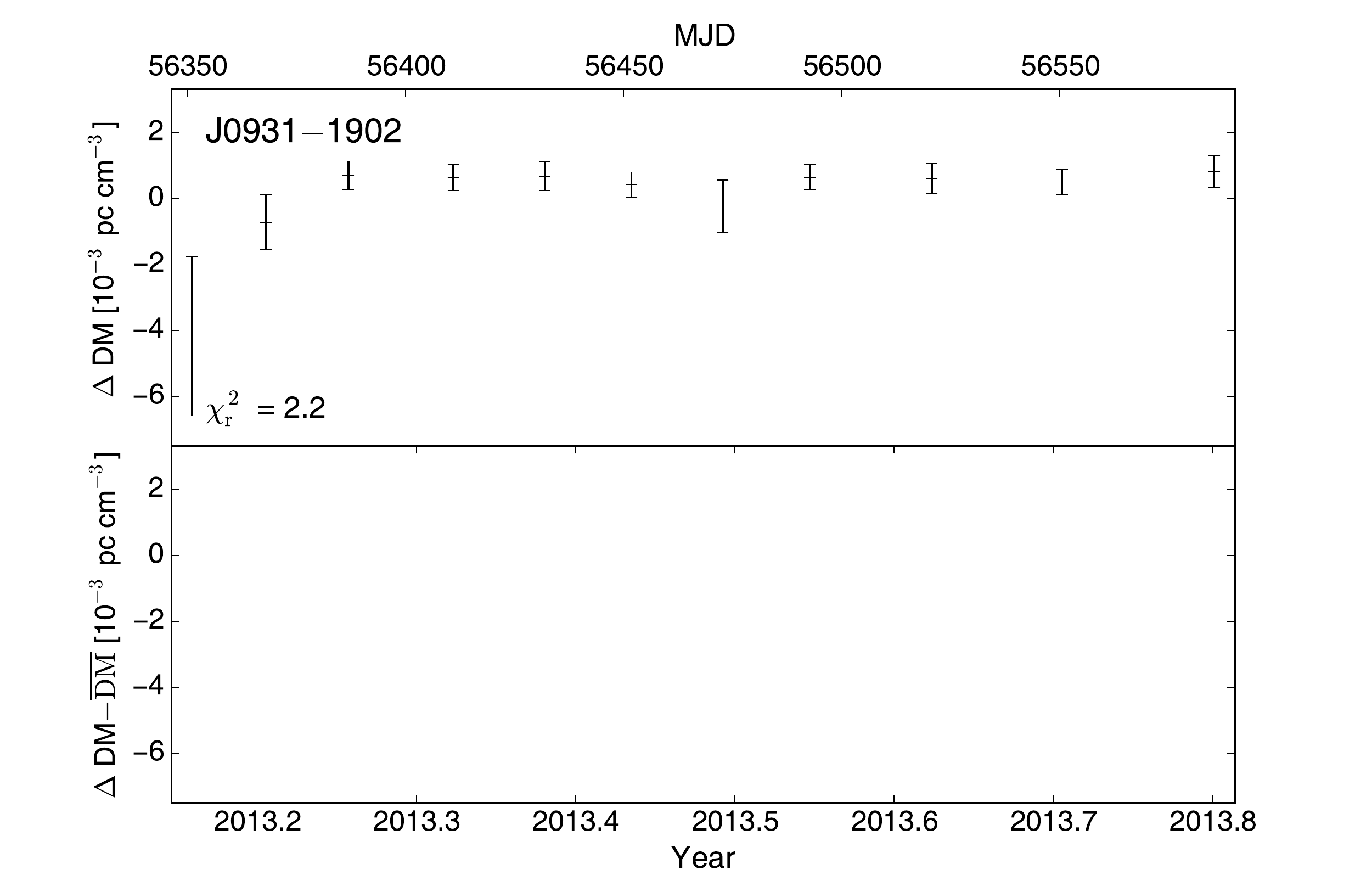}
		\includegraphics[width=0.49\textwidth]{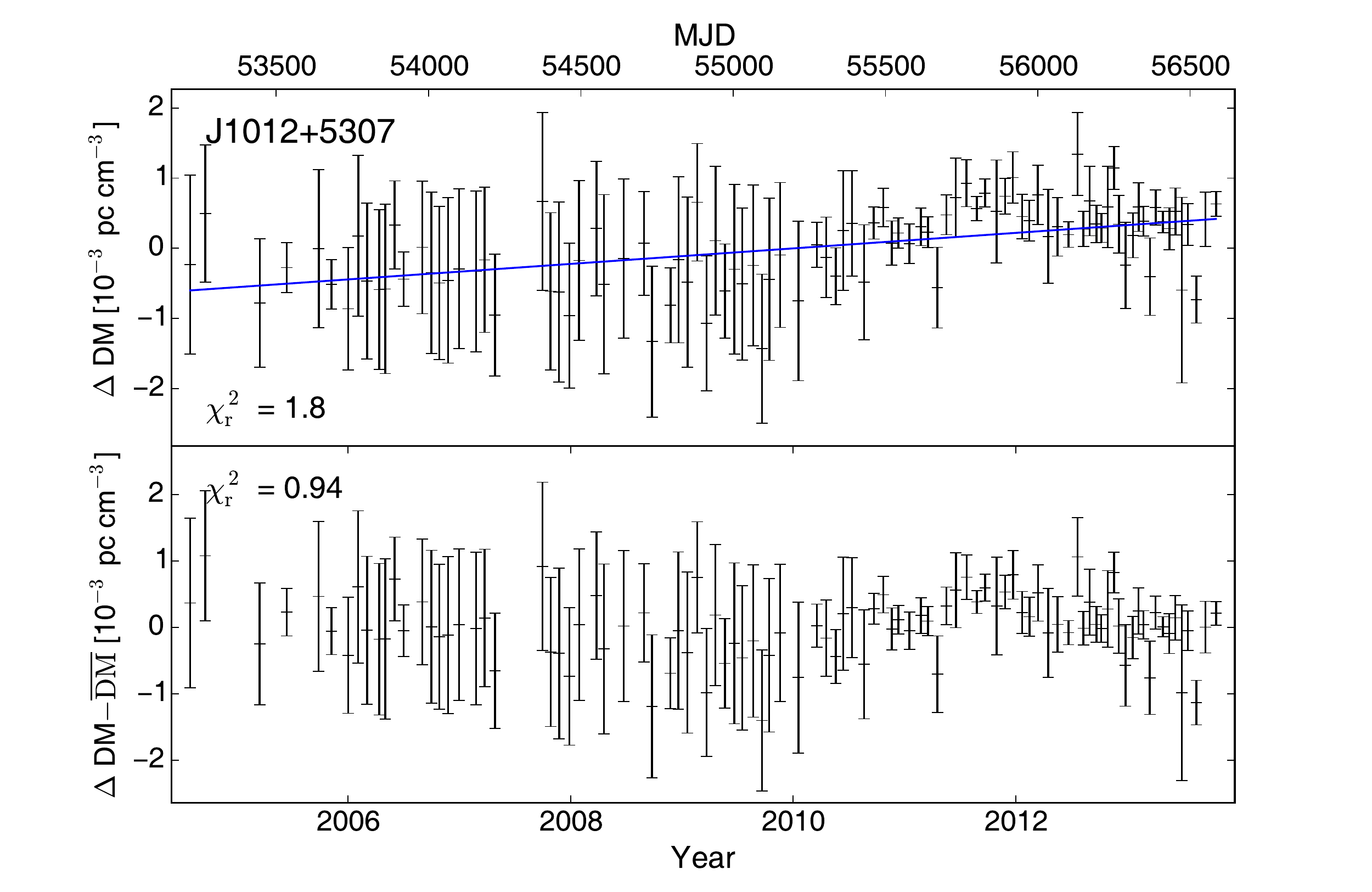}
		\includegraphics[width=0.49\textwidth]{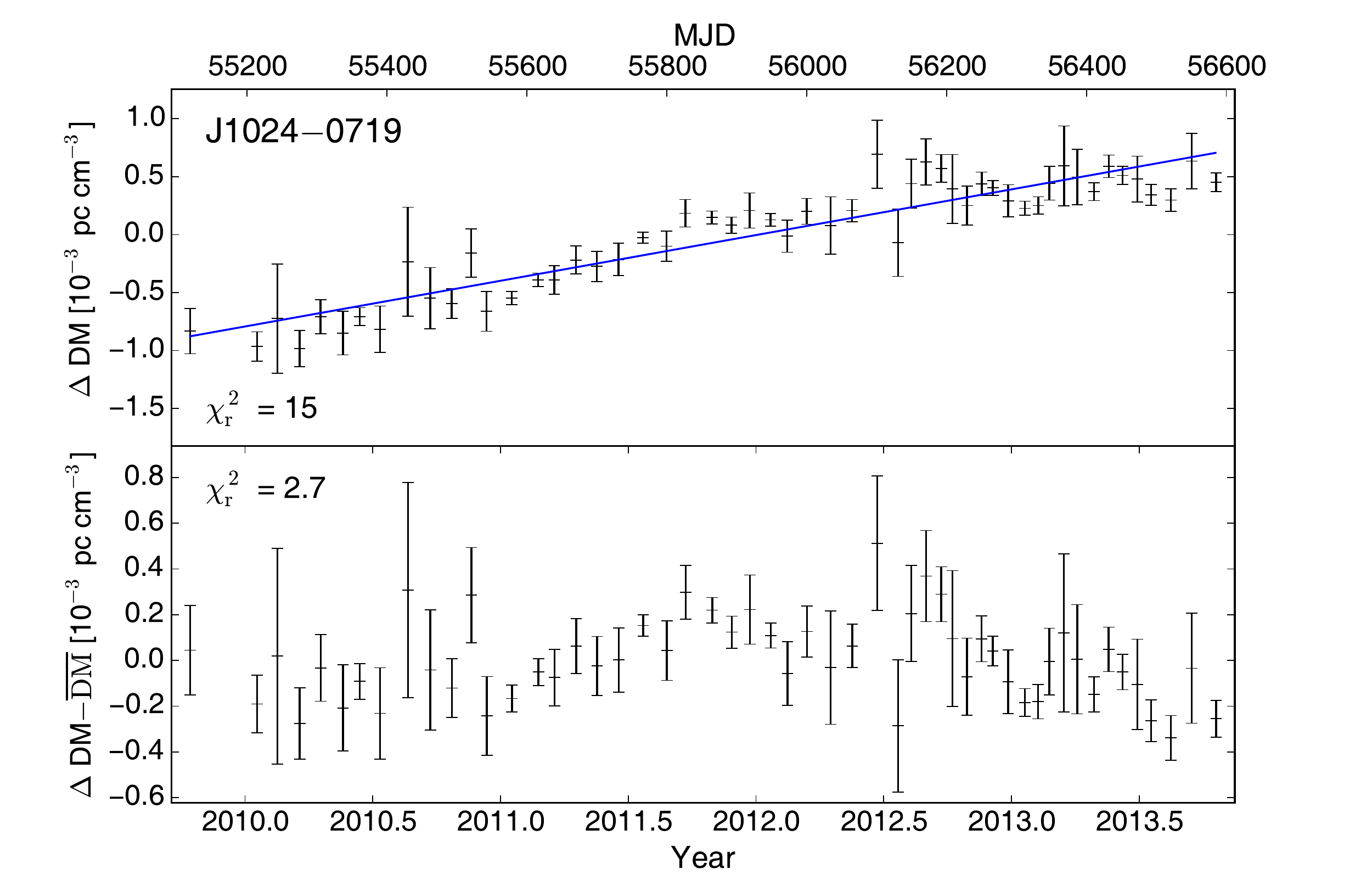}
						
	\caption{The top panel shows the DM time series with the best fit function (if applicable). The zero point for the DM variations corresponds to the fiducial DM for the data span. The error bars are $\pm1\sigma$ errors returned by TEMPO. The bottom panel shows the DM residuals after the trend has been removed from the time series; empty panels suggest no trend was found for that pulsar. The $\chi^2_r$ values before and after these fits for each pulsar appear in the top and bottom panels respectively, as well as in Tables \ref{tab:DMV} and \ref{tab:Trends}. PSR J0931--1902 has too short a data span for a trend to be determined.}
		\label{fig:trendremoval1}		
	\end{center}
\end{figure*}		
		
\begin{figure*}
	\begin{center}
		\includegraphics[width=0.49\textwidth]{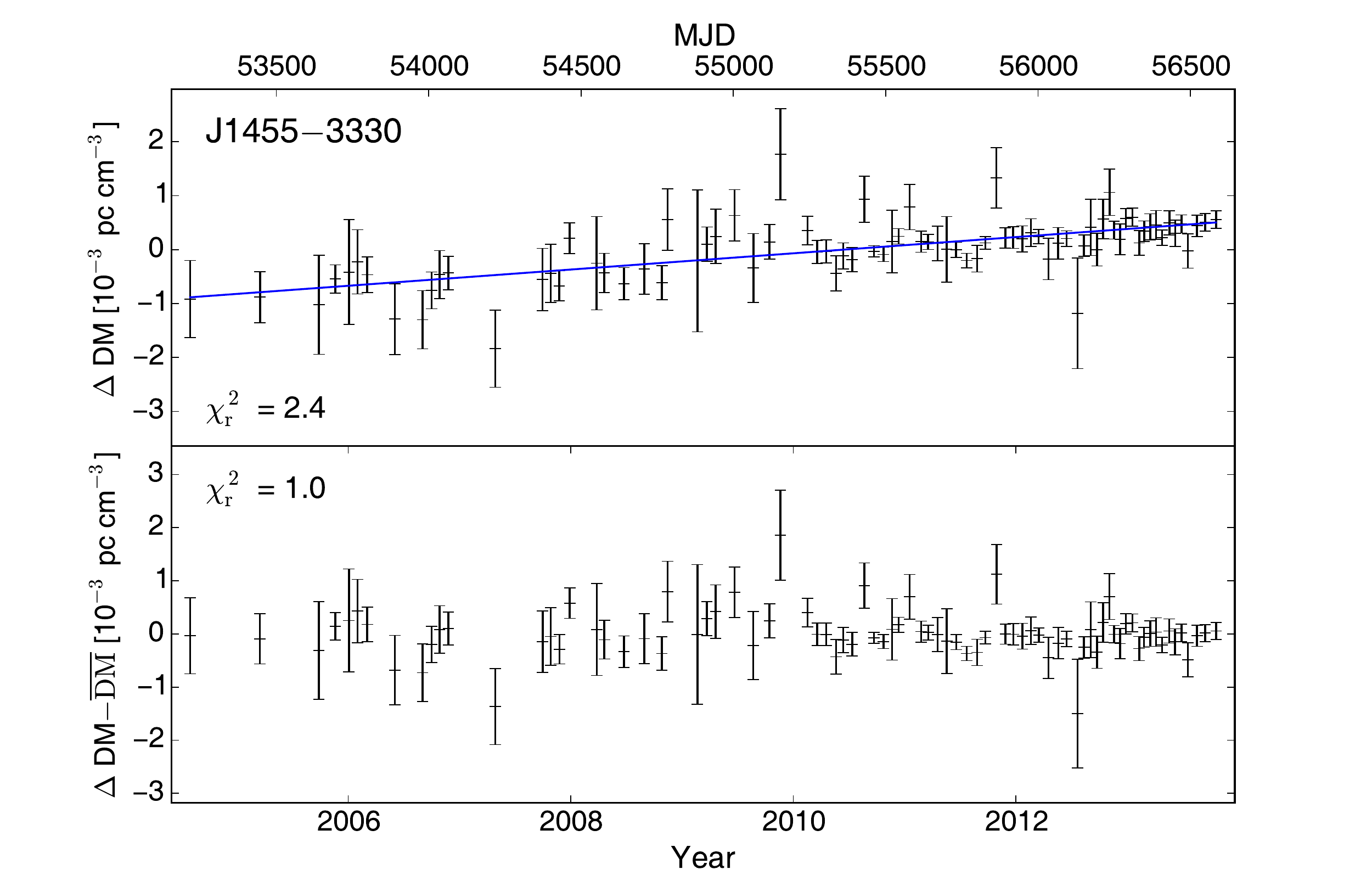}
		\includegraphics[width=0.49\textwidth]{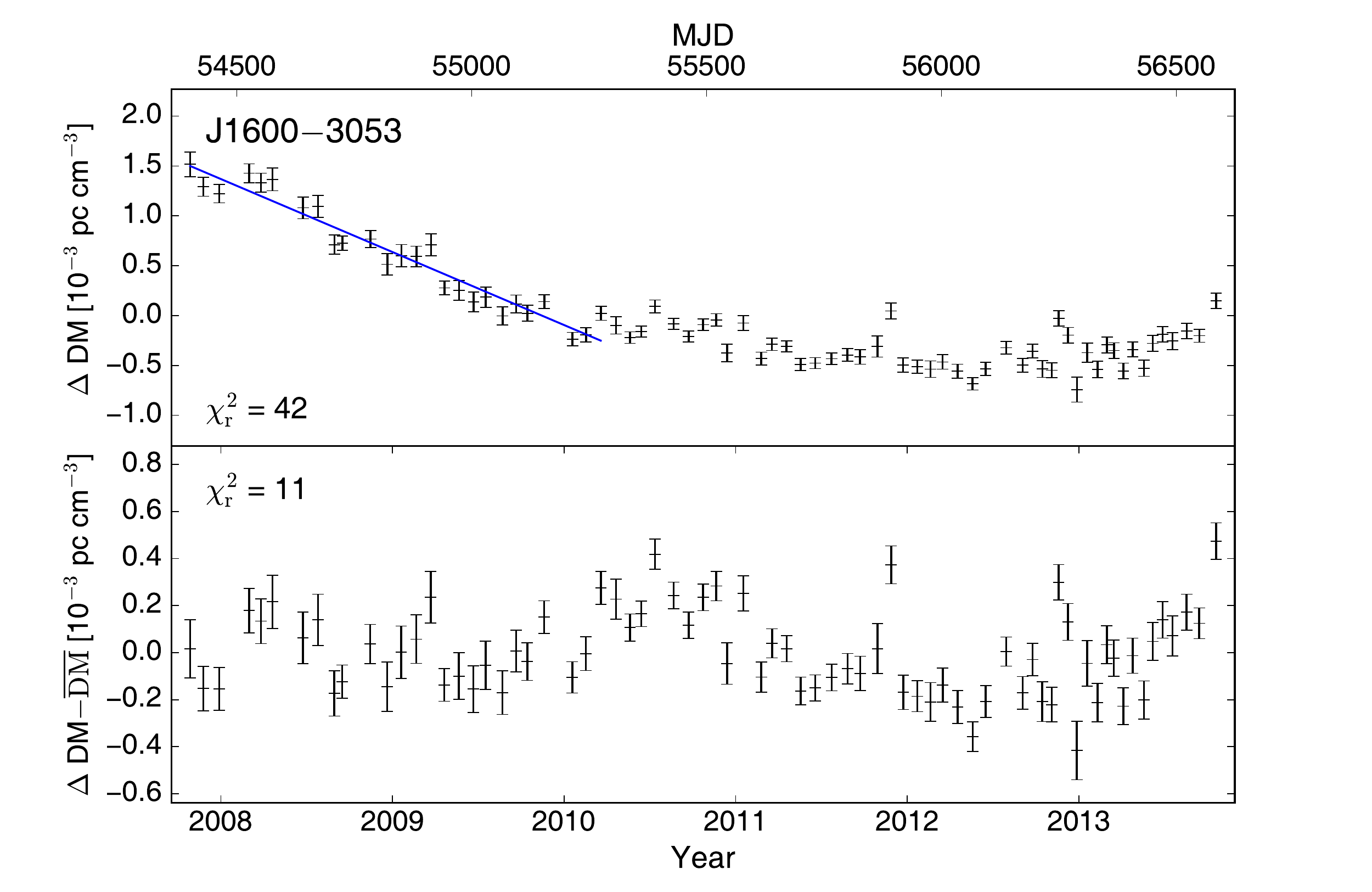}	
		\includegraphics[width=0.49\textwidth]{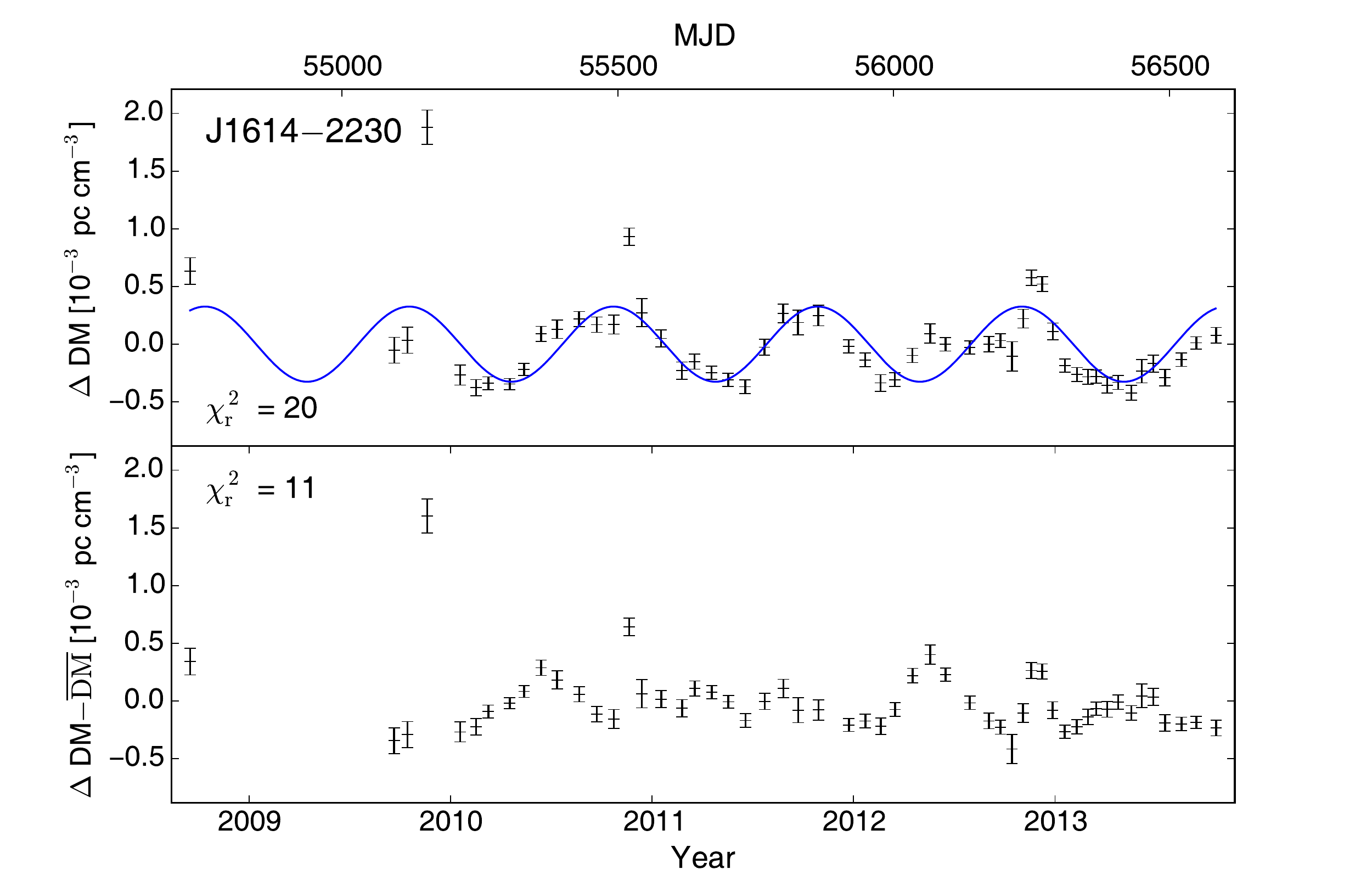}	
		\includegraphics[width=0.49\textwidth]{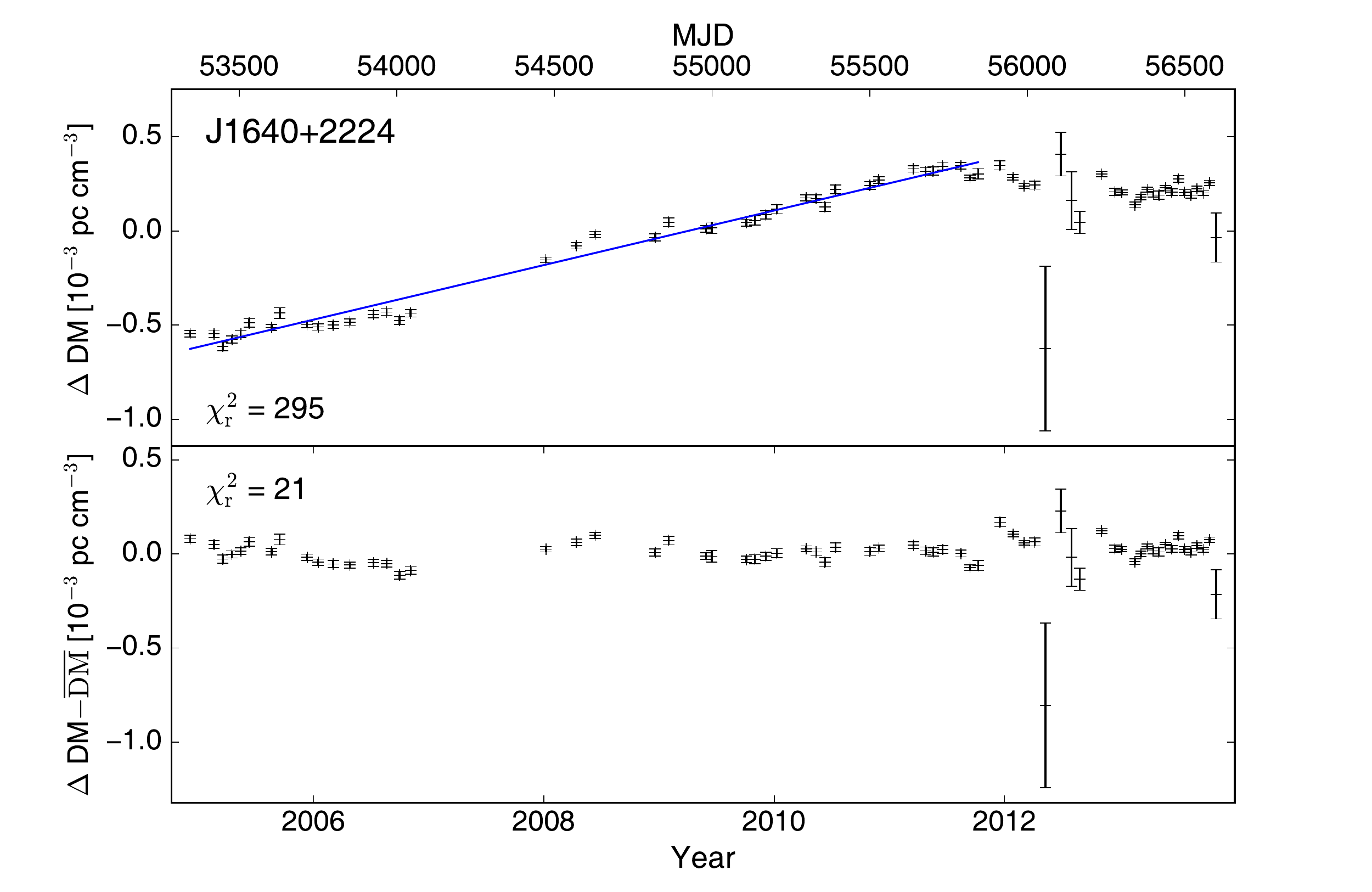}
		\includegraphics[width=0.49\textwidth]{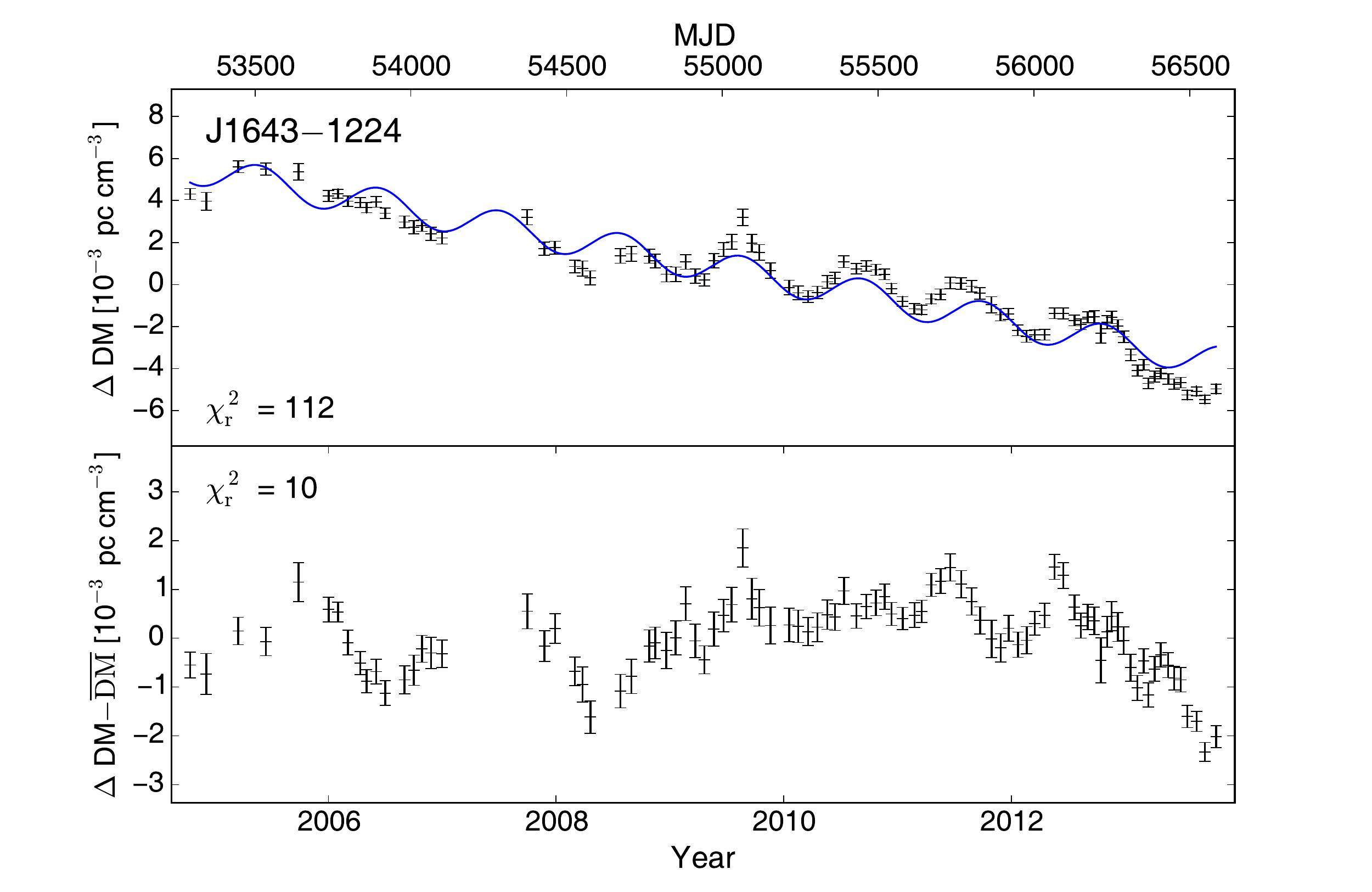}
		\includegraphics[width=0.49\textwidth]{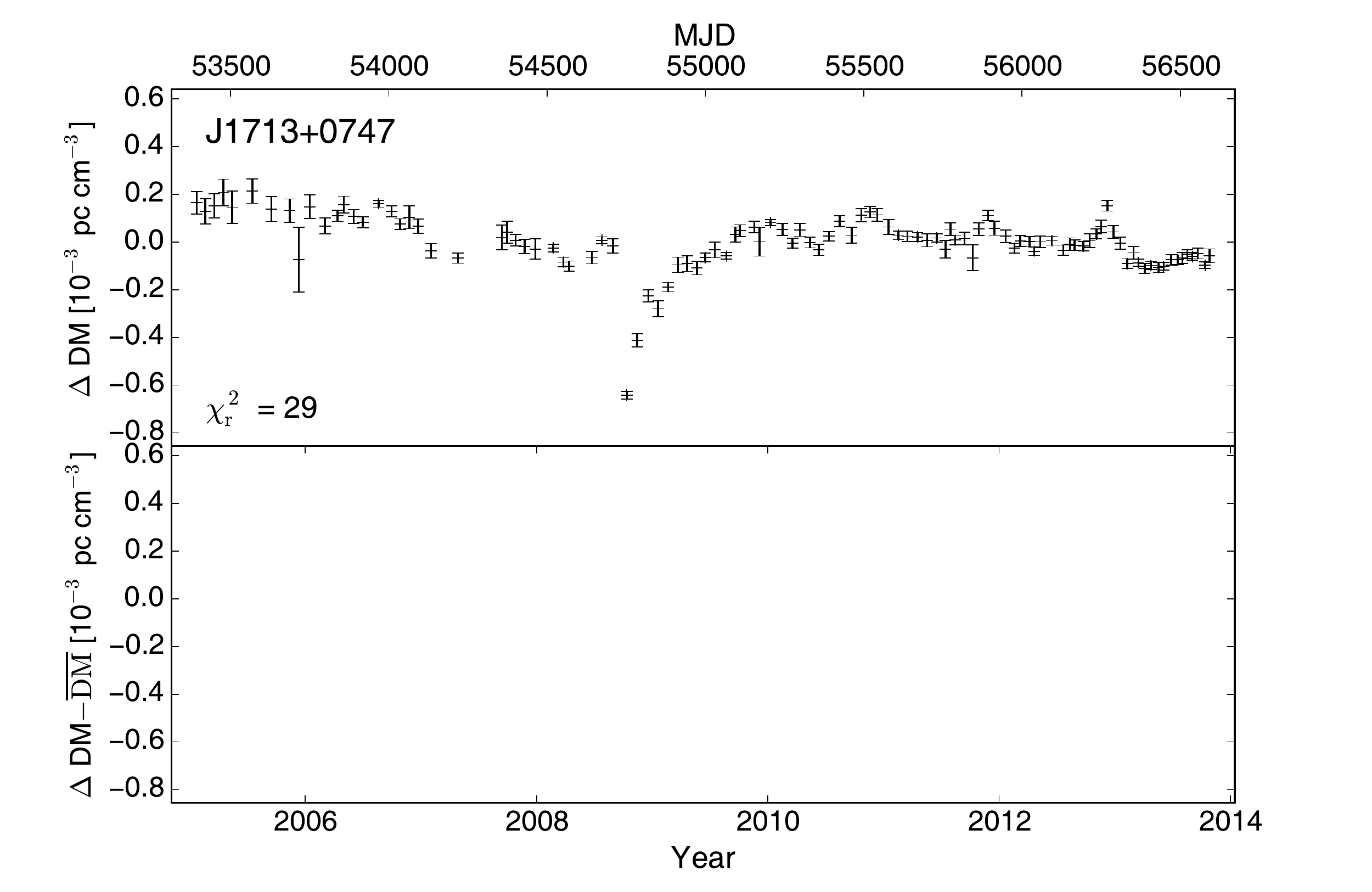}
		\includegraphics[width=0.49\textwidth]{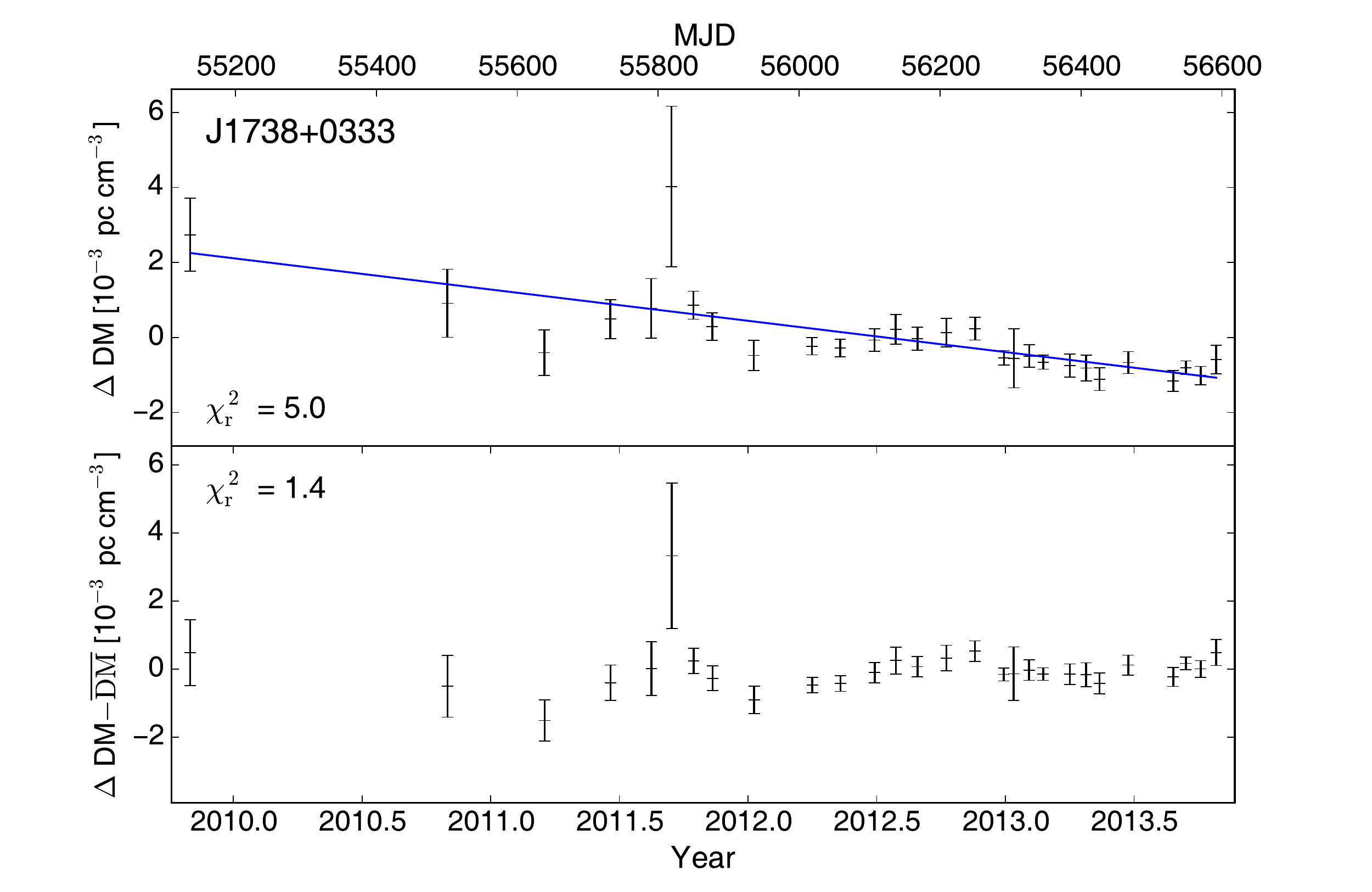}
		\includegraphics[width=0.49\textwidth]{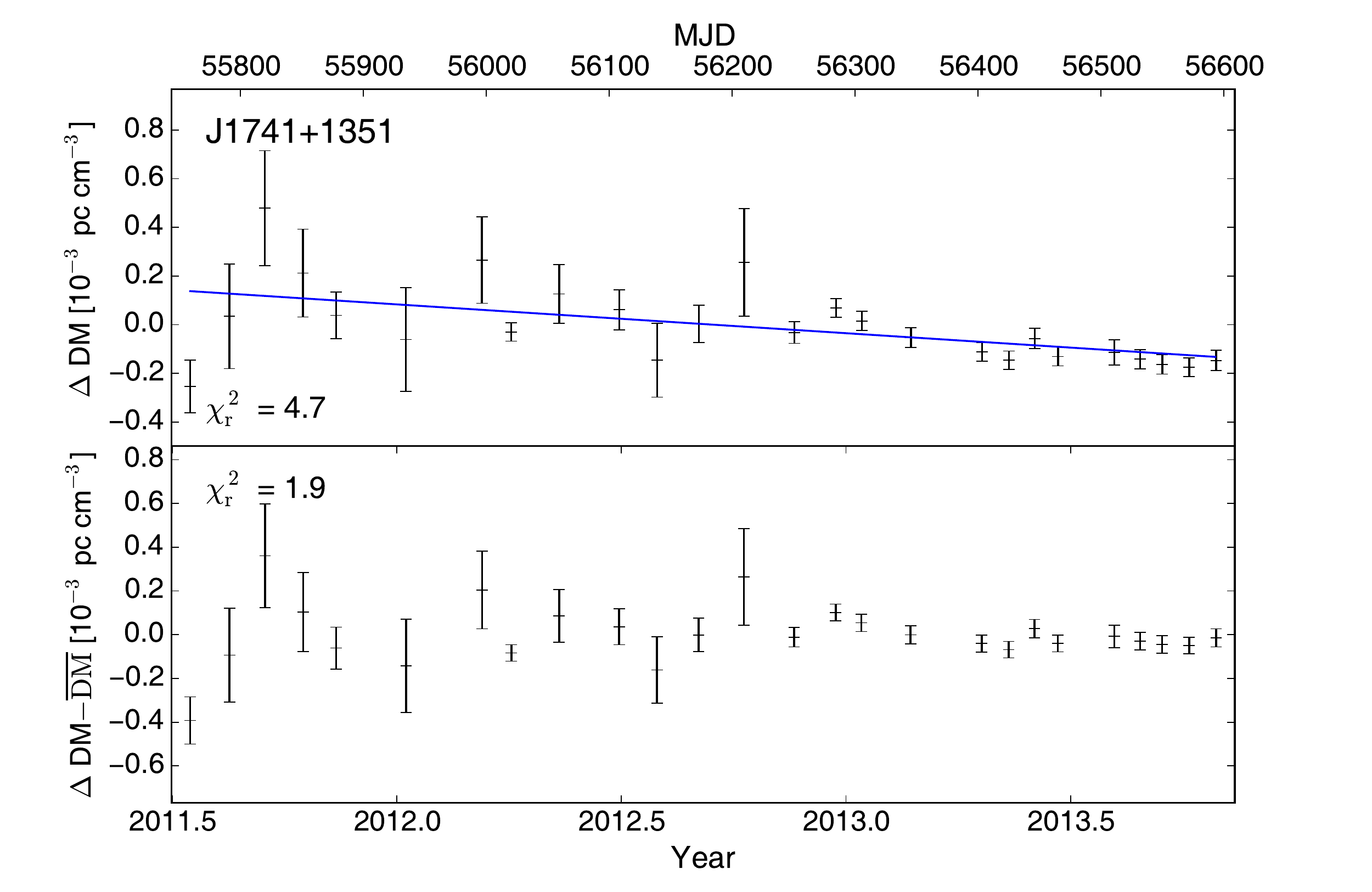}

    \caption{The top panel shows the DM time series with the best fit function (if applicable). The zero point for the DM variations corresponds to the fiducial DM for the data span. The error bars are $\pm1\sigma$ errors returned by TEMPO. The bottom panel shows the DM residuals after the trend has been removed from the time series; empty panels suggest no trend was found for that pulsar. The $\chi^2_r$ values before and after these fits for each pulsar appear in the top and bottom panels respectively, as well as in Tables \ref{tab:DMV} and \ref{tab:Trends}. PSRs J1600--3053 and J1640+2224 were not found to have significant trends in the later parts of the DM time series.}
	\label{fig:trendremoval2}		
	\end{center}
\end{figure*}
\begin{figure*}
	\begin{center}
		\includegraphics[width=0.49\textwidth]{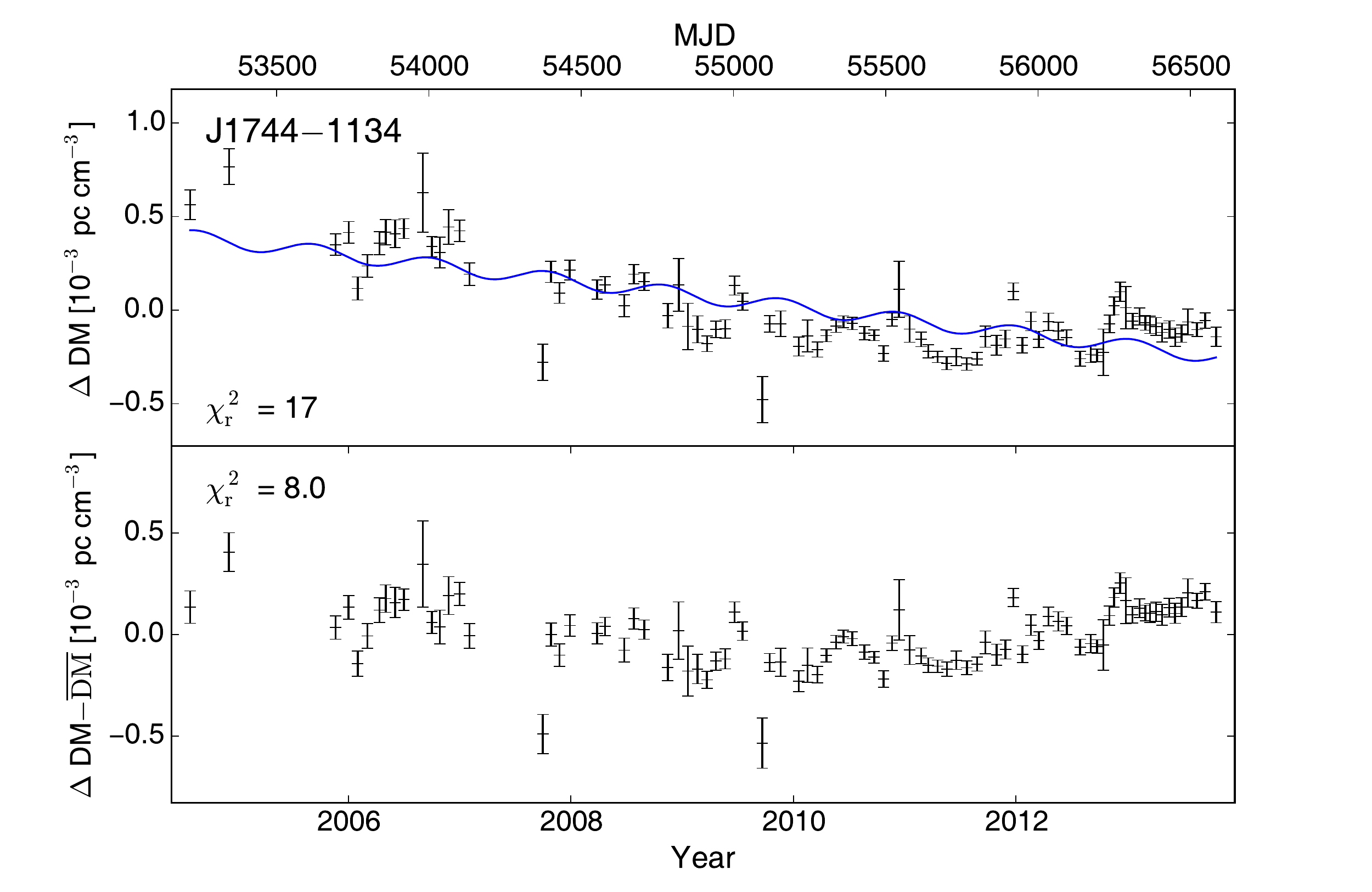}
		\includegraphics[width=0.49\textwidth]{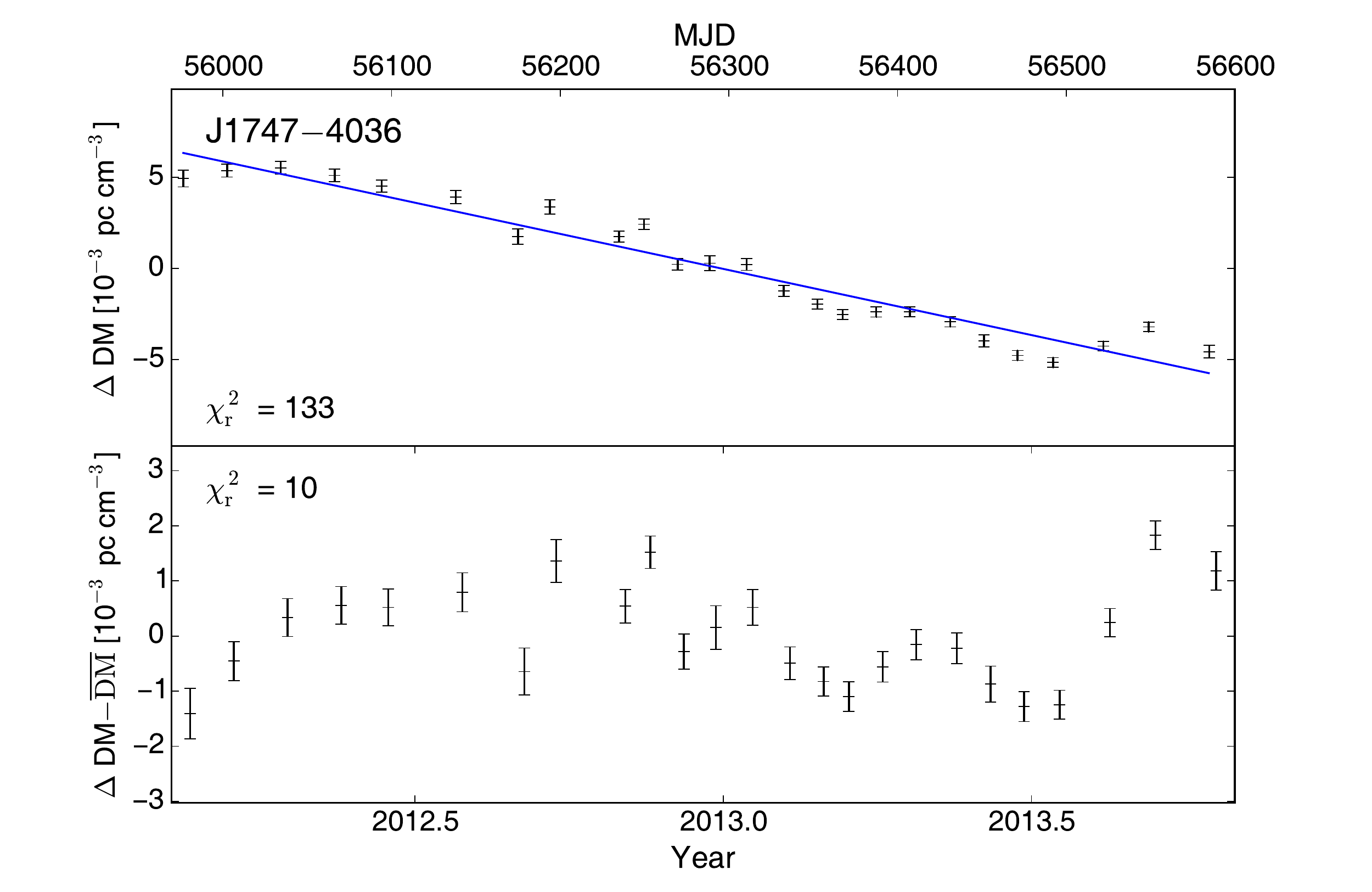}	
		\includegraphics[width=0.49\textwidth]{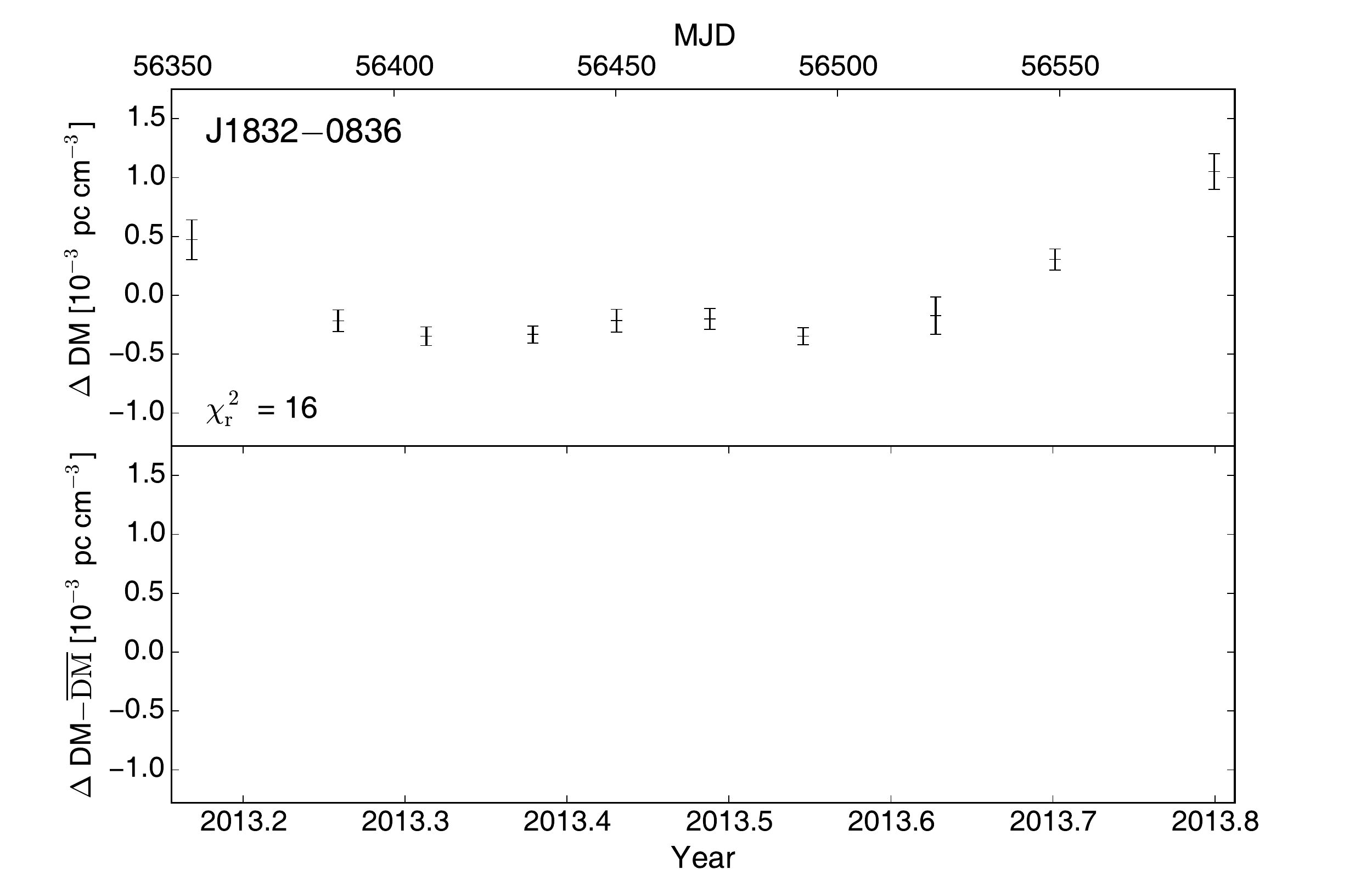}
		\includegraphics[width=0.49\textwidth]{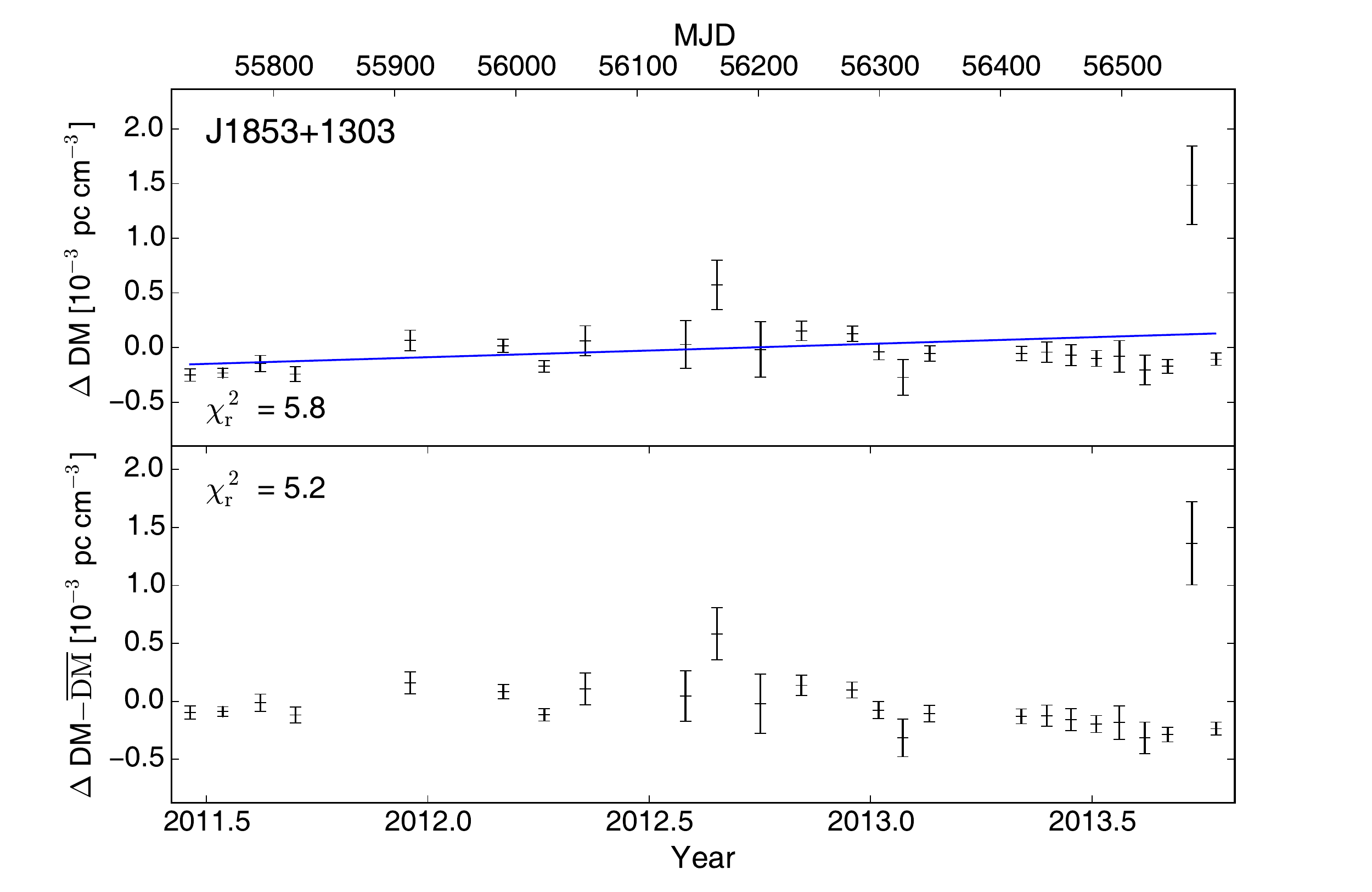}
		\includegraphics[width=0.49\textwidth]{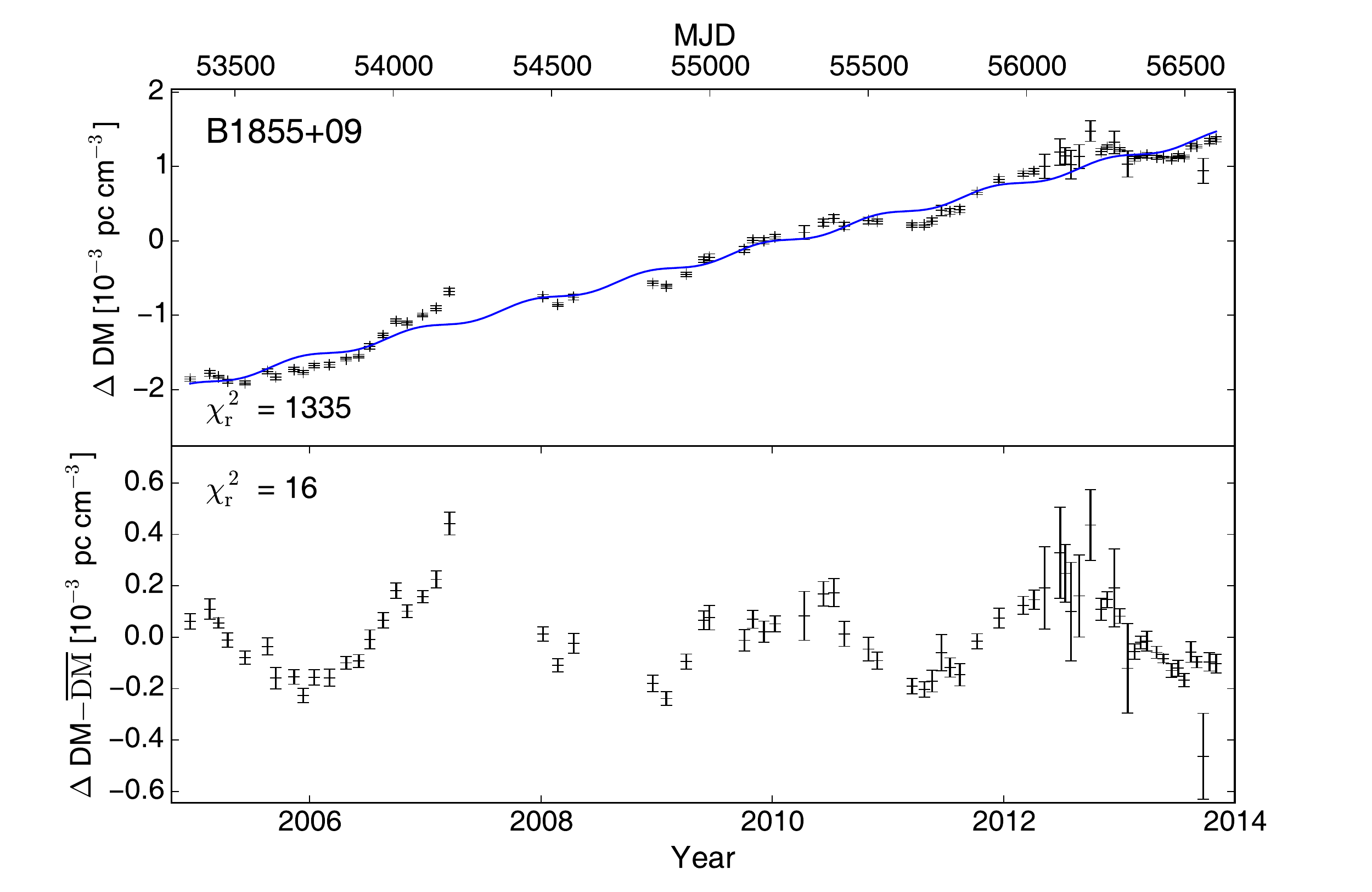}	
		\includegraphics[width=0.49\textwidth]{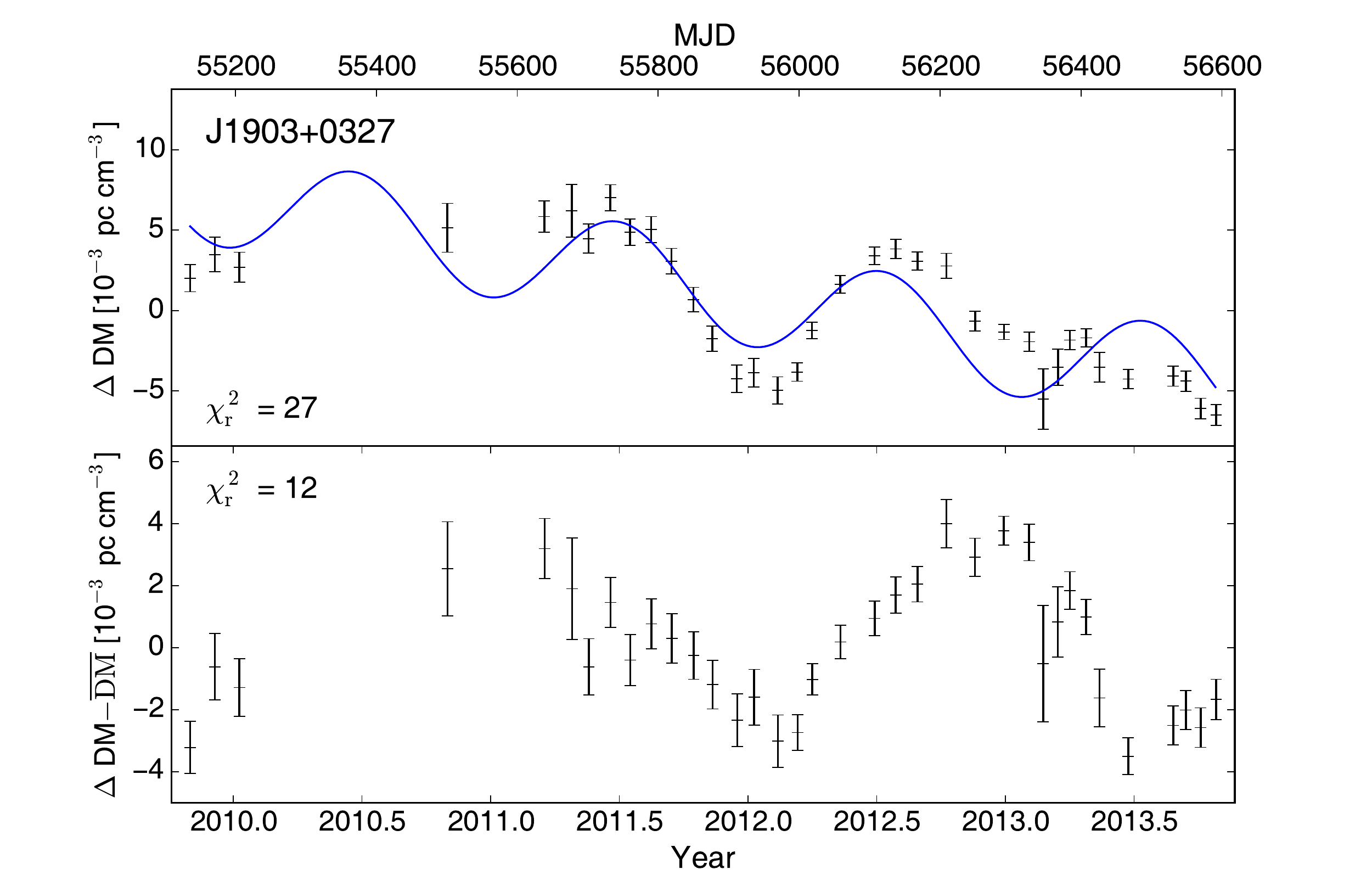}
		\includegraphics[width=0.49\textwidth]{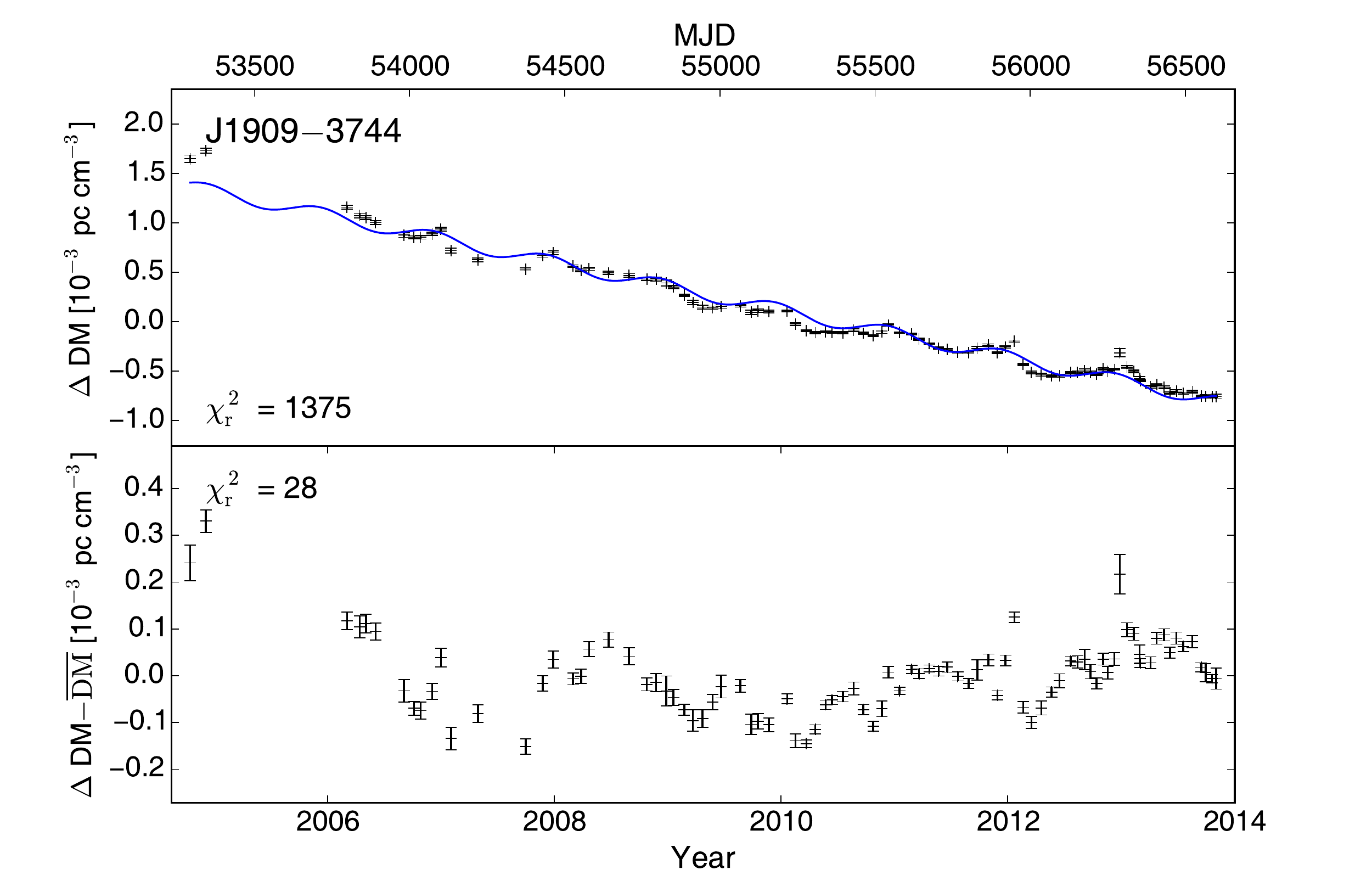}	
		\includegraphics[width=0.49\textwidth]{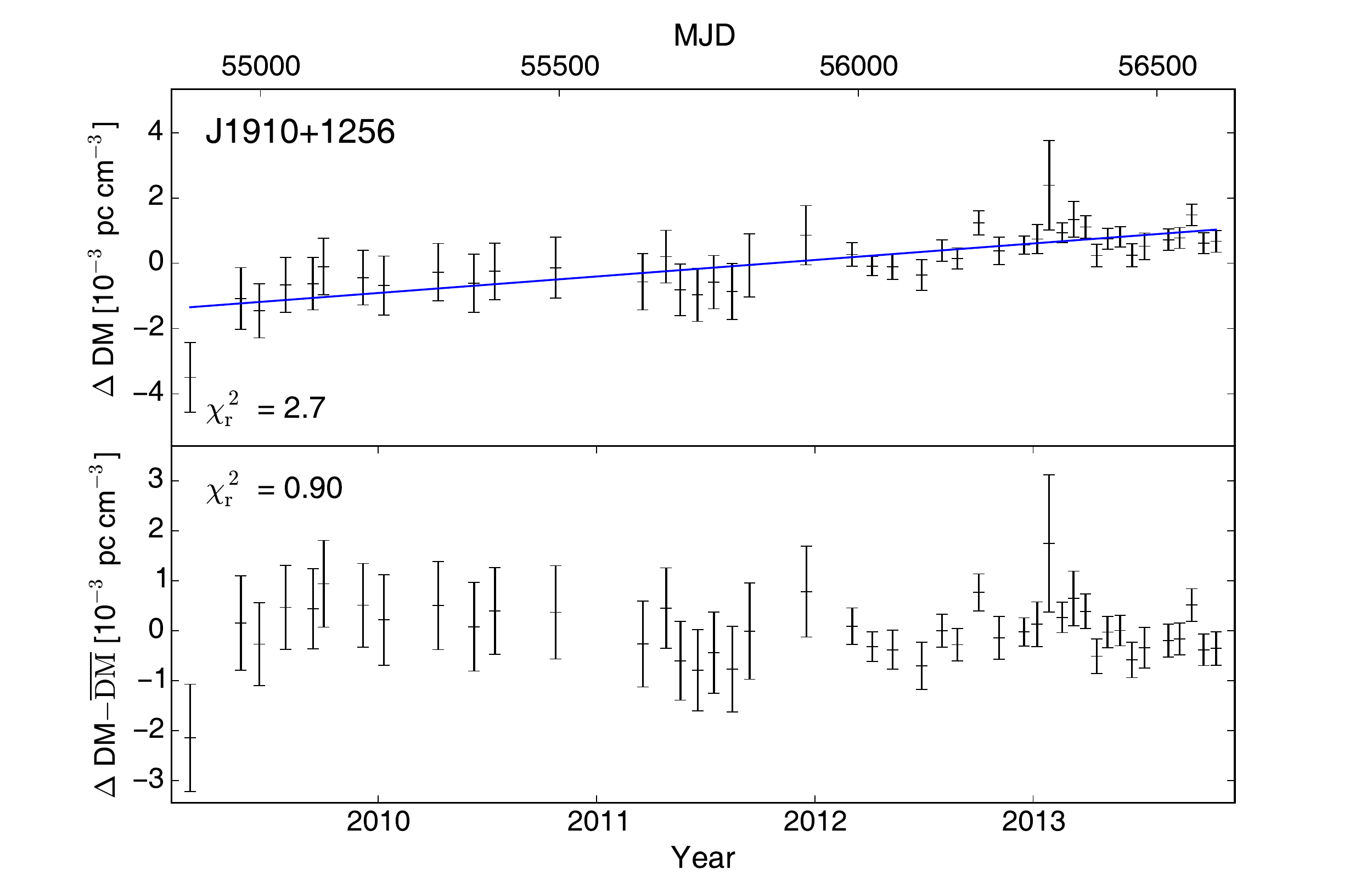}

	\caption{The top panel shows the DM time series with the best fit function (if applicable). The zero point for the DM variations corresponds to the fiducial DM for the data span. The error bars are $\pm1\sigma$ errors returned by TEMPO. The bottom panel shows the DM residuals after the trend has been removed from the time series; empty panels suggest no trend was found for that pulsar. The $\chi^2_r$ values before and after these fits for each pulsar appear in the top and bottom panels respectively, as well as in Tables \ref{tab:DMV} and \ref{tab:Trends}. PSR J1832--0836 has too short a data span for a trend to be determined.}
	\label{fig:trendremoval3}		
	\end{center}
\end{figure*}

\begin{figure*}
	\begin{center}	
		\includegraphics[width=0.49\textwidth]{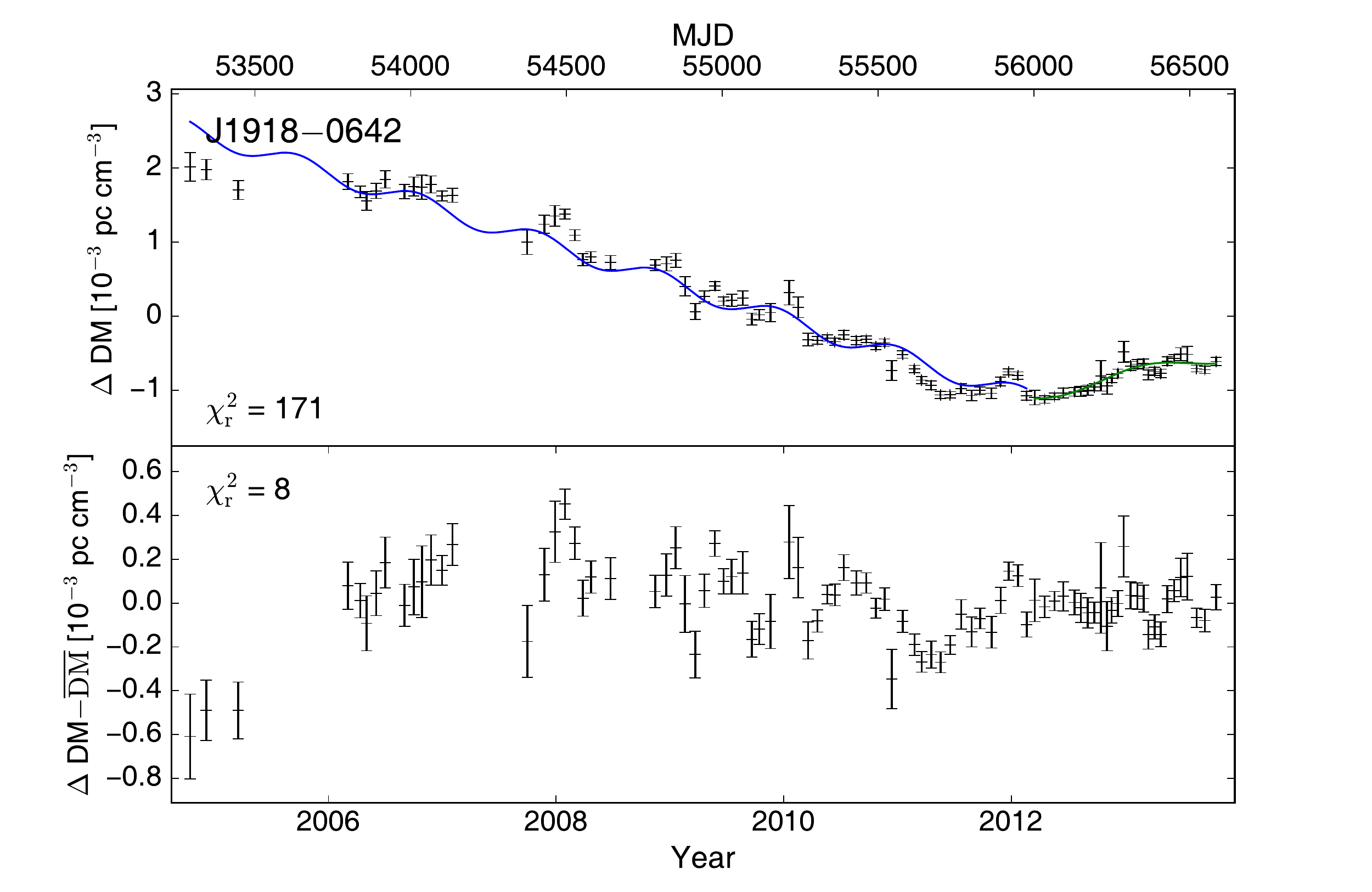}
		\includegraphics[width=0.49\textwidth]{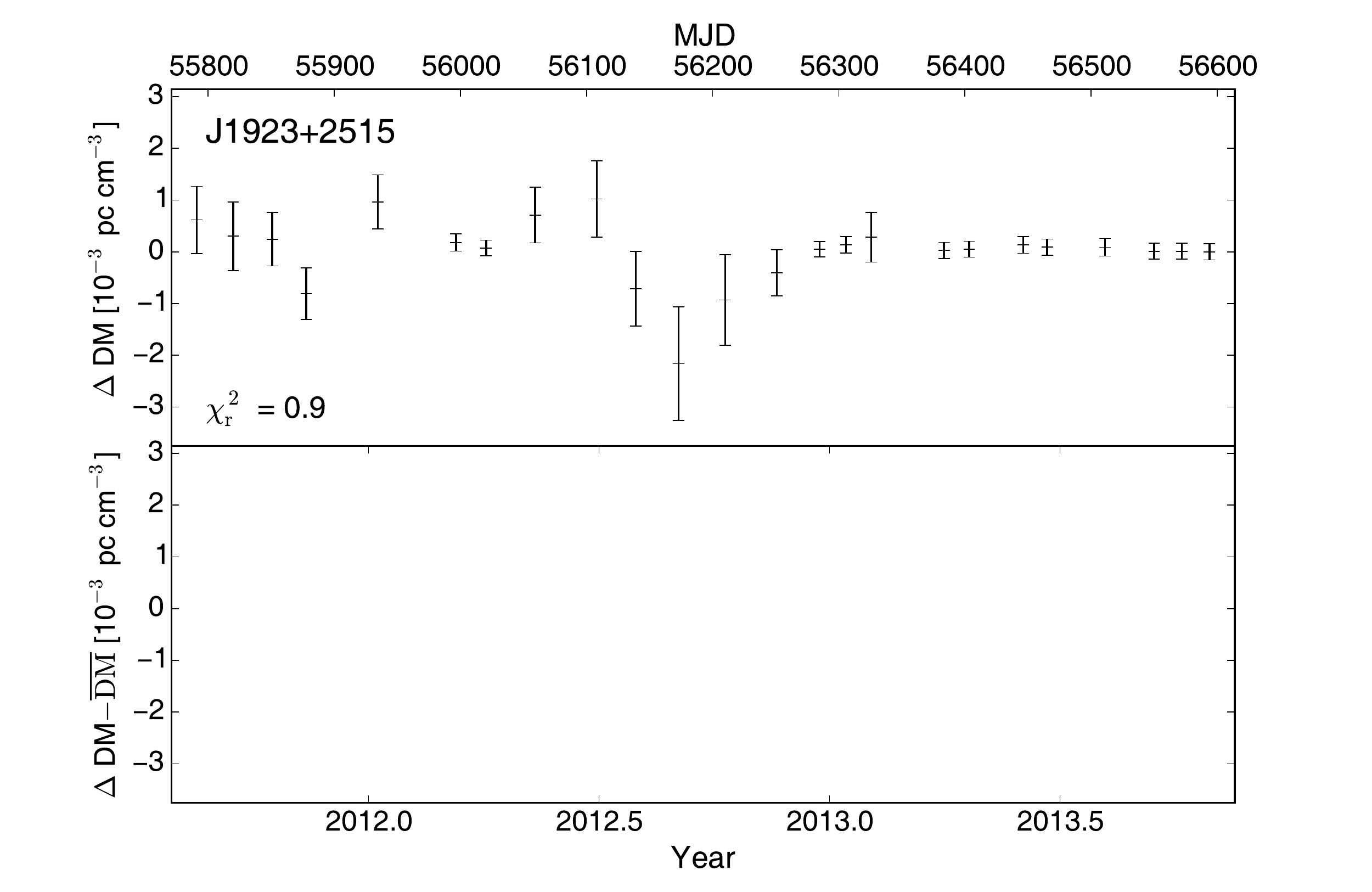}
		\includegraphics[width=0.49\textwidth]{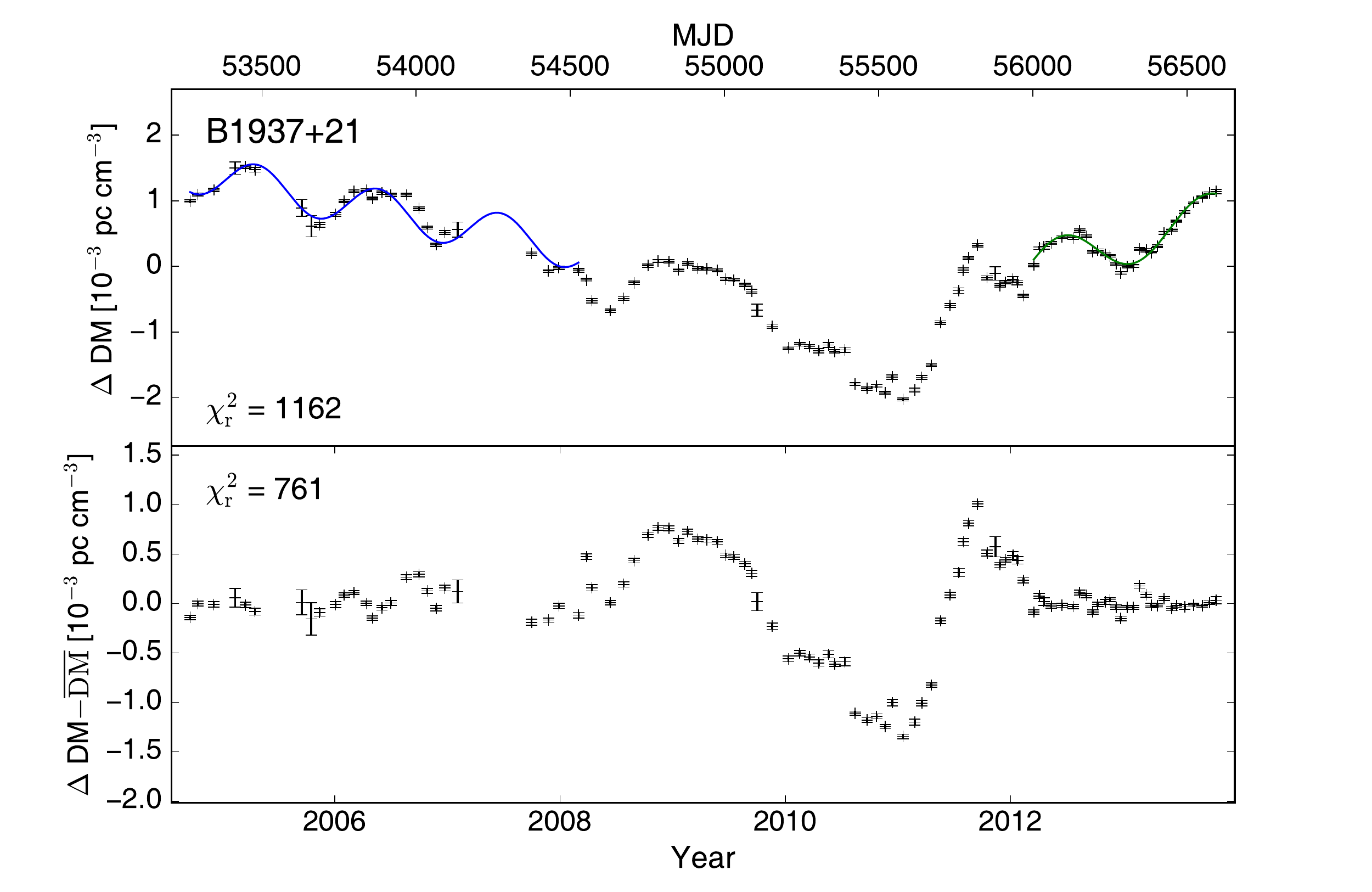}	
		\includegraphics[width=0.49\textwidth]{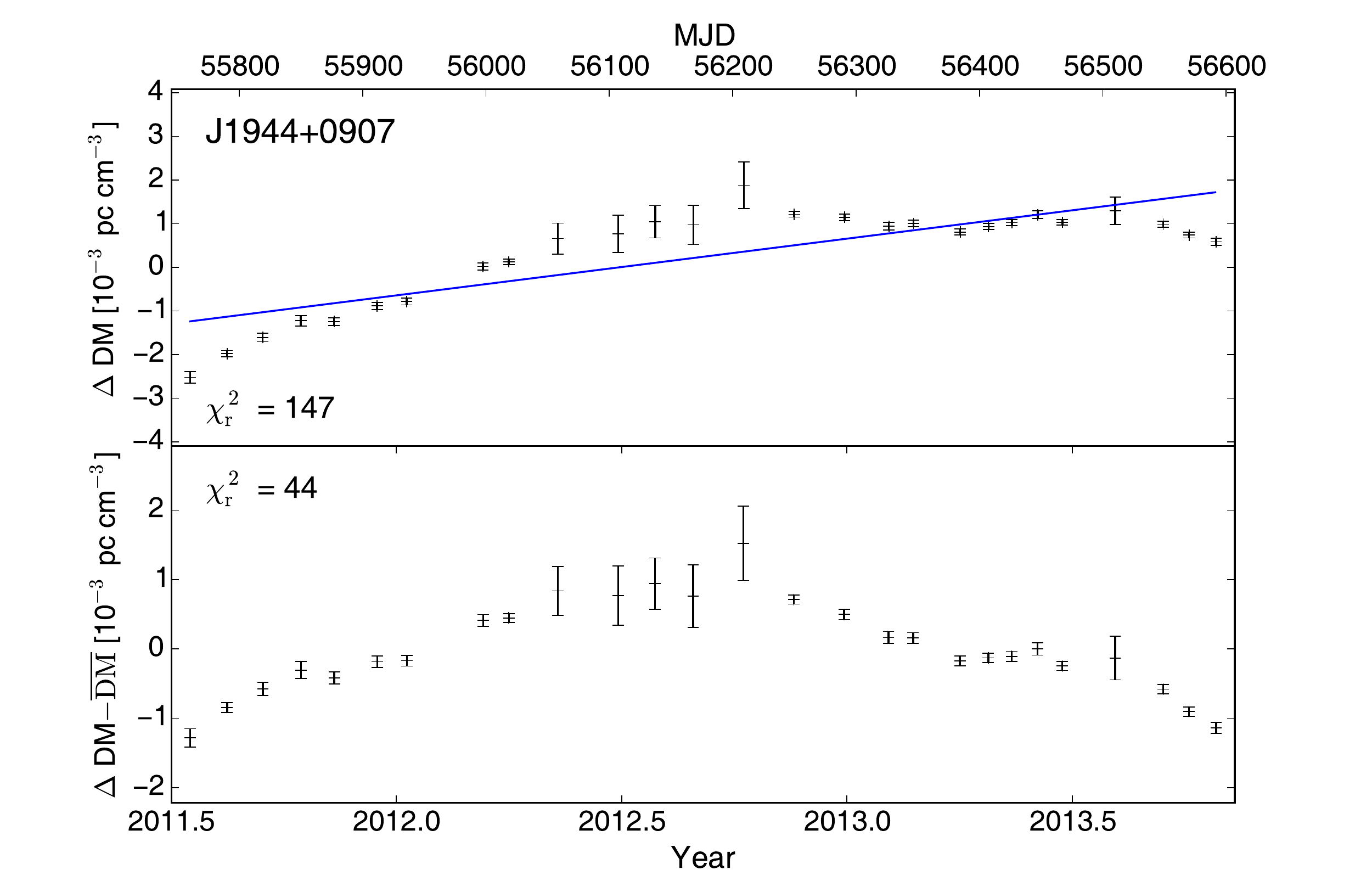}	
		\includegraphics[width=0.49\textwidth]{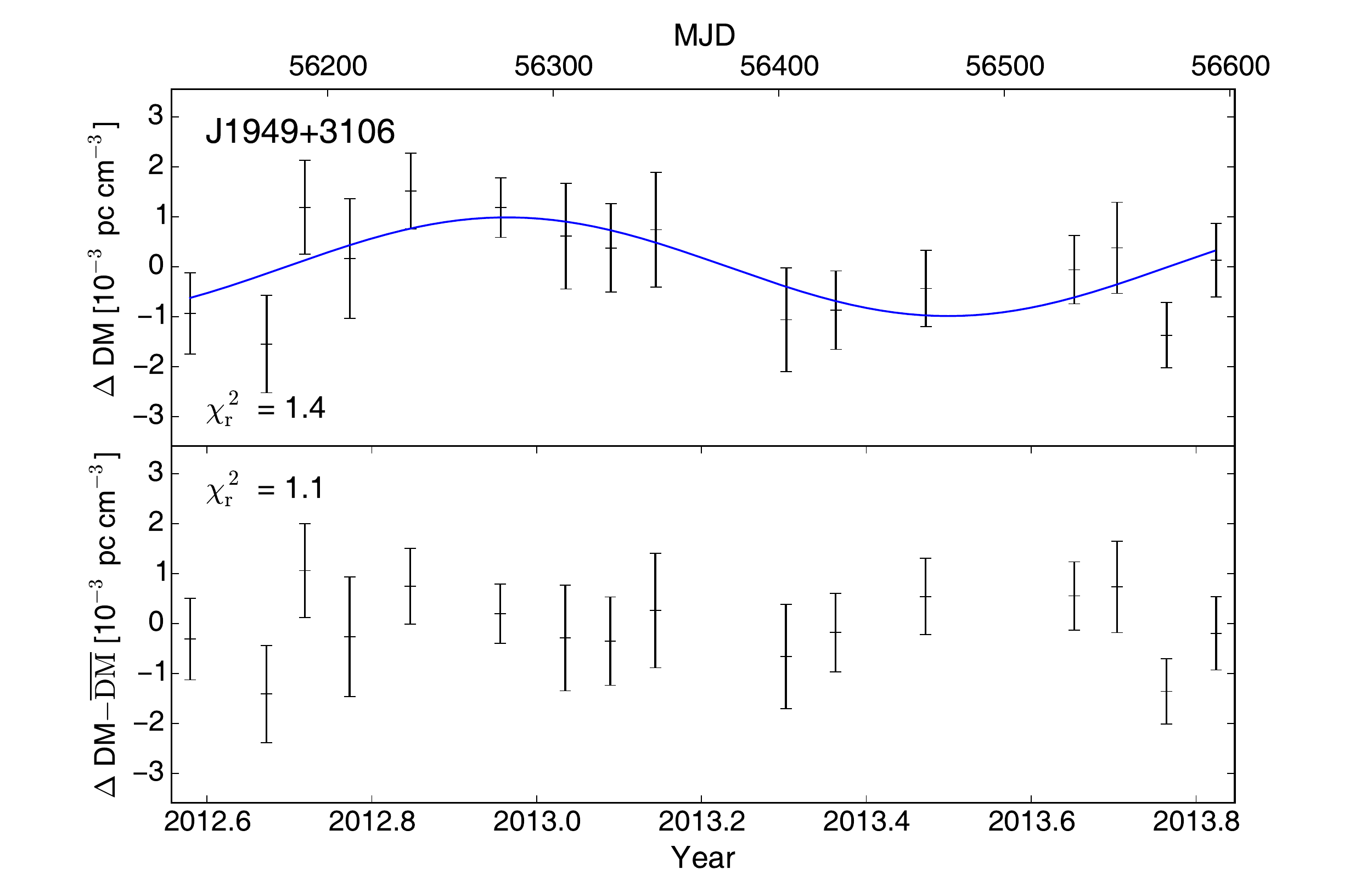}
		\includegraphics[width=0.49\textwidth]{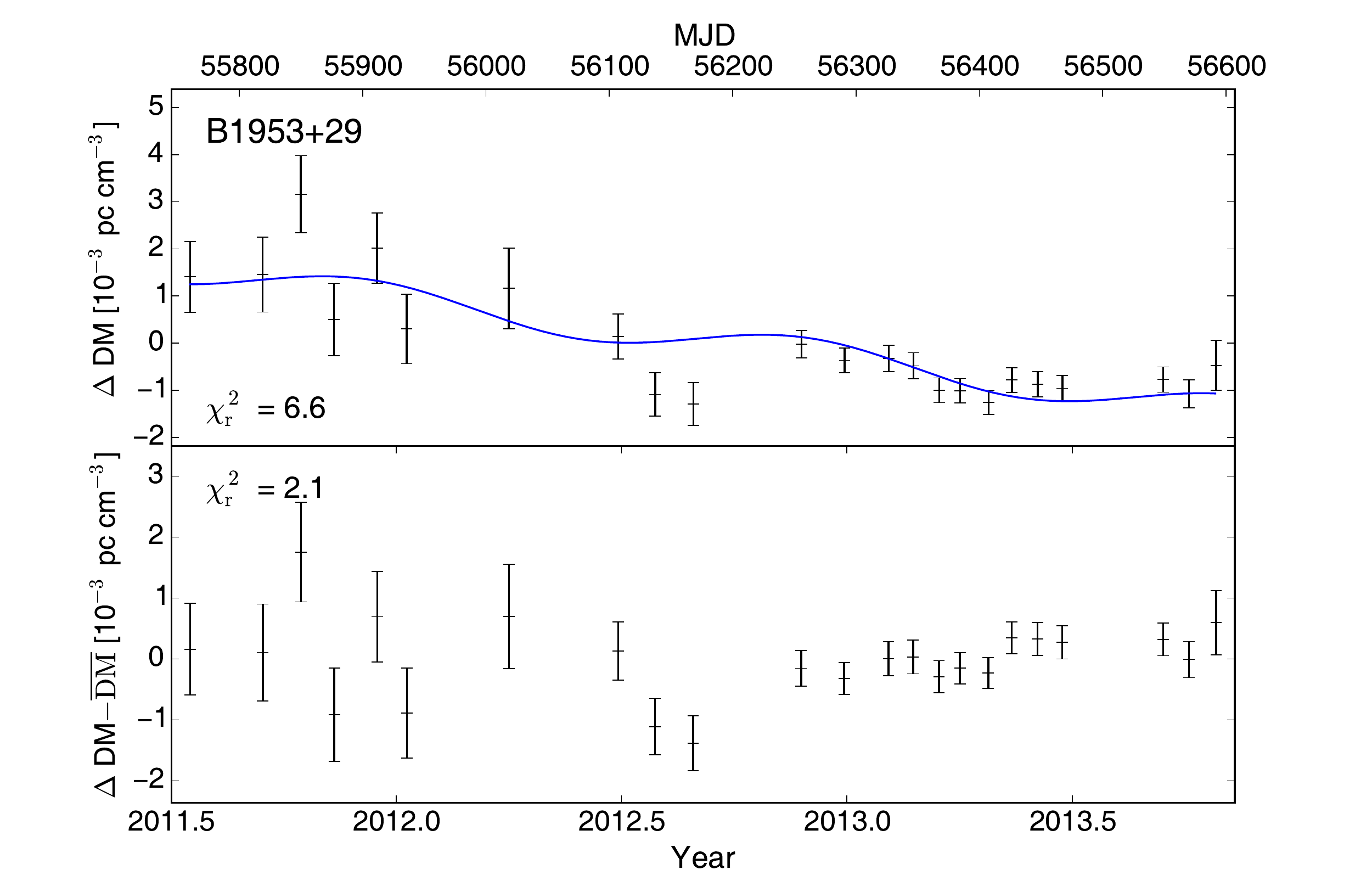}
		\includegraphics[width=0.49\textwidth]{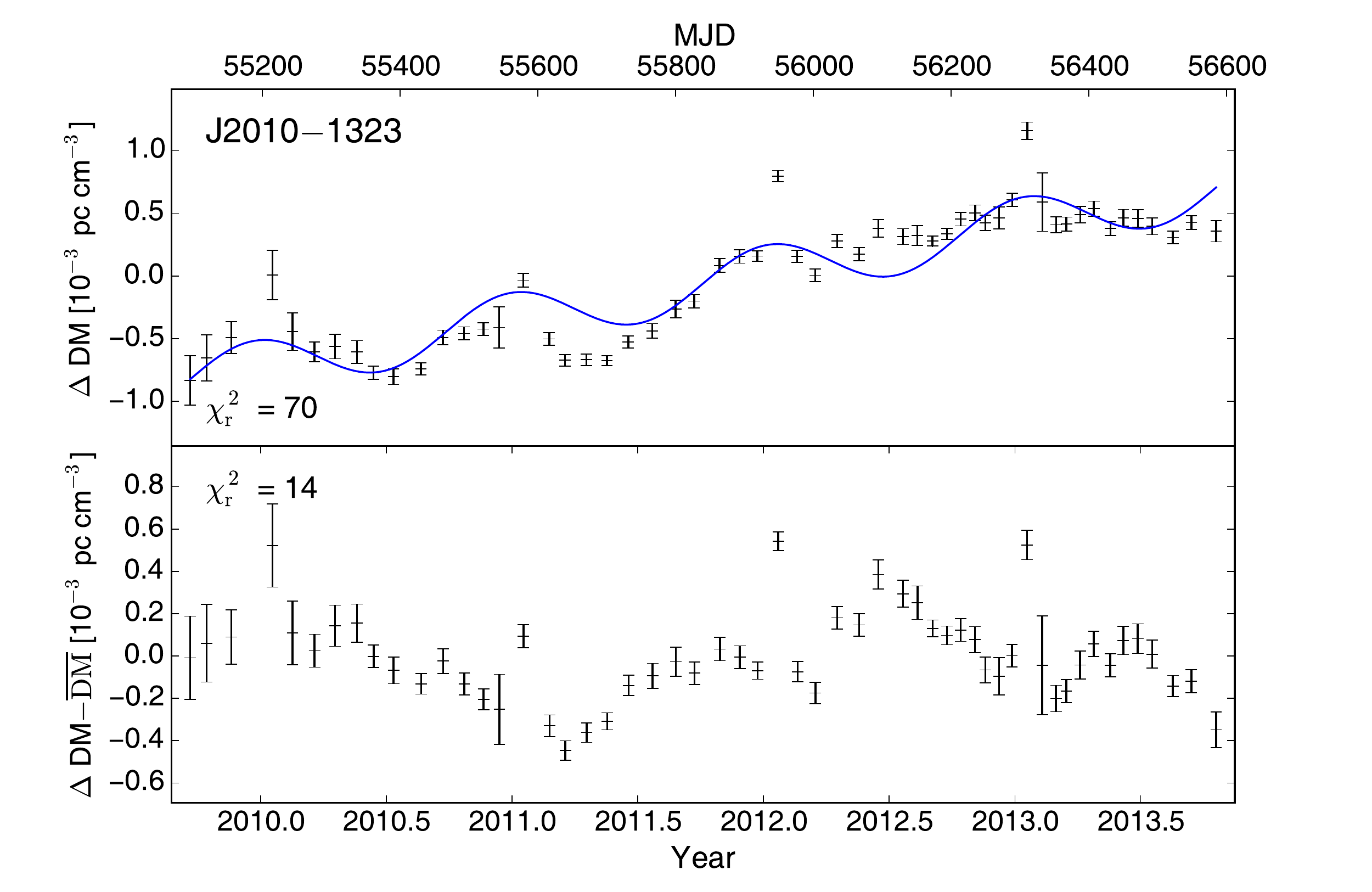}	
		\includegraphics[width=0.49\textwidth]{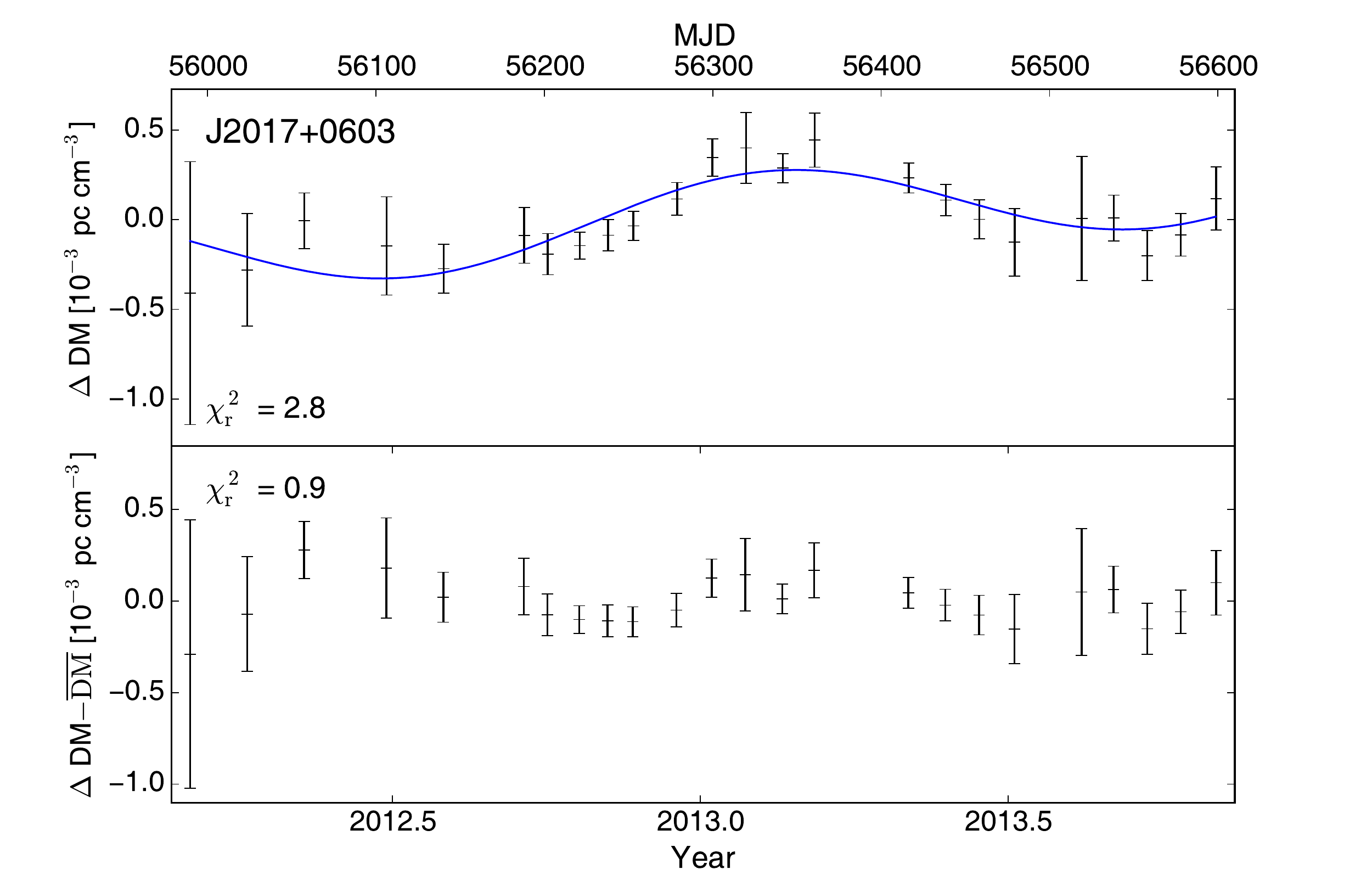}

	\caption{The top panel shows the DM time series with the best fit function (if applicable). The zero point for the DM variations corresponds to the fiducial DM for the data span. The error bars are $\pm1\sigma$ errors returned by TEMPO. The bottom panel shows the DM residuals after the trend has been removed from the time series; empty panels suggest no trend was found for that pulsar. The $\chi^2_r$ values before and after these fits for each pulsar appear in the top and bottom panels respectively, as well as in Tables \ref{tab:DMV} and \ref{tab:Trends}. PSR B1937+21 could not be fit with a periodic trend throughout the data set.}
	\label{fig:trendremoval4}		
	\end{center}
\end{figure*}

\begin{figure*}
	\begin{center}	
		\includegraphics[width=0.49\textwidth]{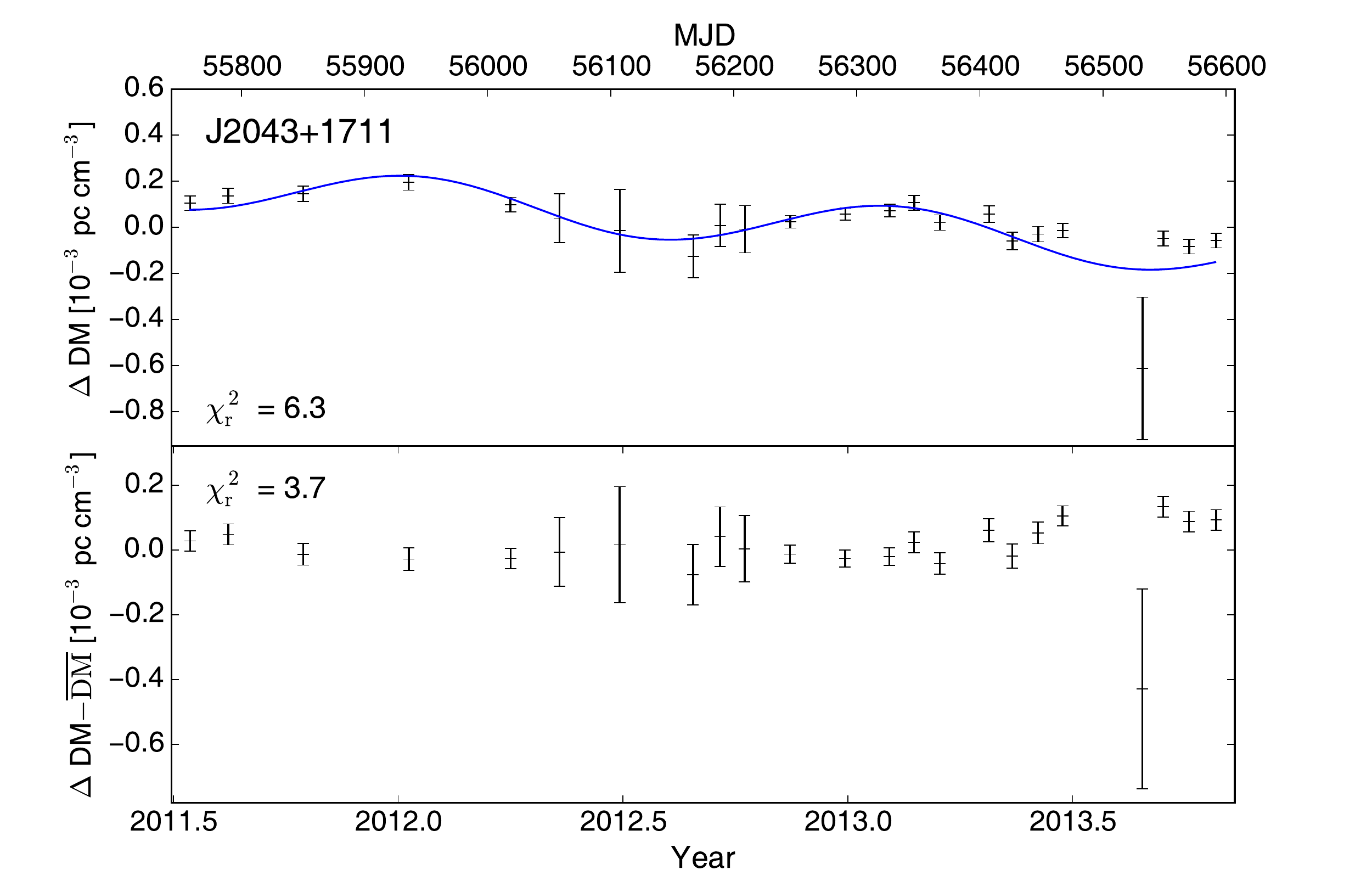}
		\includegraphics[width=0.49\textwidth]{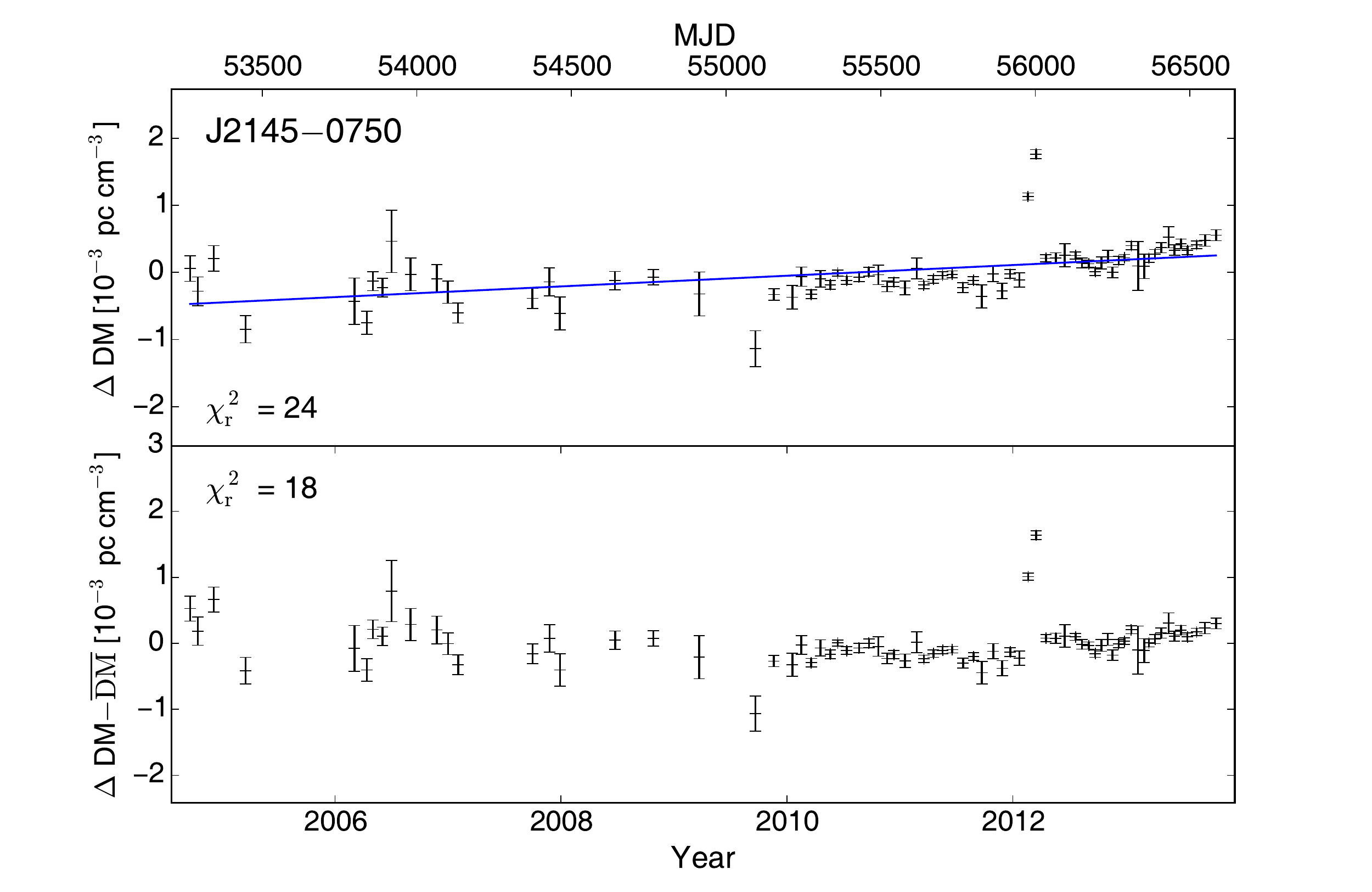}	
		\includegraphics[width=0.49\textwidth]{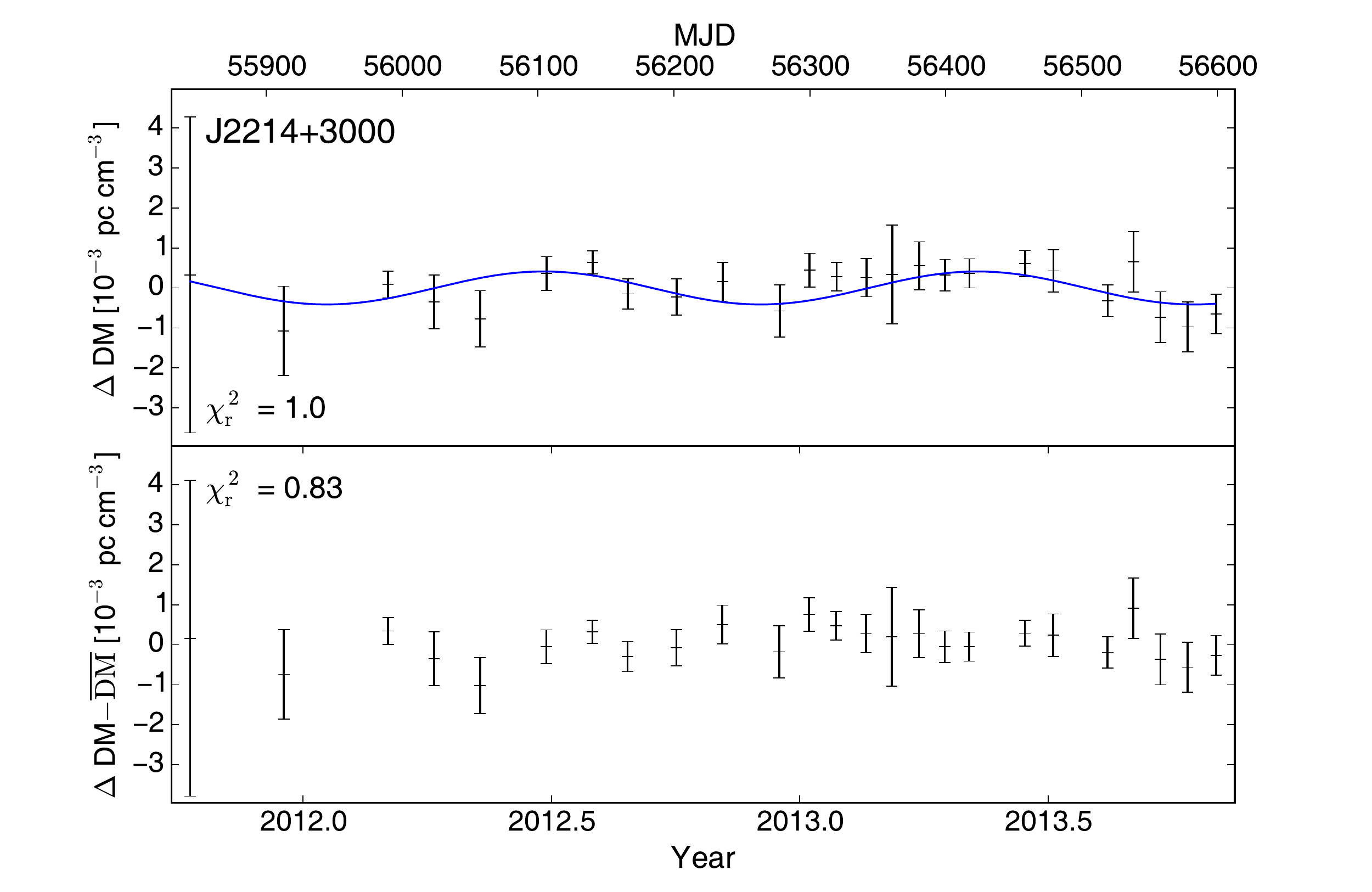}
  	    \includegraphics[width=0.49\textwidth]{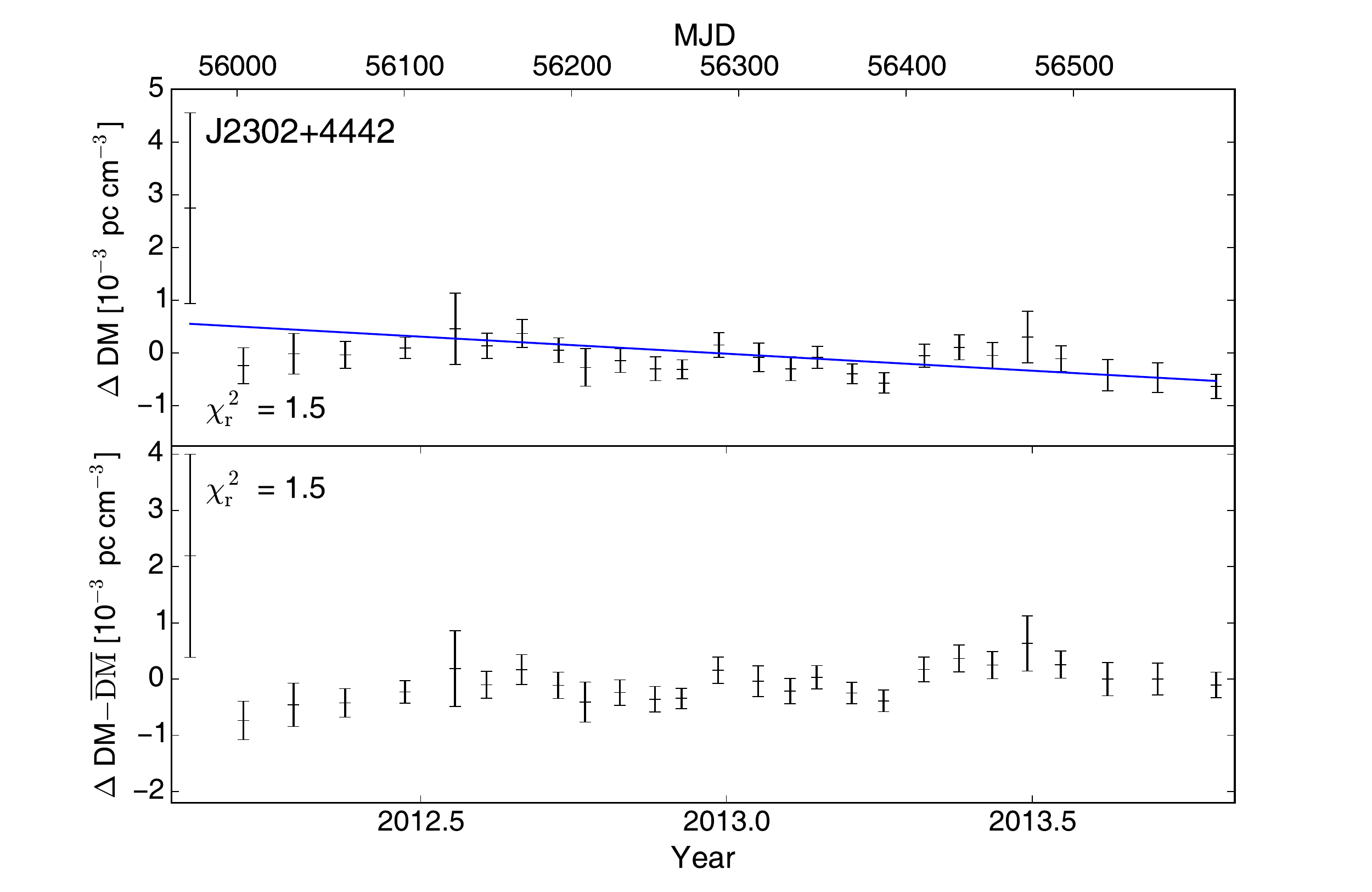}
	    \includegraphics[width=0.49\textwidth]{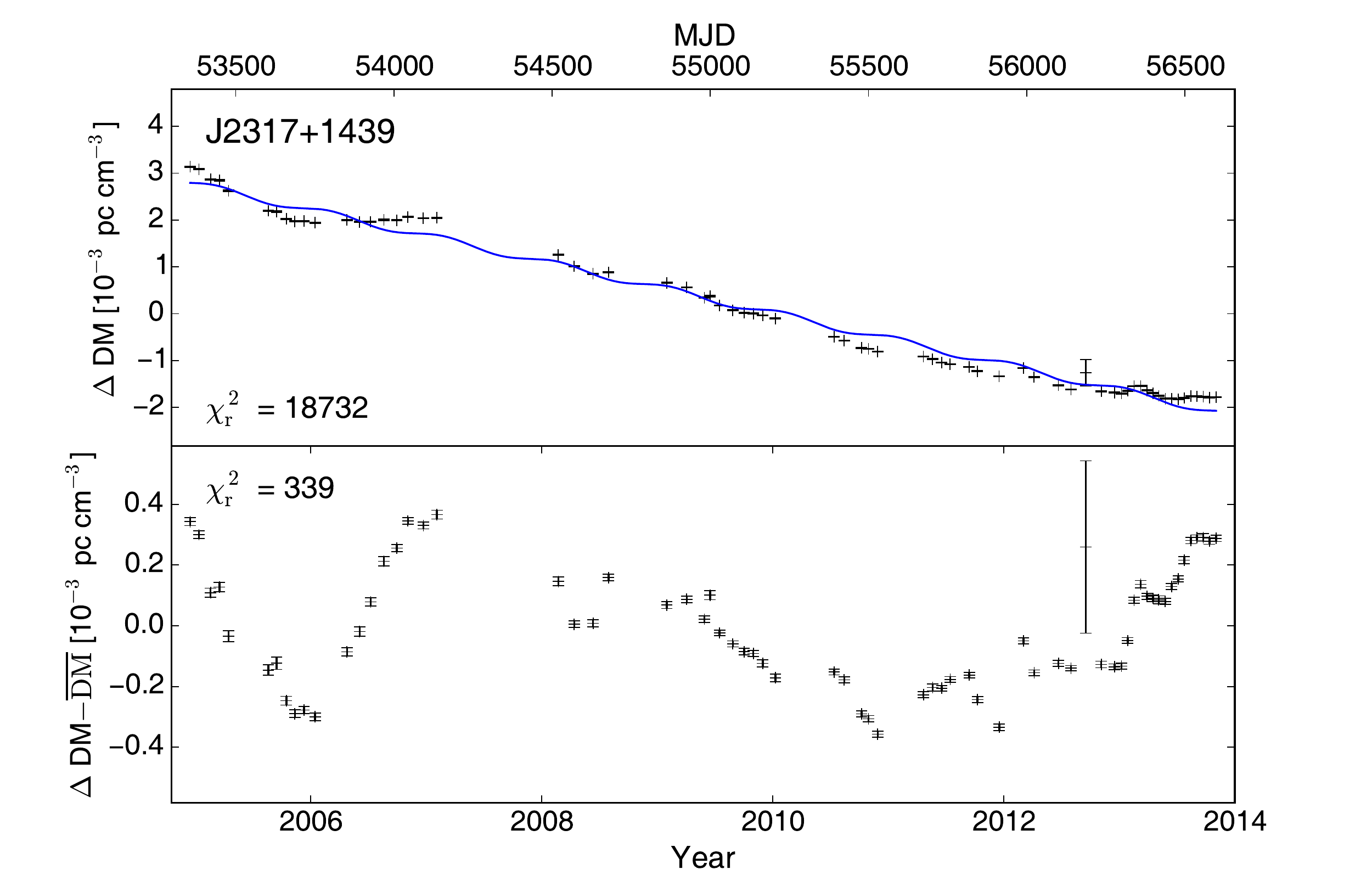}	
	 \caption{The top panel shows the DM time series with the best fit function (if applicable). The zero point for the DM variations corresponds to the fiducial DM for the data span. The error bars are $\pm1\sigma$ errors returned by TEMPO. The bottom panel shows the DM residuals after the trend has been removed from the time series; empty panels suggest no trend was found for that pulsar. The $\chi^2_r$ values before and after these fits for each pulsar appear in the top and bottom panels respectively, as well as in Tables \ref{tab:DMV} and \ref{tab:Trends}.}
	\label{fig:trendremoval5}		
	\end{center}
\end{figure*}

%% file: Figure7.tex
\begin{figure*}
	\begin{center}
		\includegraphics[trim=0cm 1.95cm 0.4cm 0.4cm,clip,width=0.45\textwidth]{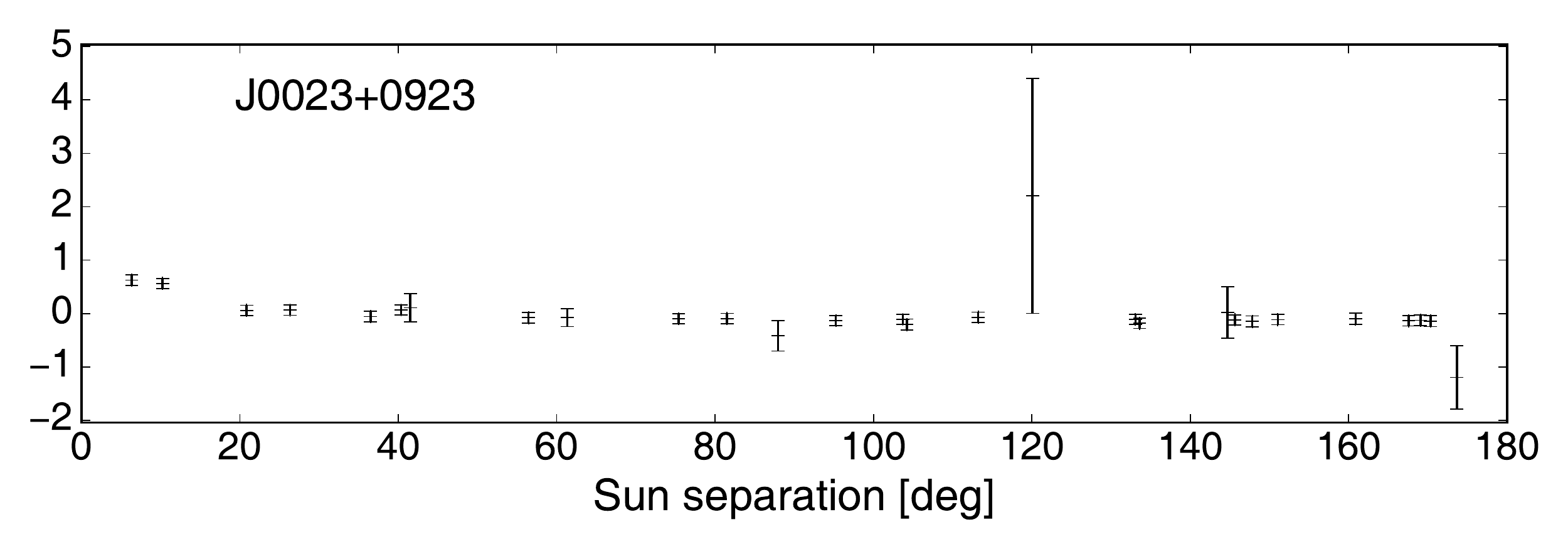}
		\includegraphics[trim=0cm 1.95cm 0.4cm 0.4cm,clip,width=0.45\textwidth]{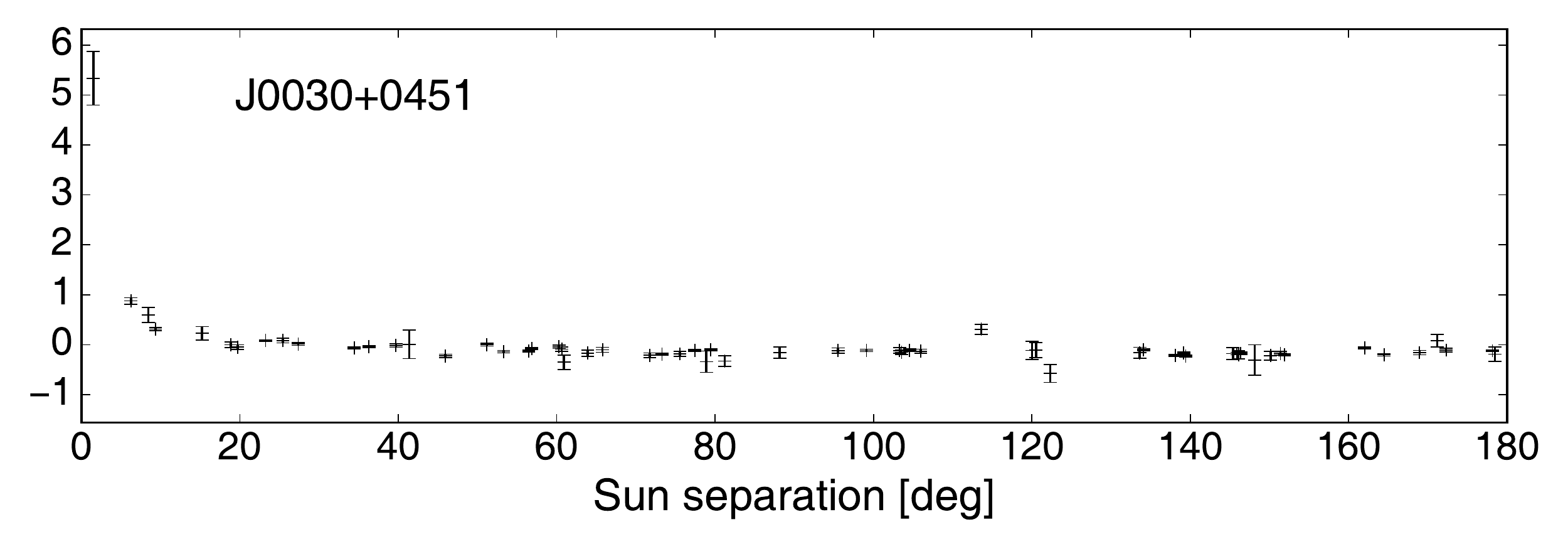}		
		\includegraphics[trim=0.5cm 1.4cm 0.4cm 0.4cm,clip,width=0.45\textwidth]{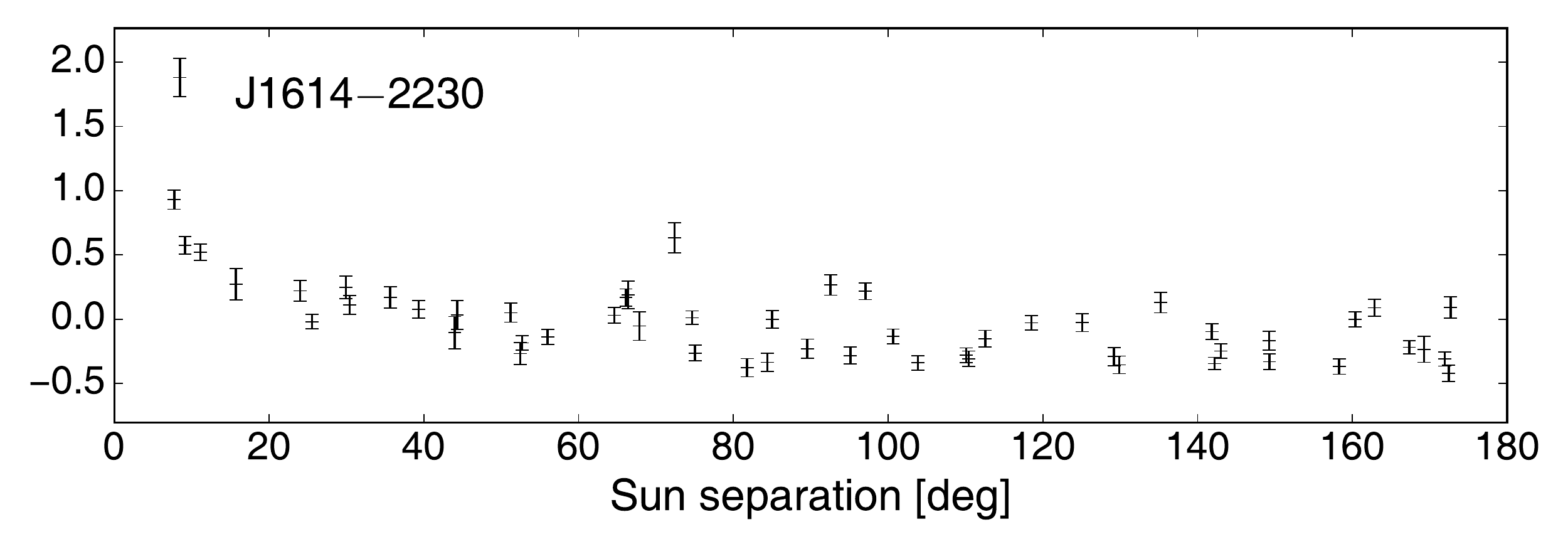}
		\includegraphics[trim=0.5cm 1.4cm 0.4cm 0.4cm,clip,width=0.45\textwidth]{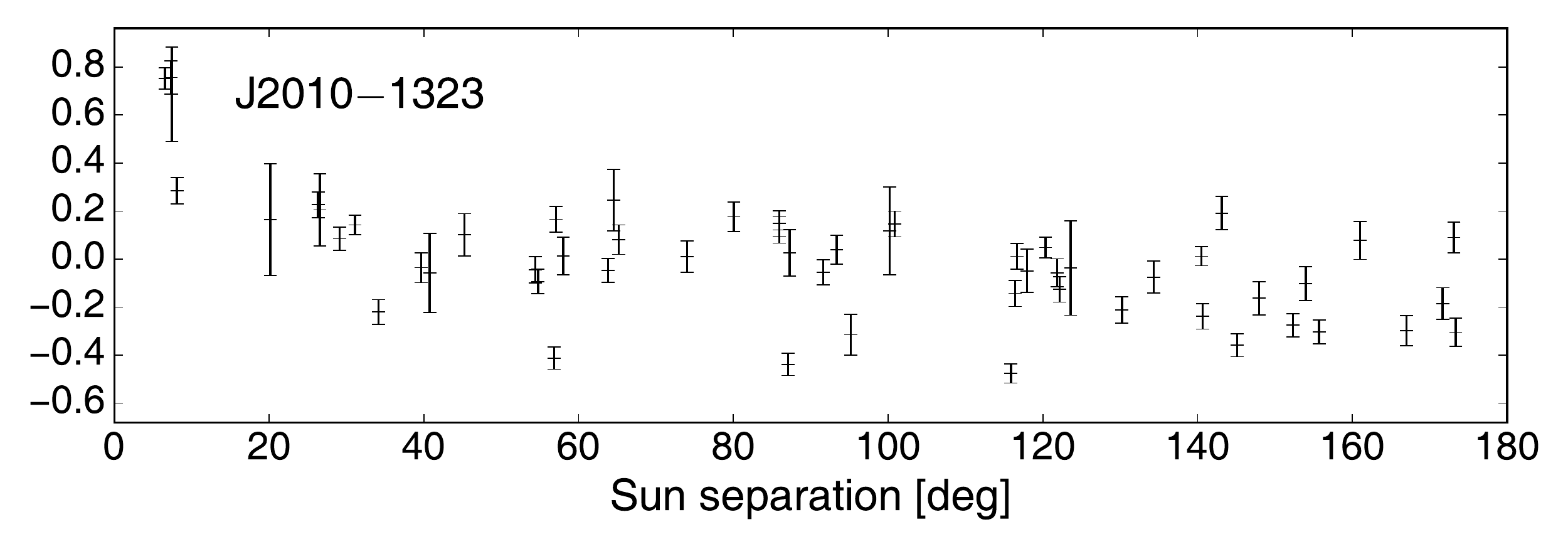}

\leavevmode\smash{\makebox[0pt]{\hspace{2em}
  \rotatebox[origin=l]{90}{\hspace{2.5em}
    $\Delta$DM [10$^{-3}$~pc~cm$^{-3}$]}%
}}\hspace{0pt plus 1filll}\null

Pulsar--Sun separation (degrees)		

	\caption{DM variations with respect to the solar position angle. Those with linear trends identified in Table \ref{tab:Trends} have had the linear trend subtracted in order to better identify any correlation in the DM as a function of solar angle.}
	\label{fig:sunpos}		
	\end{center}
\end{figure*}

%% file: Figures8to11.tex
\begin{figure*}
	\begin{center}
		\includegraphics[trim=0cm 0cm 0cm 0cm, clip,width=0.4\textwidth]{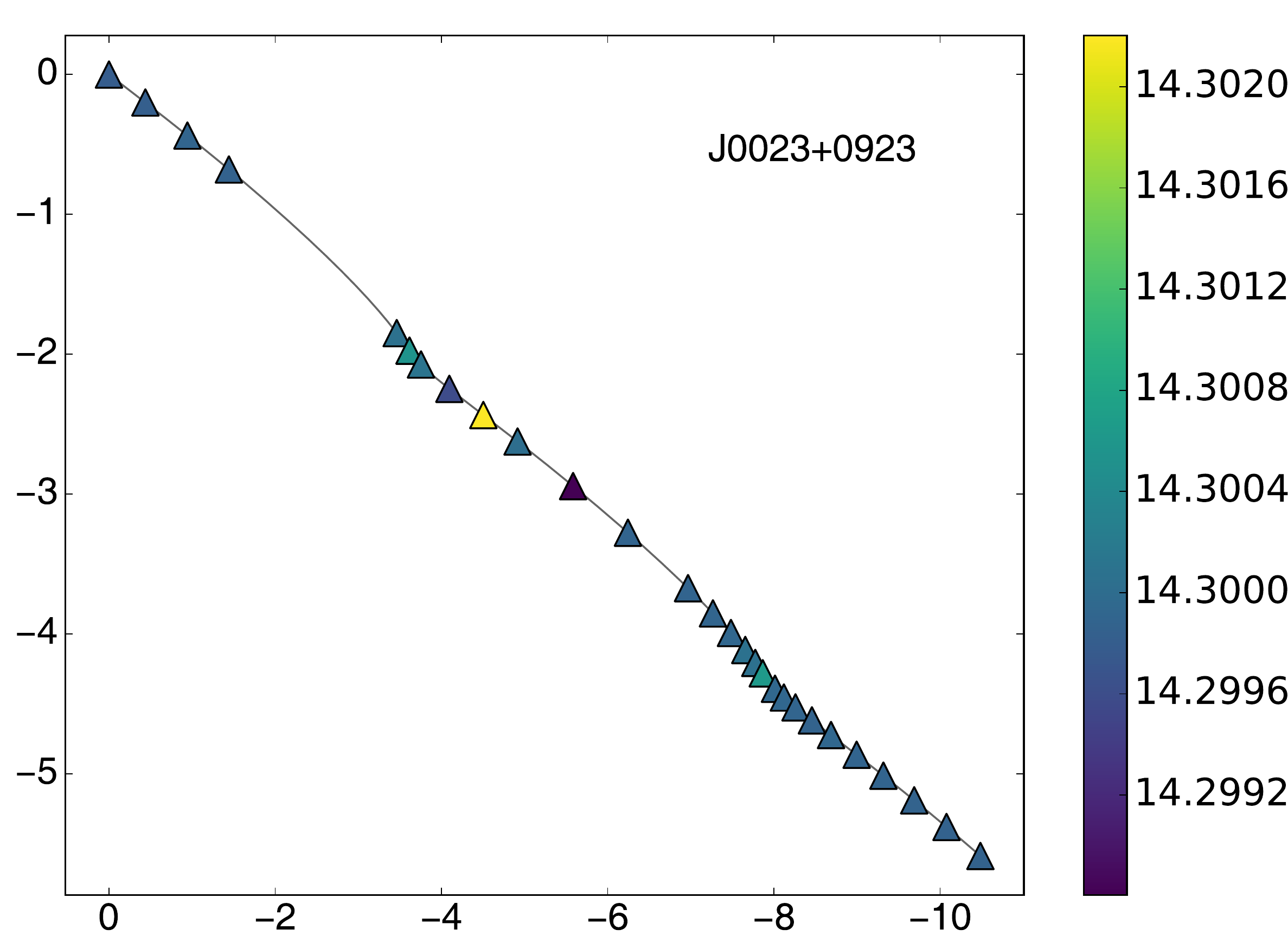}
		\includegraphics[trim=0cm 0cm 0cm 0cm, clip,width=0.4\textwidth]{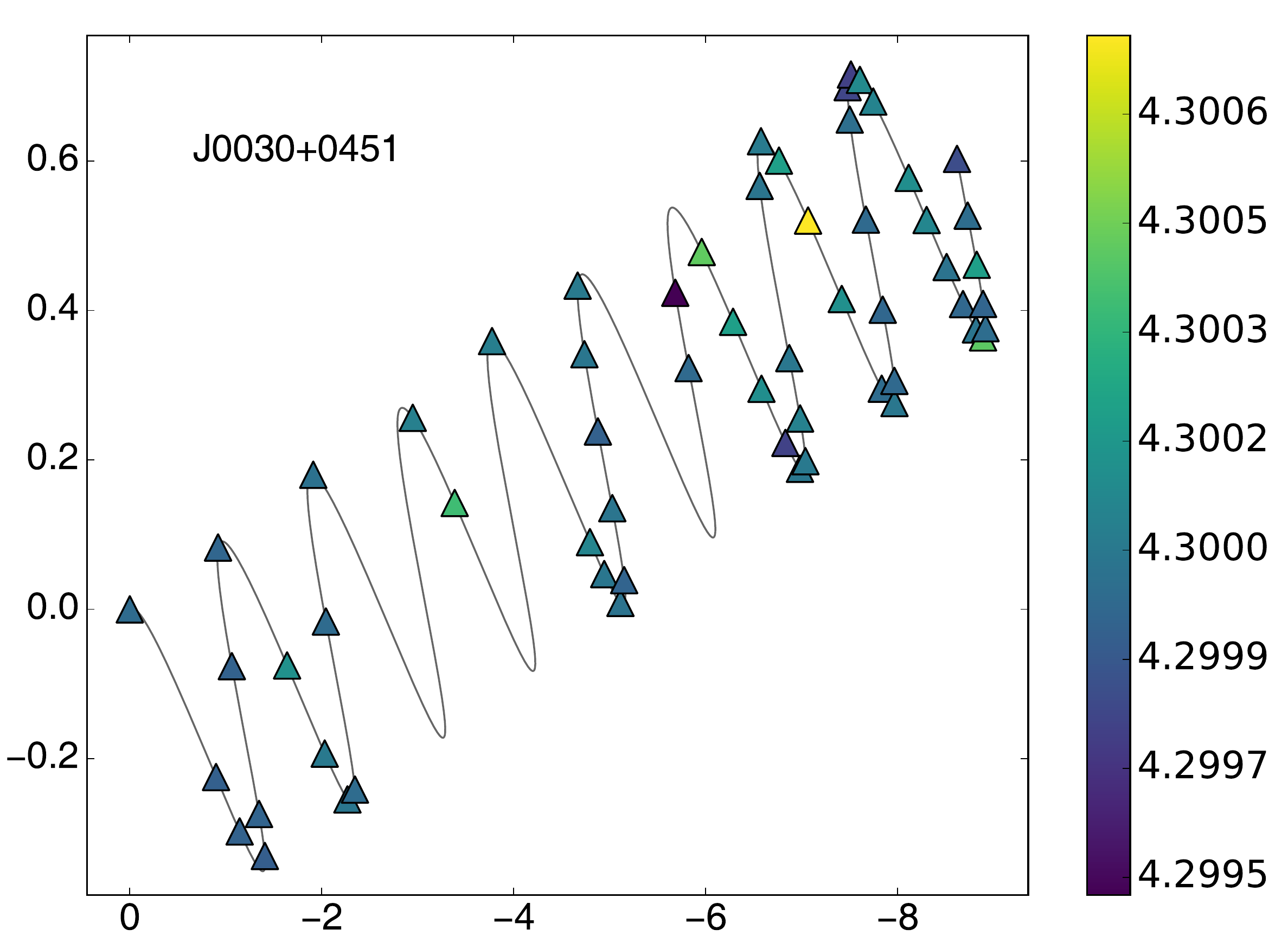}
		\includegraphics[trim=0cm 0cm 0cm 0cm, clip,width=0.4\textwidth]{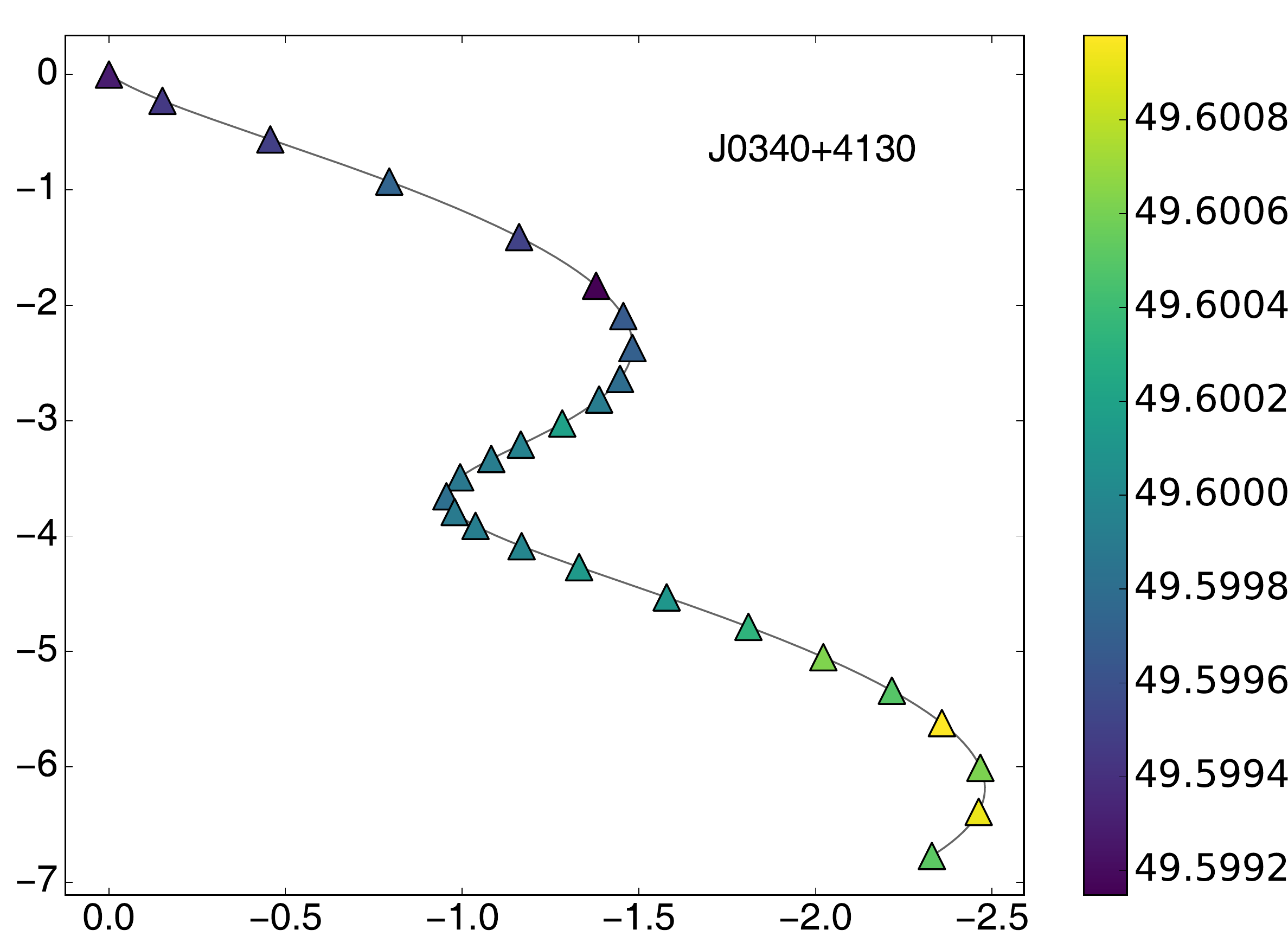}
		\includegraphics[trim=0cm 0cm 0cm 0cm, clip,width=0.4\textwidth]{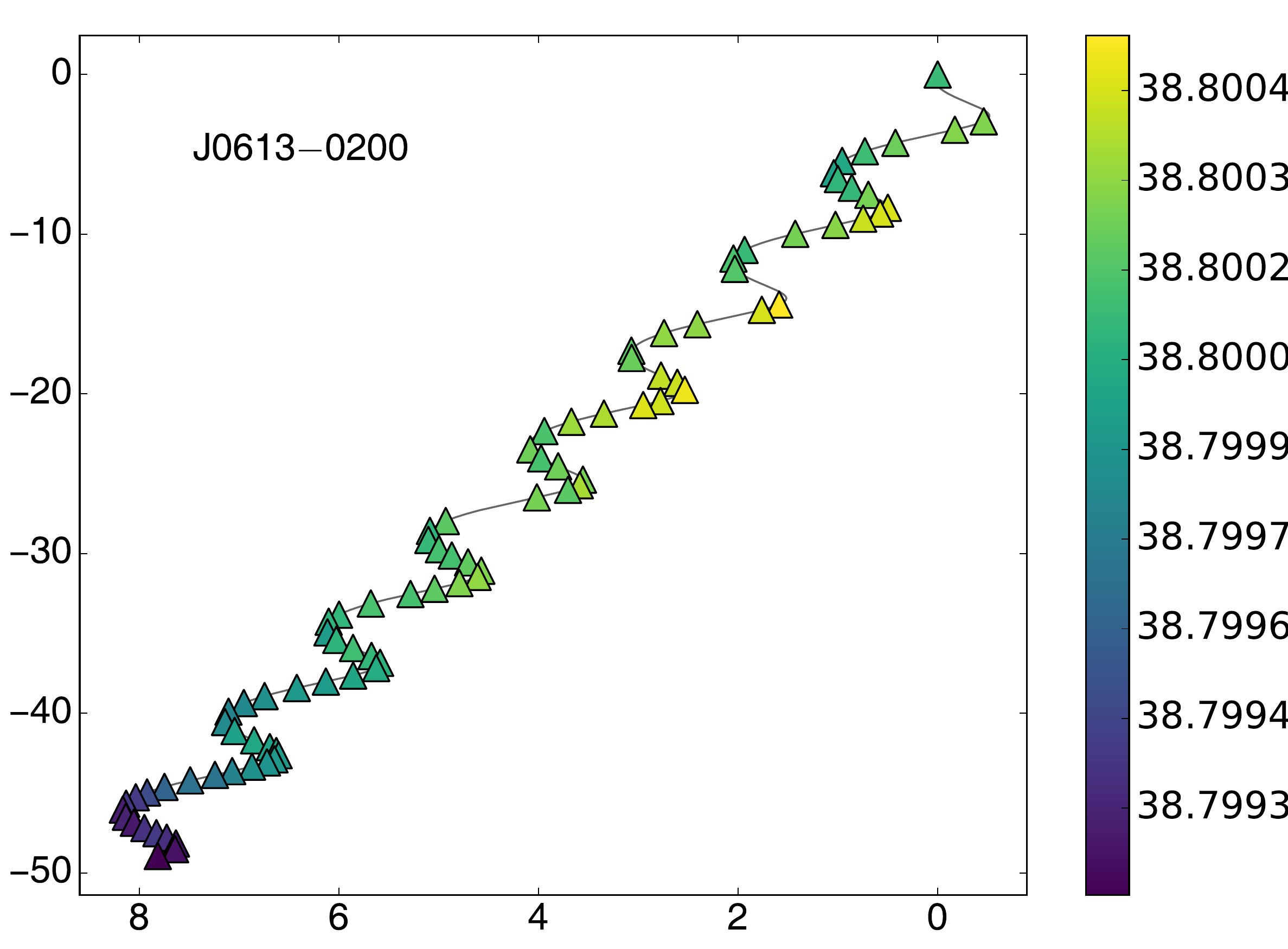}	
		\includegraphics[trim=0cm 0cm 0cm 0cm, clip,width=0.4\textwidth]{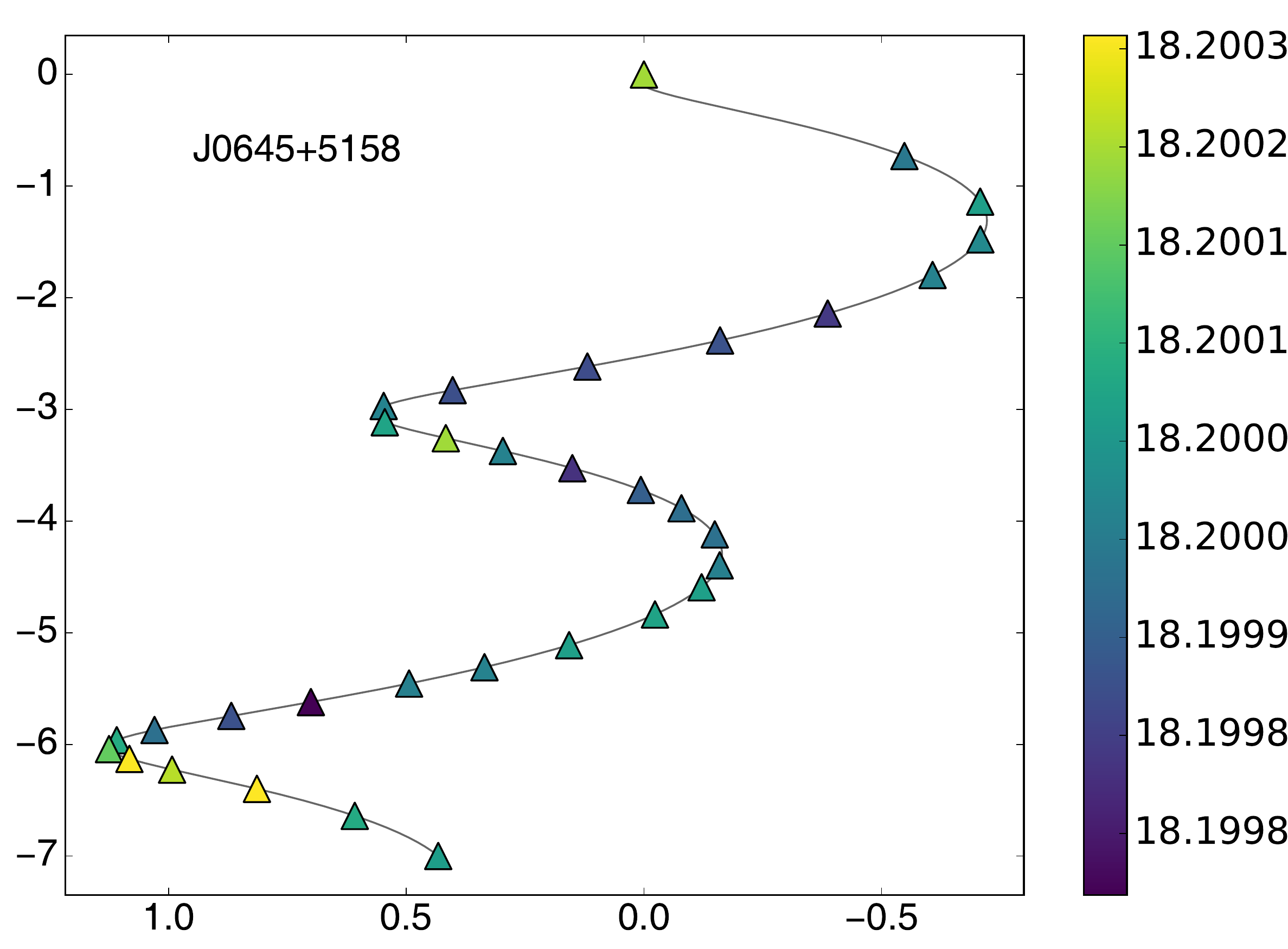}
		\includegraphics[trim=0cm 0cm 0cm 0cm, clip,width=0.4\textwidth]{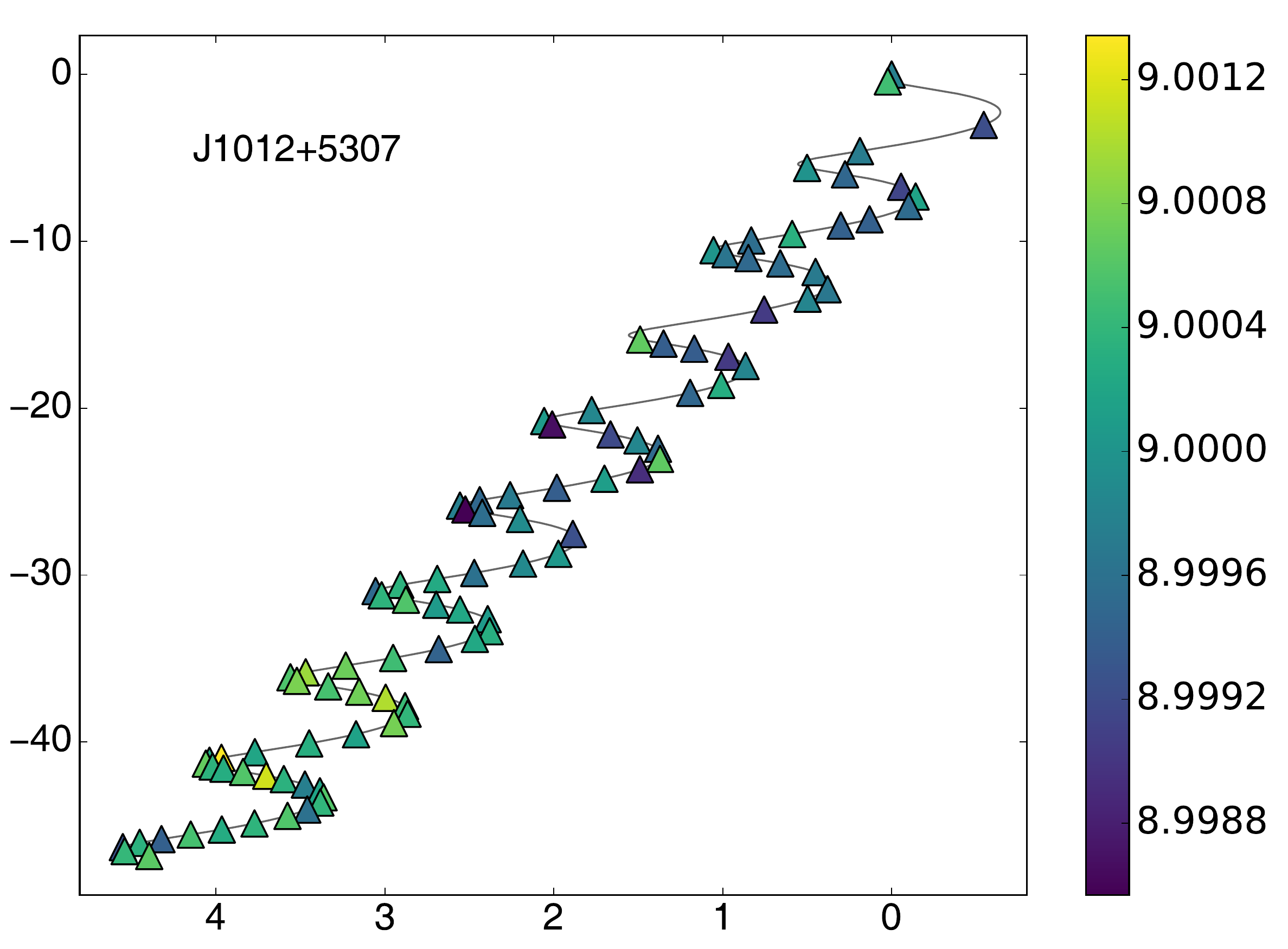}
		\includegraphics[trim=0cm 0cm 0cm 0cm, clip,width=0.4\textwidth]{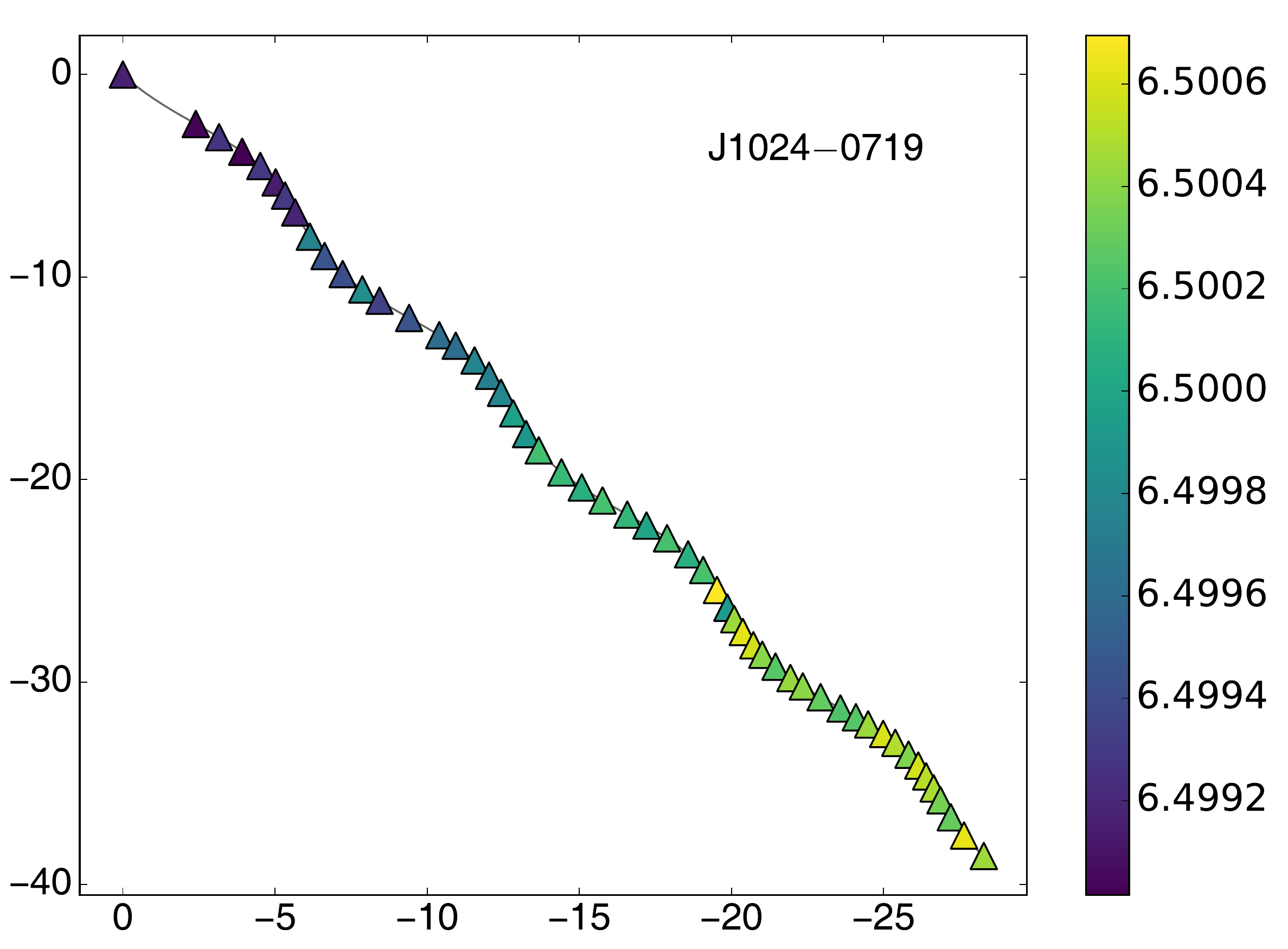}
		\includegraphics[trim=0cm 0cm 0cm 0cm, clip,width=0.4\textwidth]{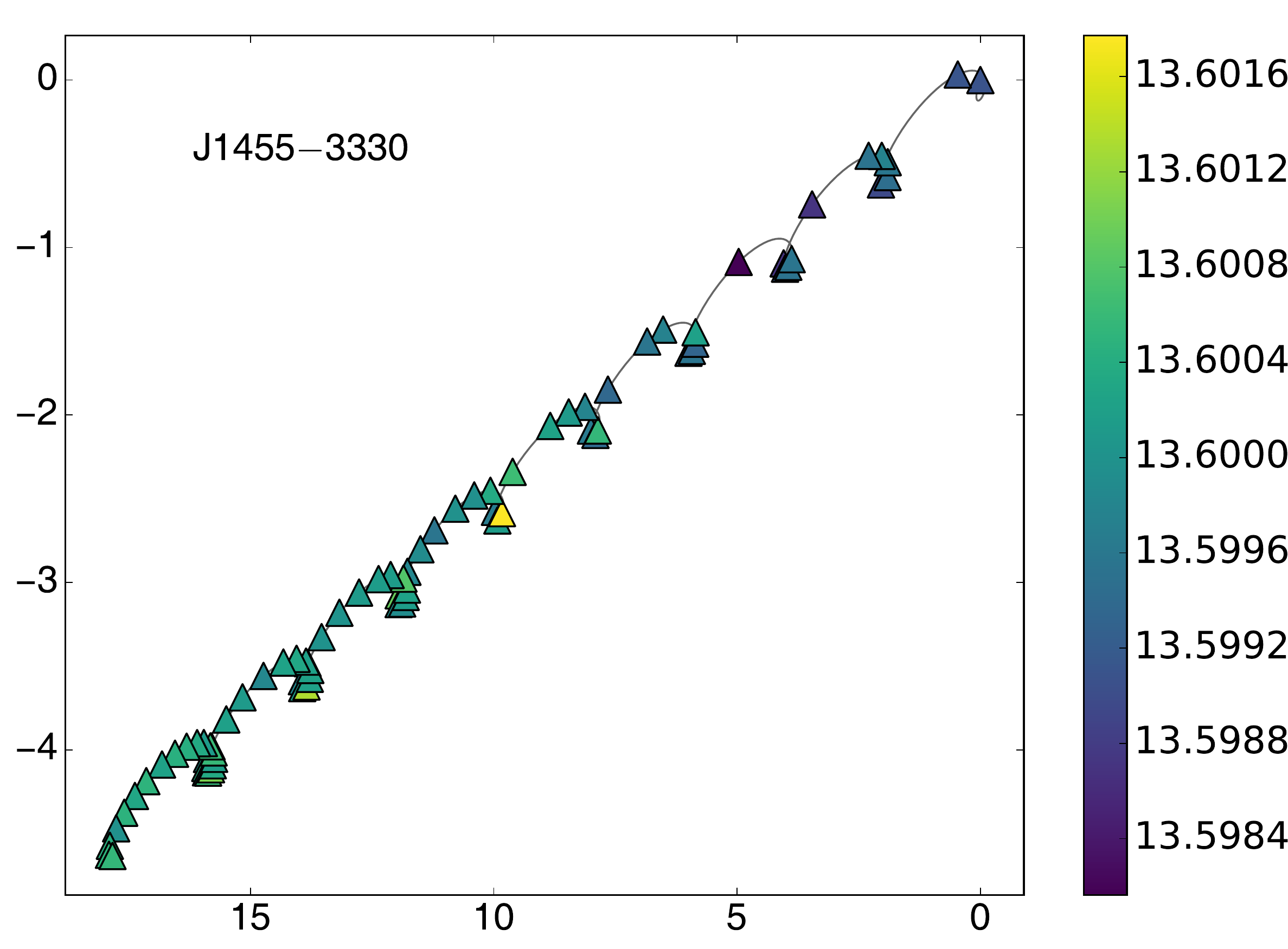}
		
  \leavevmode\smash{\makebox[0pt]{\hspace{5em}
  \rotatebox[origin=l]{90}{\hspace{30em}
    $\Delta_Y$ (AU)}%
}}\hspace{0pt plus 1filll}\null
\leavevmode\smash{\makebox[0pt]{\hspace{47em}
  \rotatebox[origin=l]{90}{\hspace{30em}
    DM (pc cm$^{-3}$)}%
}}\hspace{0pt plus 1filll}\null

$\Delta_X$ (AU)		
	\caption{MSP trajectories are plotted with color mapping the DM at each epoch. The trajectories are calculated assuming that all the free electrons along the LOS are sitting in a phase screen halfway between the Earth and the MSP; the trajectory is then the projected motion of the pulsar on the phase screen. The axes depict the space traversed at the phase screen in AU in the RA and Dec directions.
	The pulsar's motion starts at (0,0). Pulsars closer to the ecliptic will show a tighter sinusoid than those further away. The trajectory plot can be used to show limited localized structure.}
		\label{fig:trajectory1}		
	\end{center}
\end{figure*}		
		
\begin{figure*}
	\begin{center}	
		\includegraphics[trim=0cm 0cm 0cm 0cm, clip,width=0.4\textwidth]{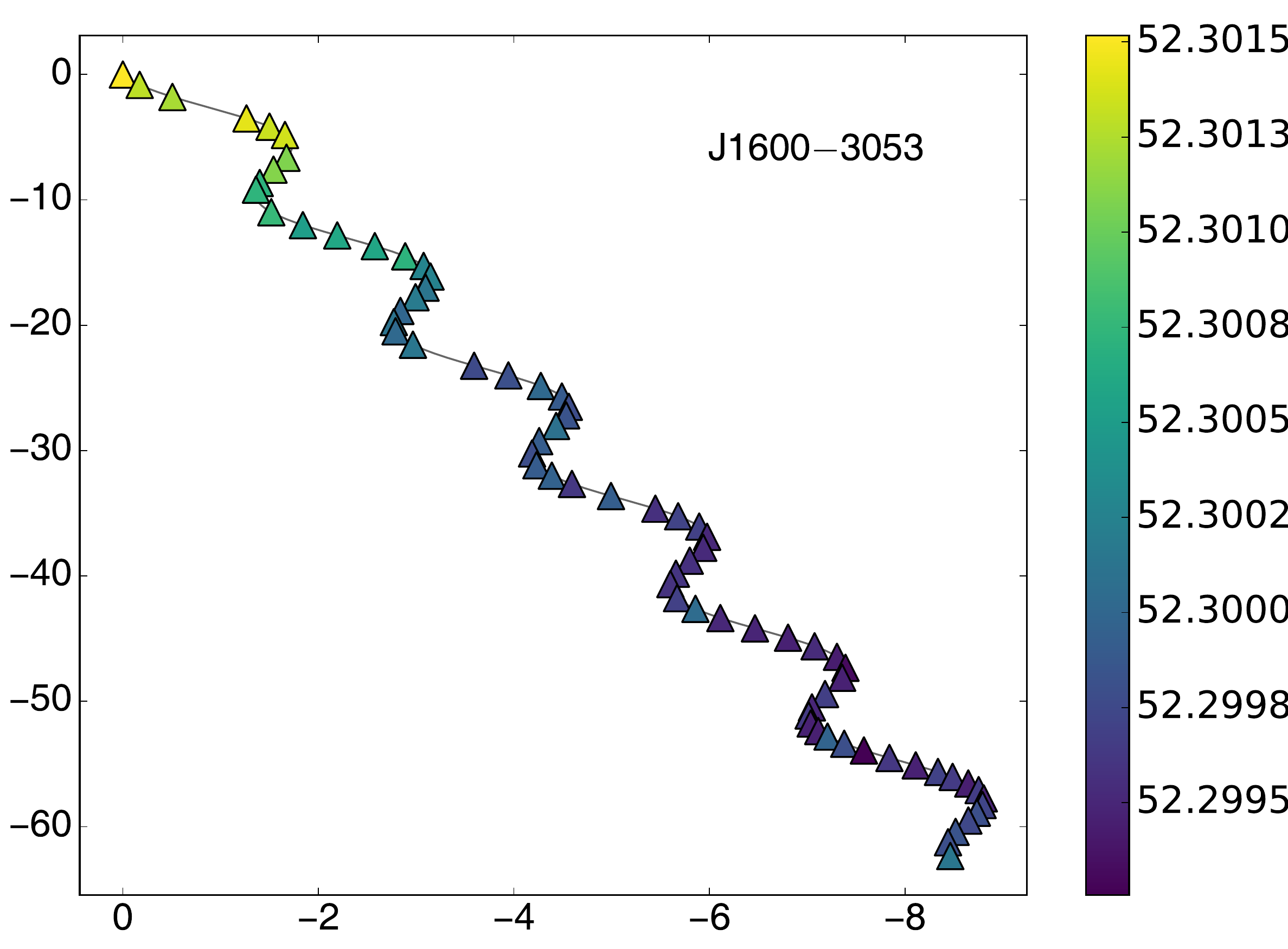}	
		\includegraphics[trim=0cm 0cm 0cm 0cm, clip,width=0.4\textwidth]{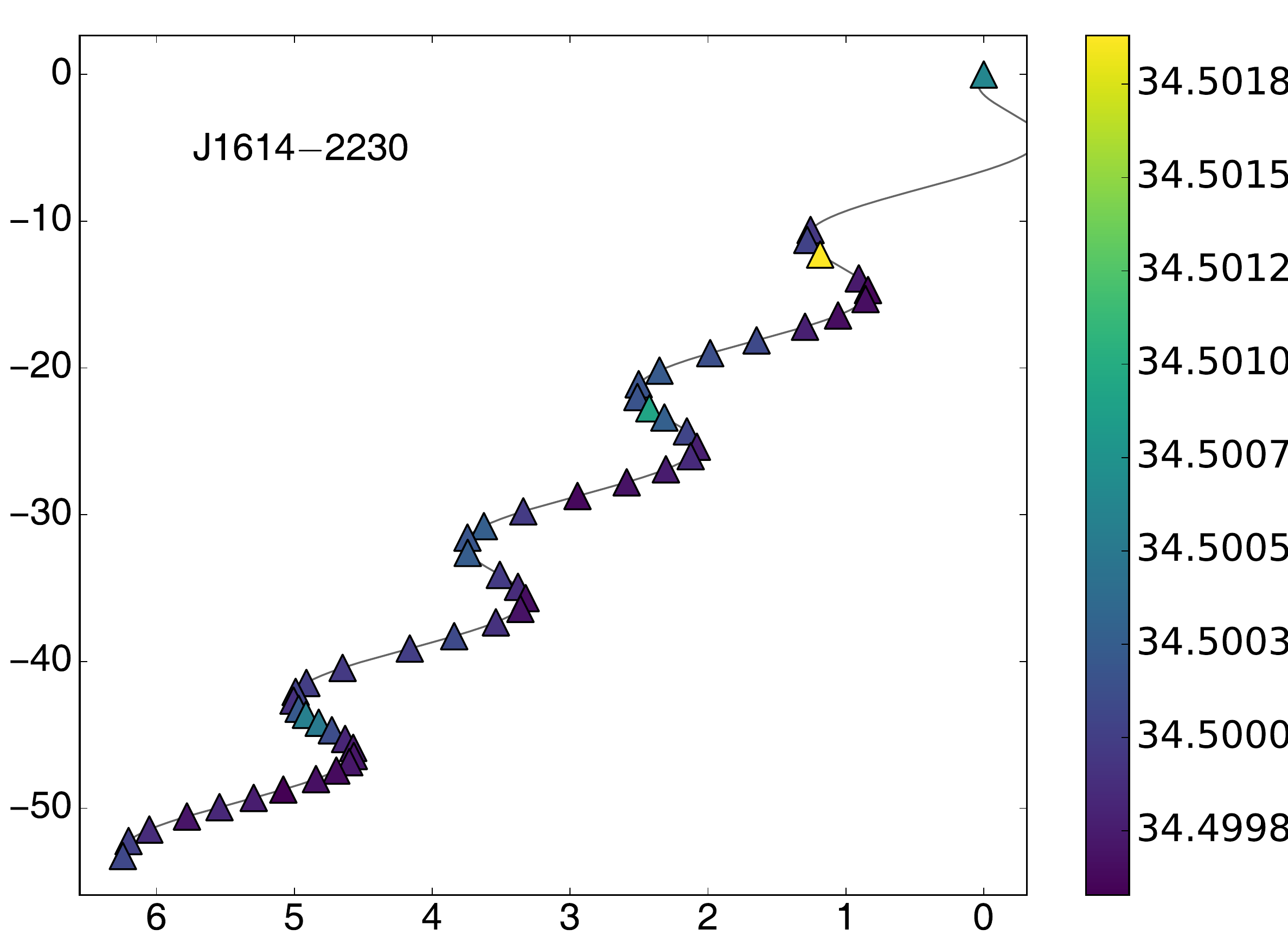}	
		\includegraphics[trim=0cm 0cm 0cm 0cm, clip,width=0.4\textwidth]{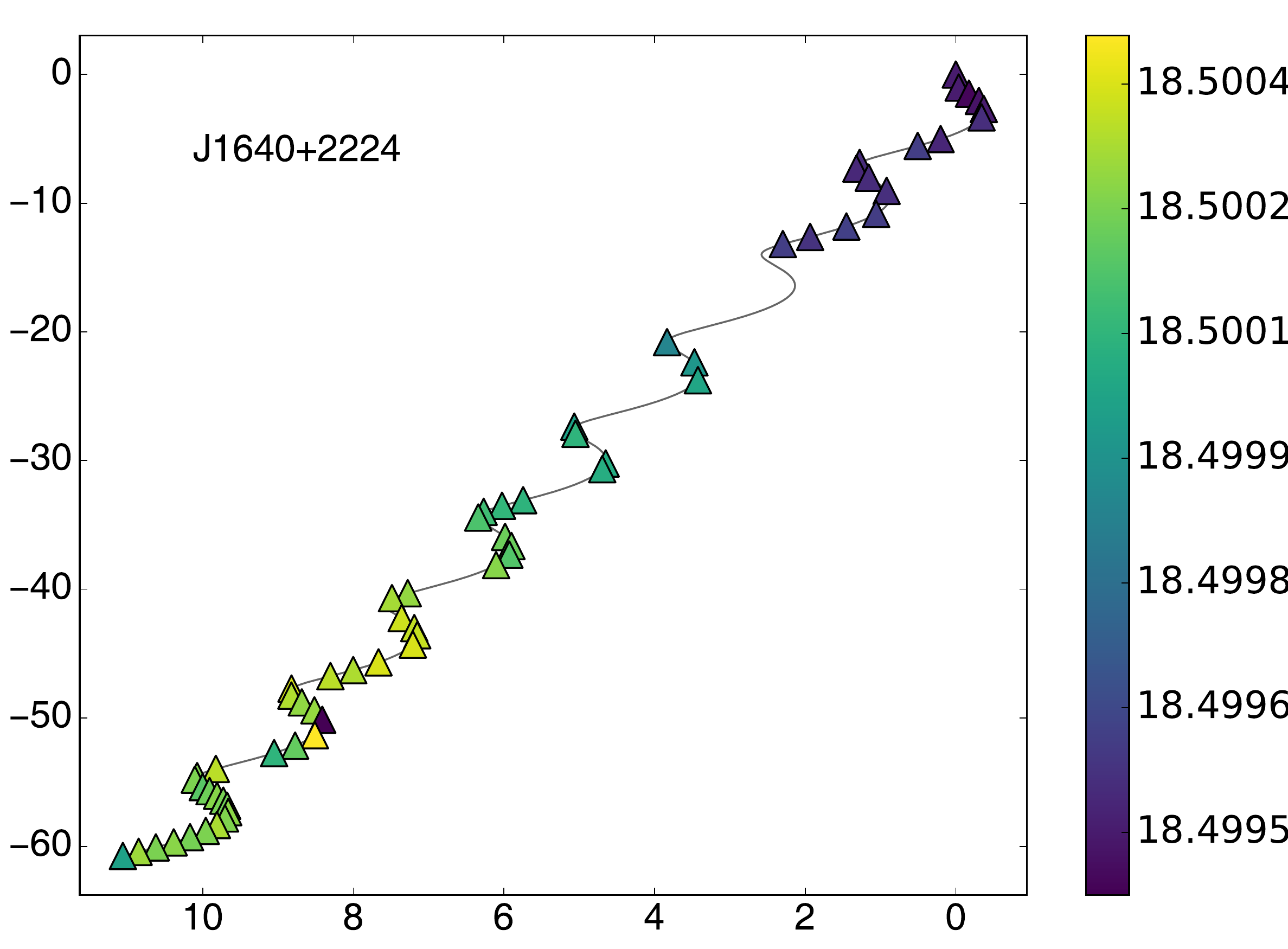}	
		\includegraphics[trim=0cm 0cm 0cm 0cm, clip,width=0.4\textwidth]{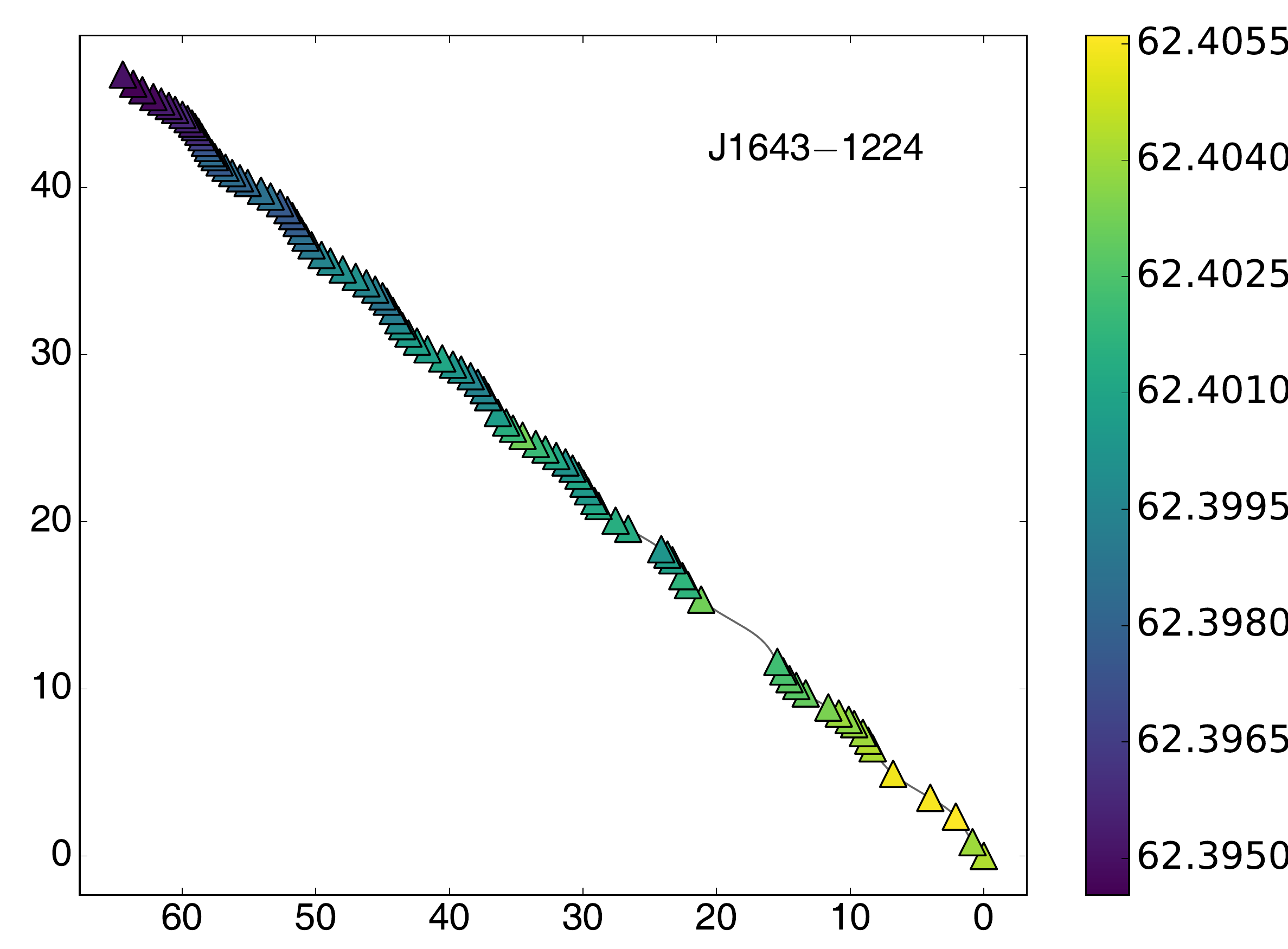}
		\includegraphics[trim=0cm 0cm 0cm 0cm, clip,width=0.4\textwidth]{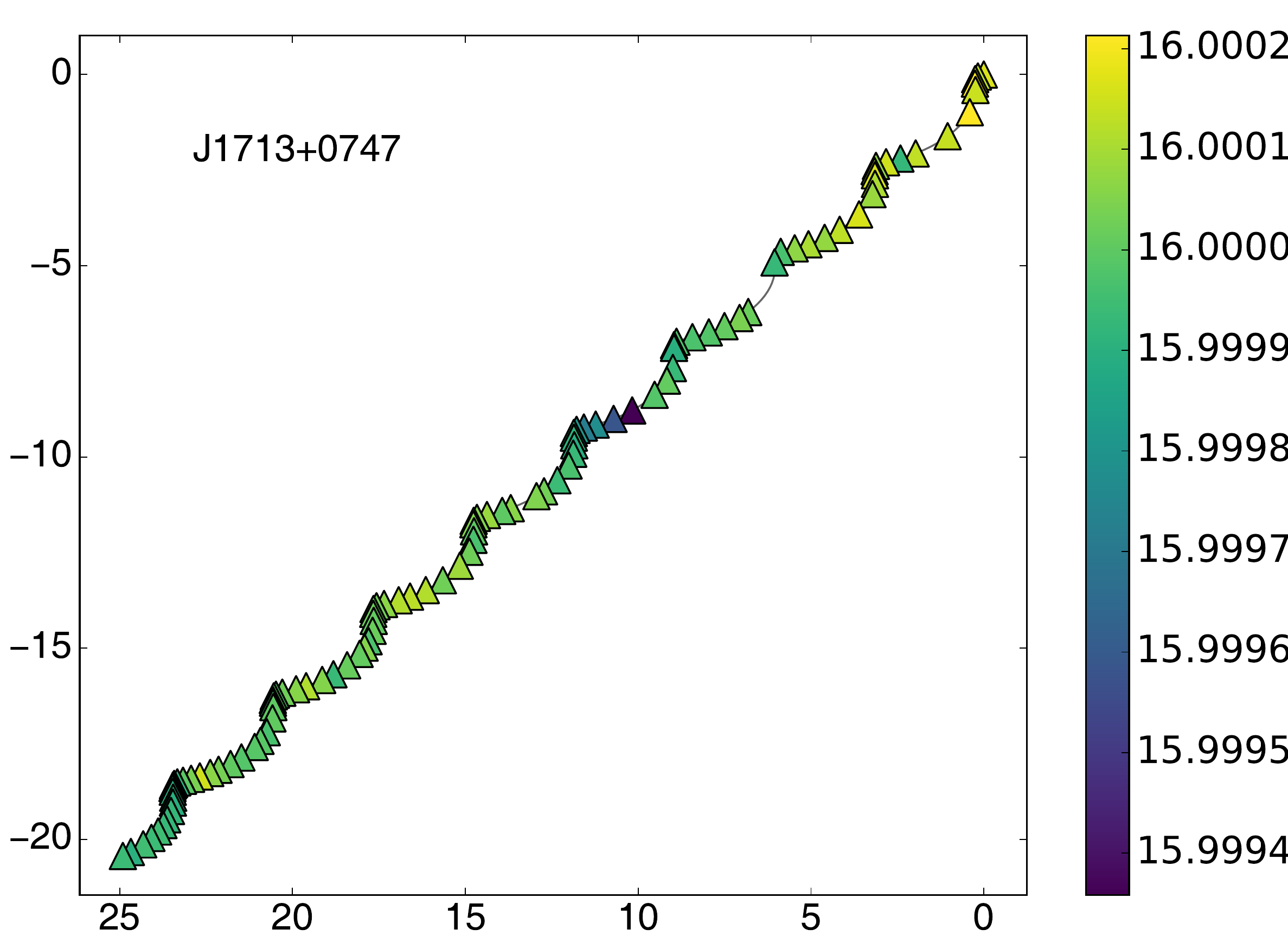}	
		\includegraphics[trim=0cm 0cm 0cm 0cm, clip,width=0.4\textwidth]{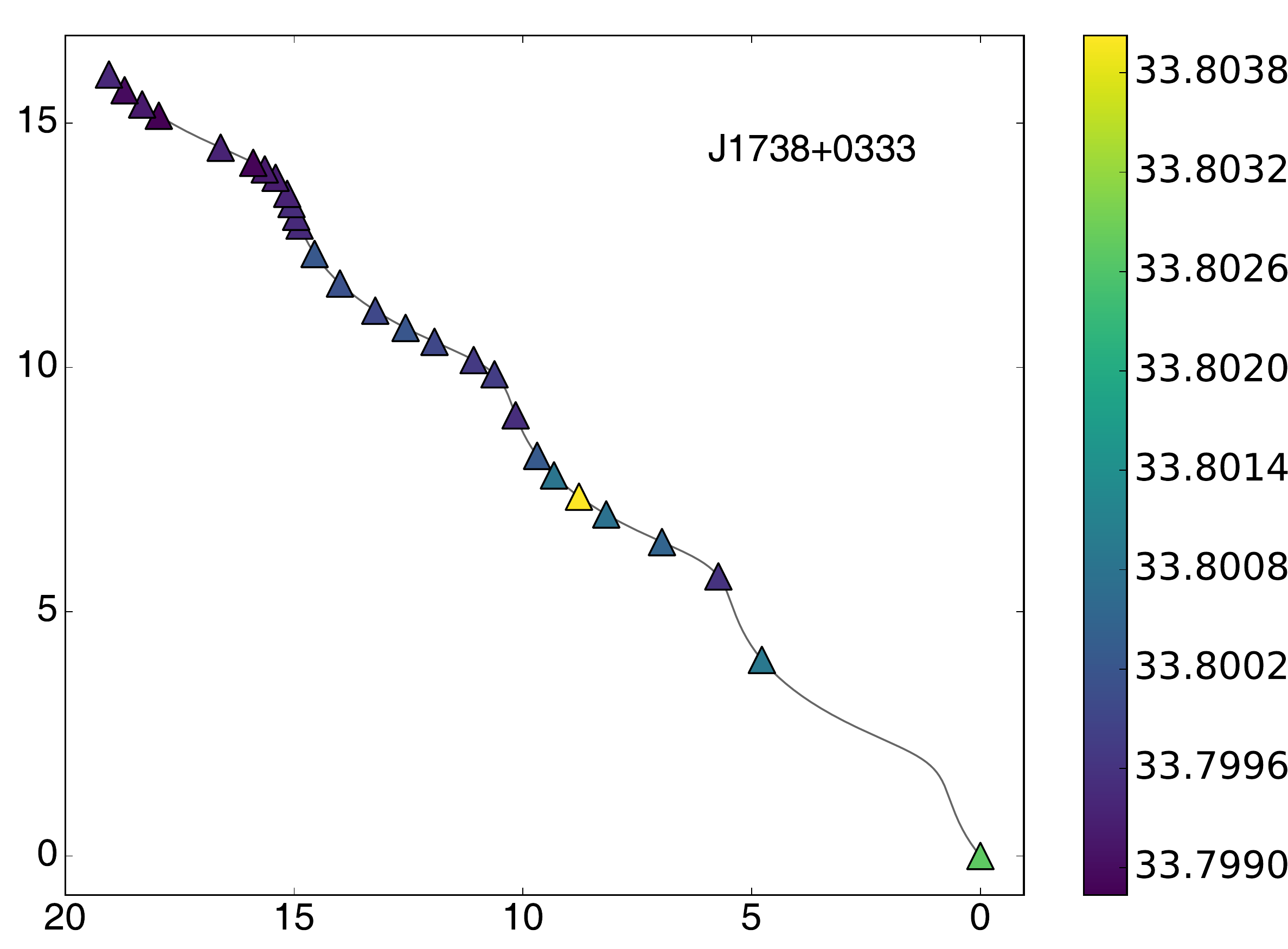}
		\includegraphics[trim=0cm 0cm 0cm 0cm, clip,width=0.4\textwidth]{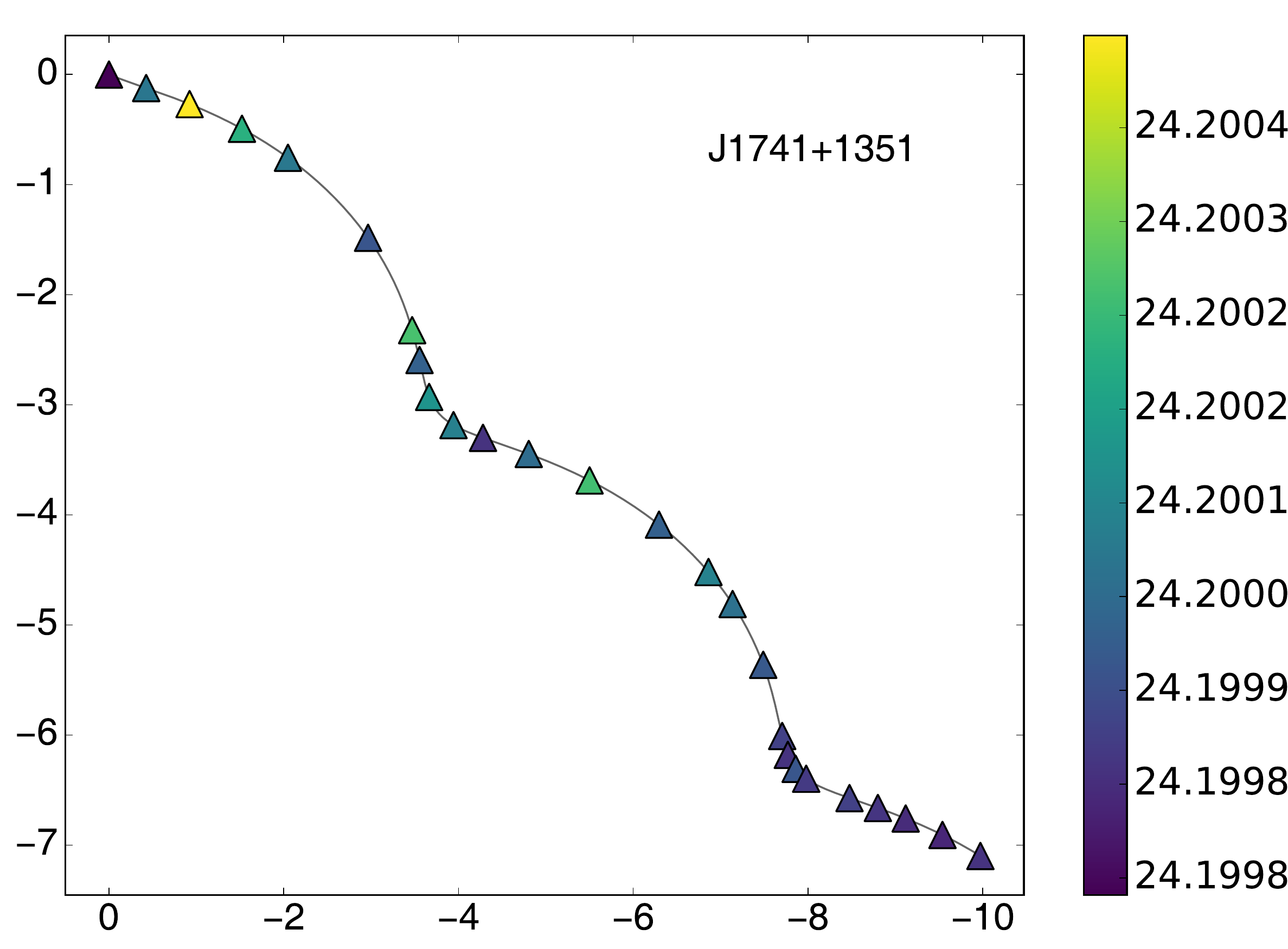}
		\includegraphics[trim=0cm 0cm 0cm 0cm, clip,width=0.4\textwidth]{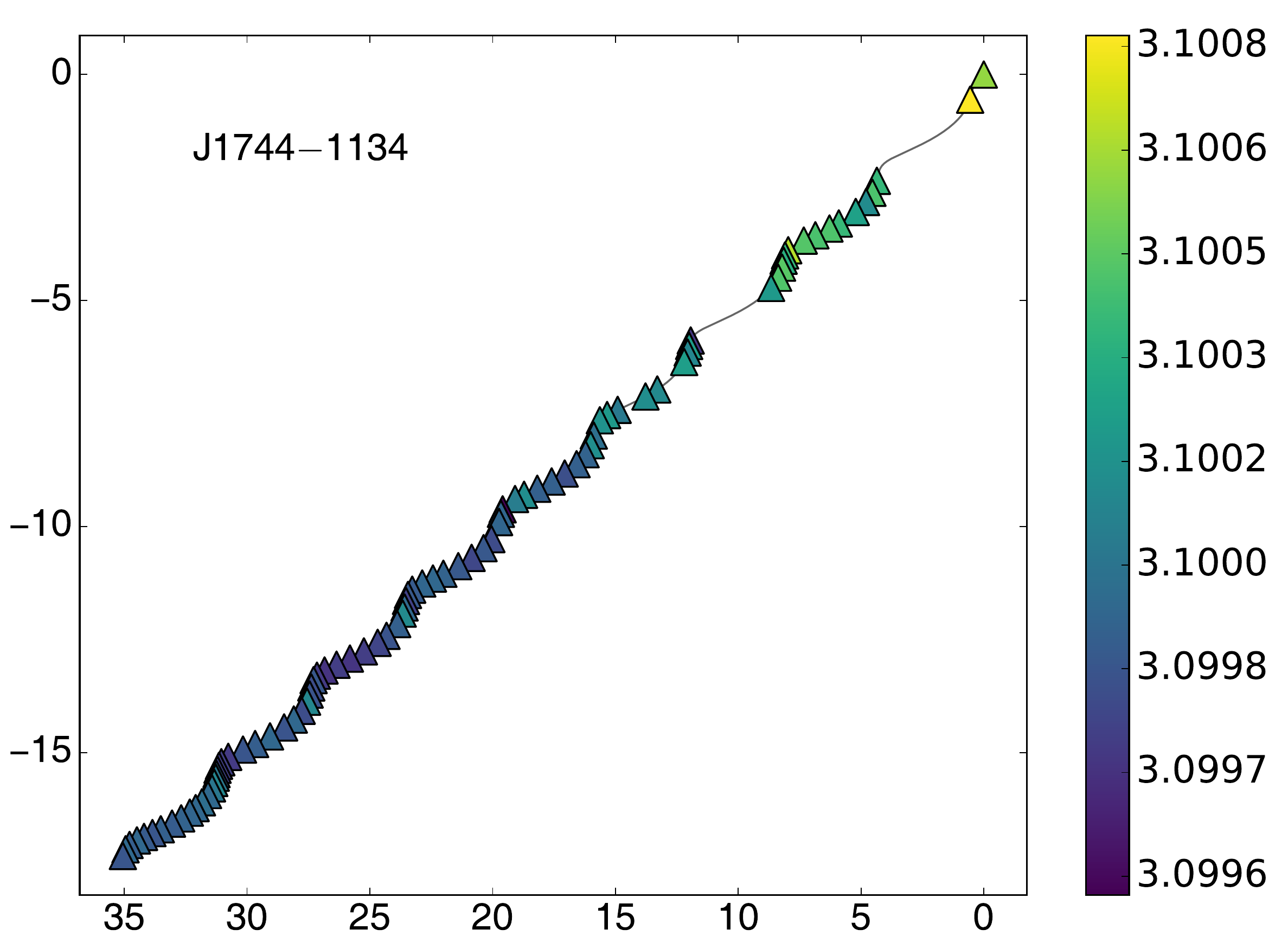}	
		
  \leavevmode\smash{\makebox[0pt]{\hspace{5em}
  \rotatebox[origin=l]{90}{\hspace{30em}
    $\Delta_Y$ (AU)}%
}}\hspace{0pt plus 1filll}\null
\leavevmode\smash{\makebox[0pt]{\hspace{47em}
  \rotatebox[origin=l]{90}{\hspace{30em}
    DM (pc cm$^{-3}$)}%
}}\hspace{0pt plus 1filll}\null

$\Delta_X$ (AU)			
		\caption{MSP trajectories are plotted with color mapping the DM at each epoch. 
		The trajectories are calculated assuming that all the free electrons along the LOS are sitting in a phase screen halfway between the Earth and the MSP; the trajectory is then the projected motion of the pulsar on the phase screen. The axes depict the space traversed at the phase screen in AU in the RA and Dec directions.
		The pulsar's motion starts at (0,0). Pulsars closer to the ecliptic will show a tighter sinusoid than those further away. The trajectory plot can be used to show limited localized structure.}
	\label{fig:trajectory2}		
	\end{center}
\end{figure*}
\begin{figure*}
	\begin{center}	
		\includegraphics[trim=0cm 0cm 0cm 0cm, clip,width=0.4\textwidth]{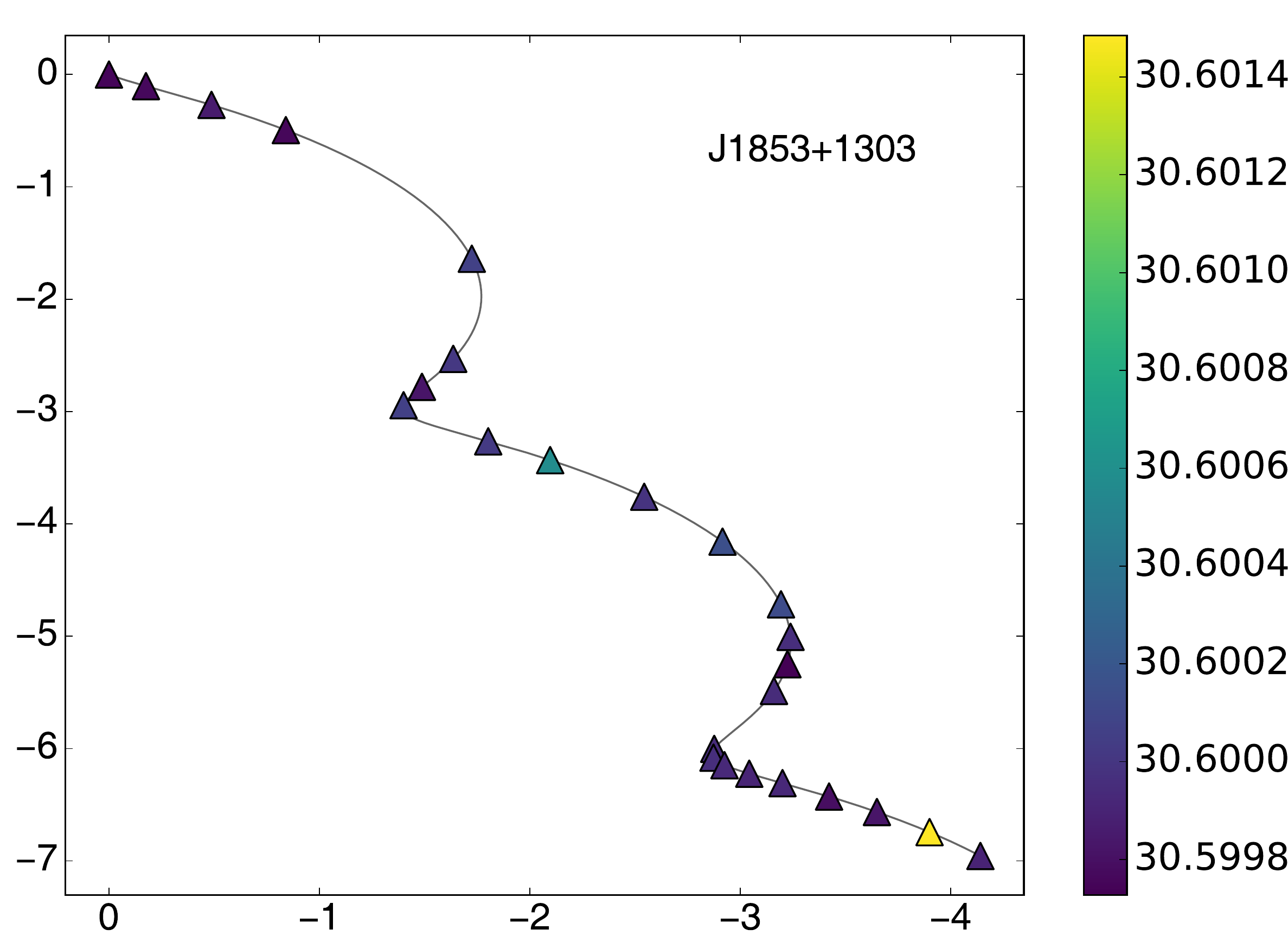}
		\includegraphics[trim=0cm 0cm 0cm 0cm, clip,width=0.4\textwidth]{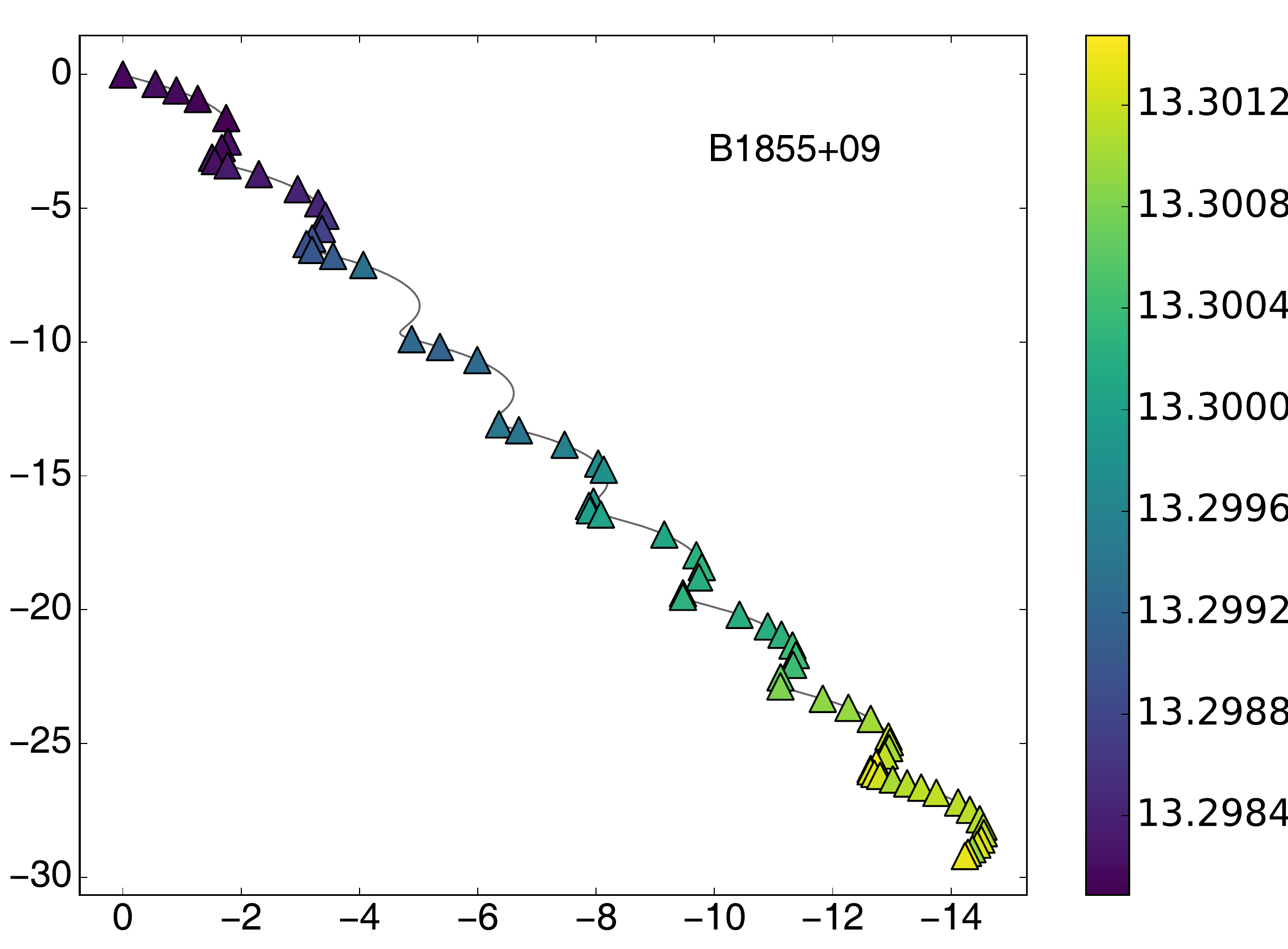}	
		\includegraphics[trim=0cm 0cm 0cm 0cm, clip,width=0.4\textwidth]{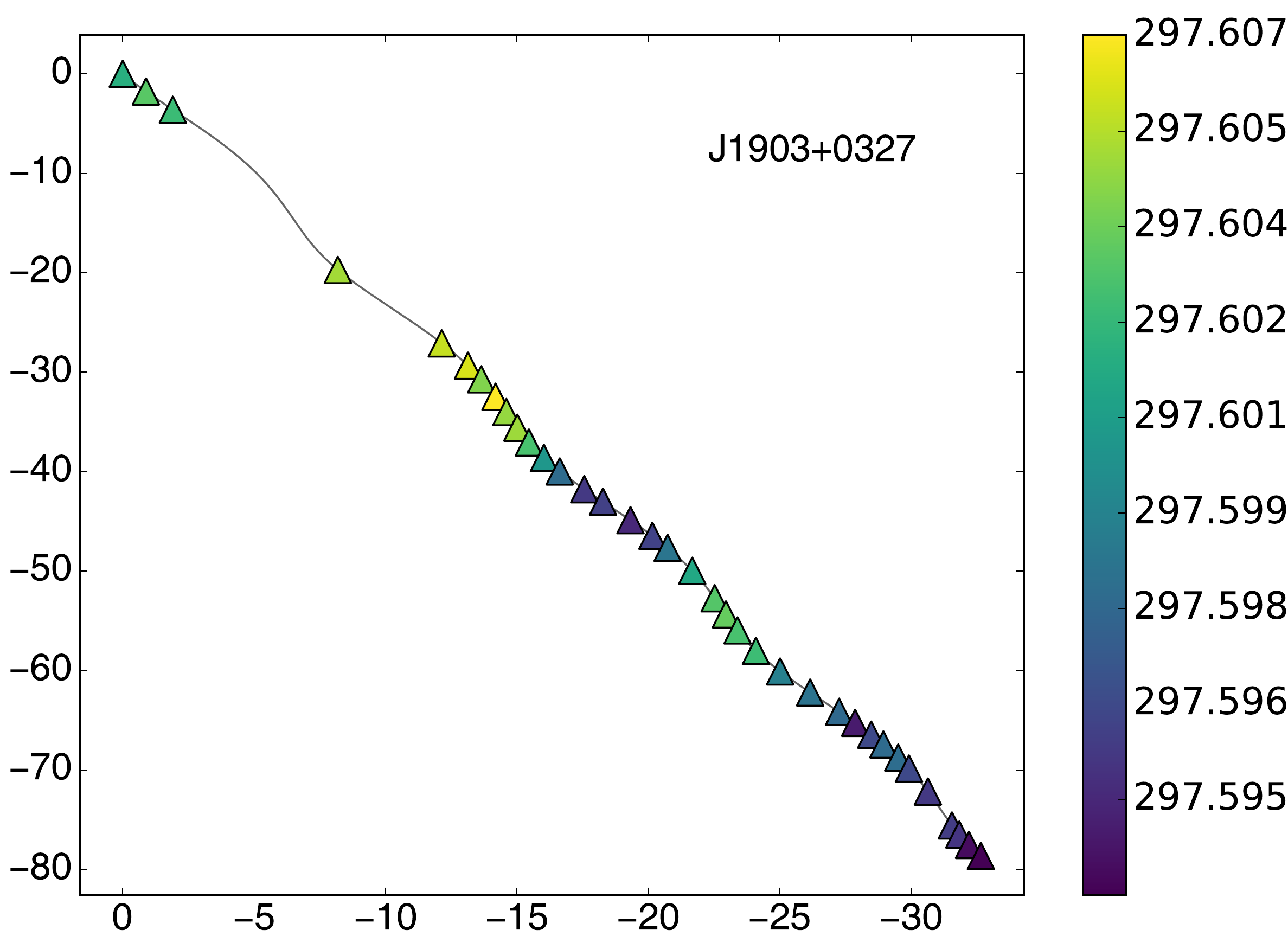}	
		\includegraphics[trim=0cm 0cm 0cm 0cm, clip,width=0.4\textwidth]{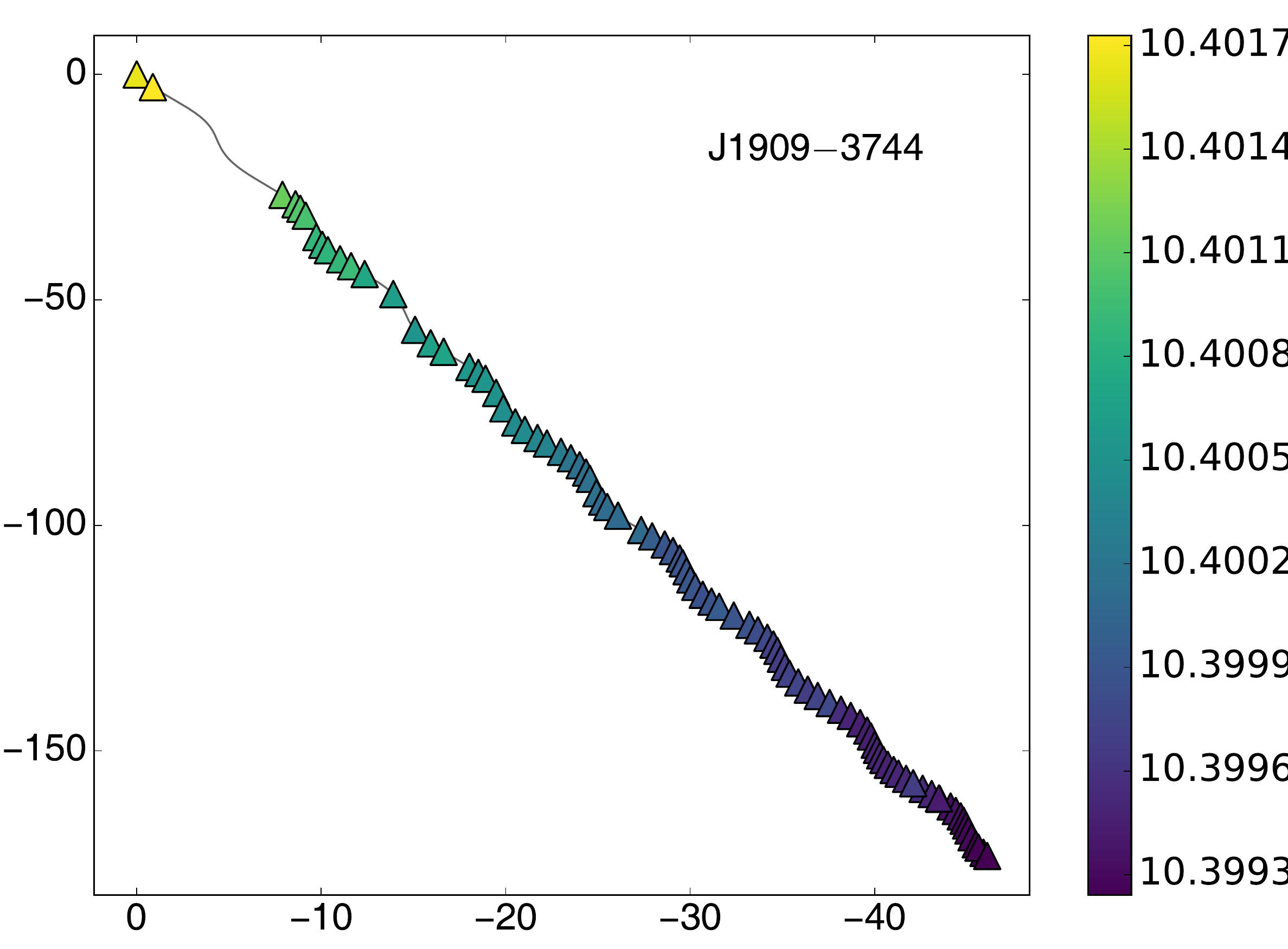}
		\includegraphics[trim=0cm 0cm 0cm 0cm, clip,width=0.4\textwidth]{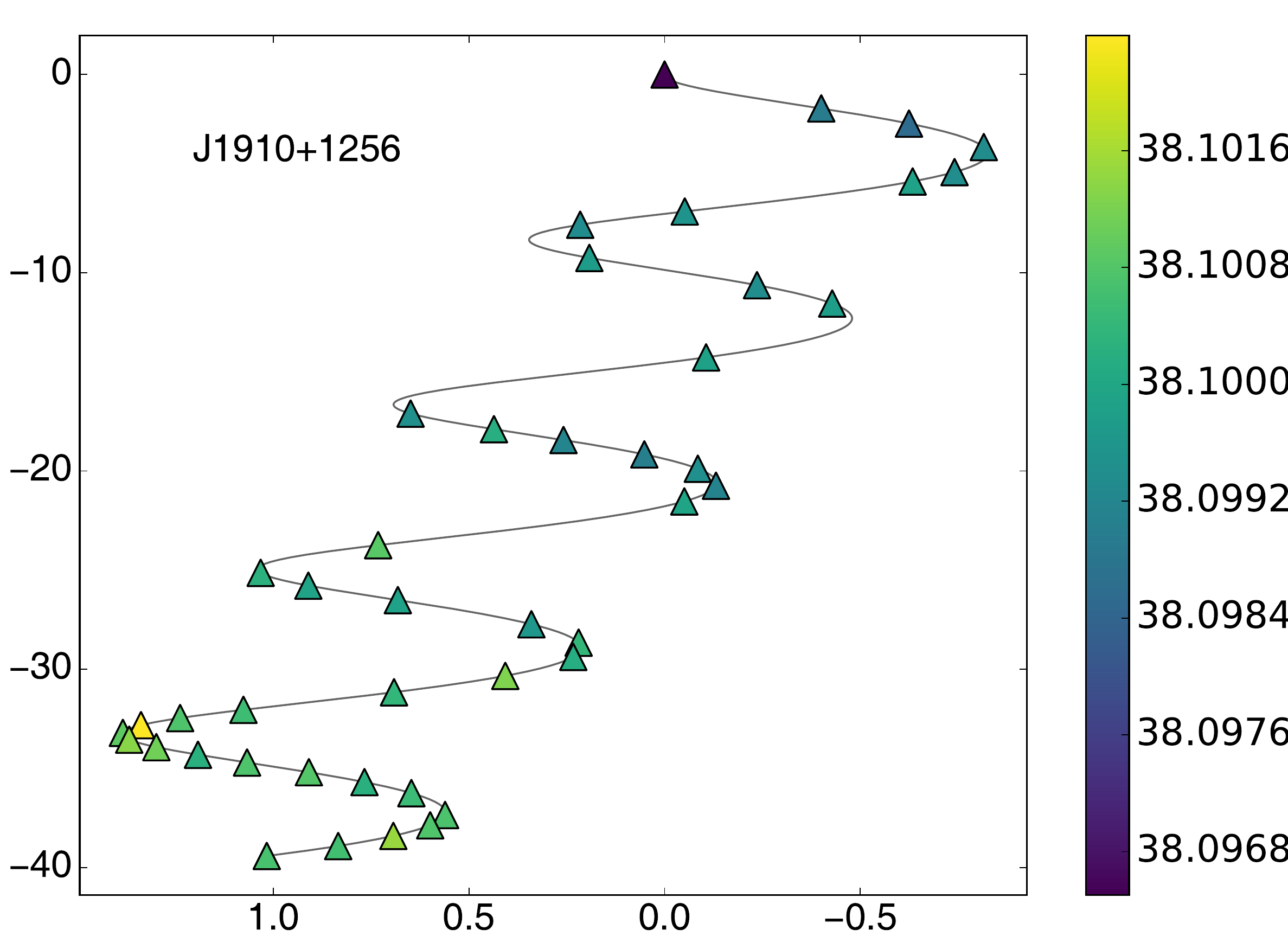}	
		\includegraphics[trim=0cm 0cm 0cm 0cm, clip,width=0.4\textwidth]{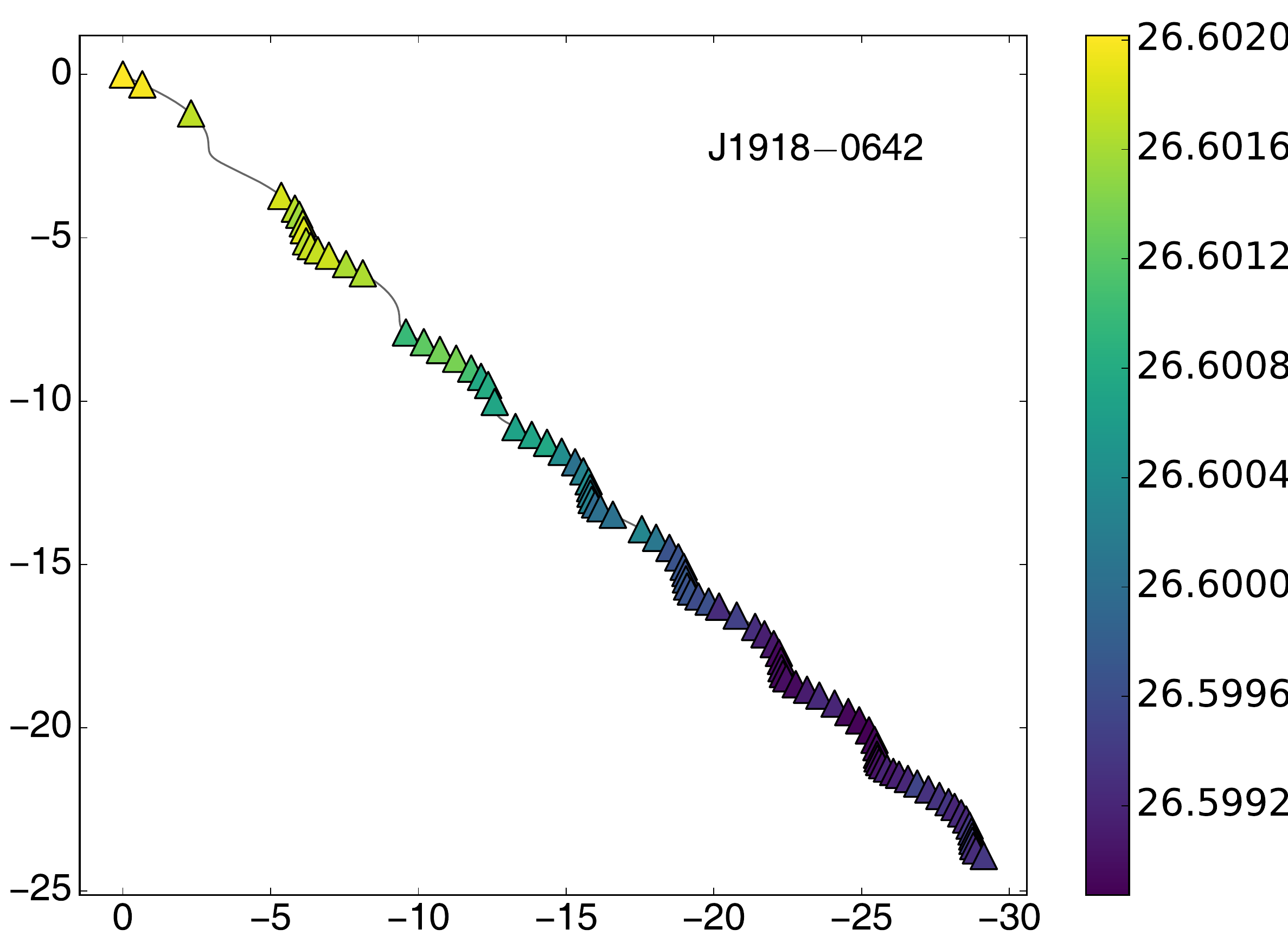}
		\includegraphics[trim=0cm 0cm 0cm 0cm, clip,width=0.4\textwidth]{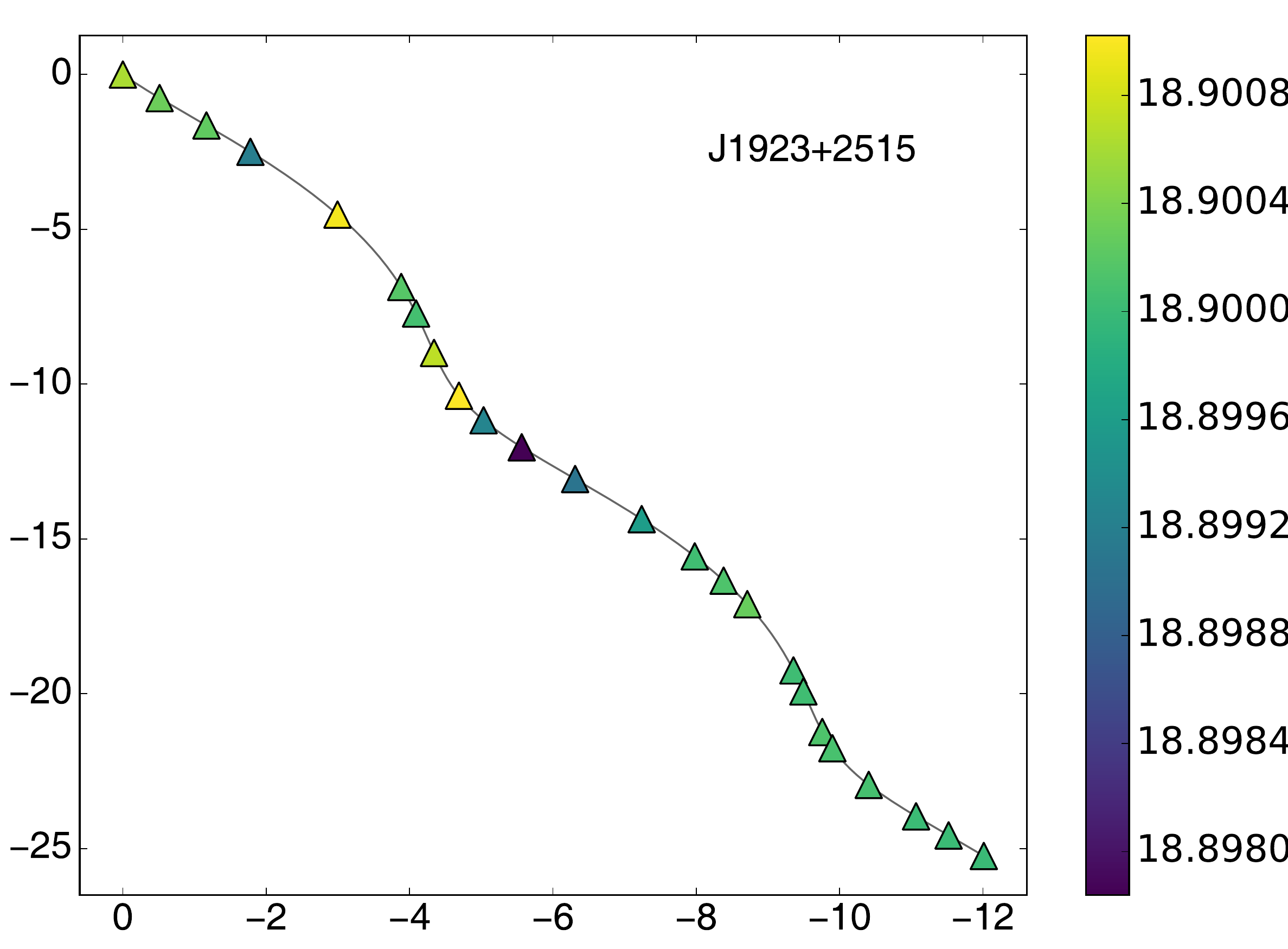}
		\includegraphics[trim=0cm 0cm 0cm 0cm, clip,width=0.4\textwidth]{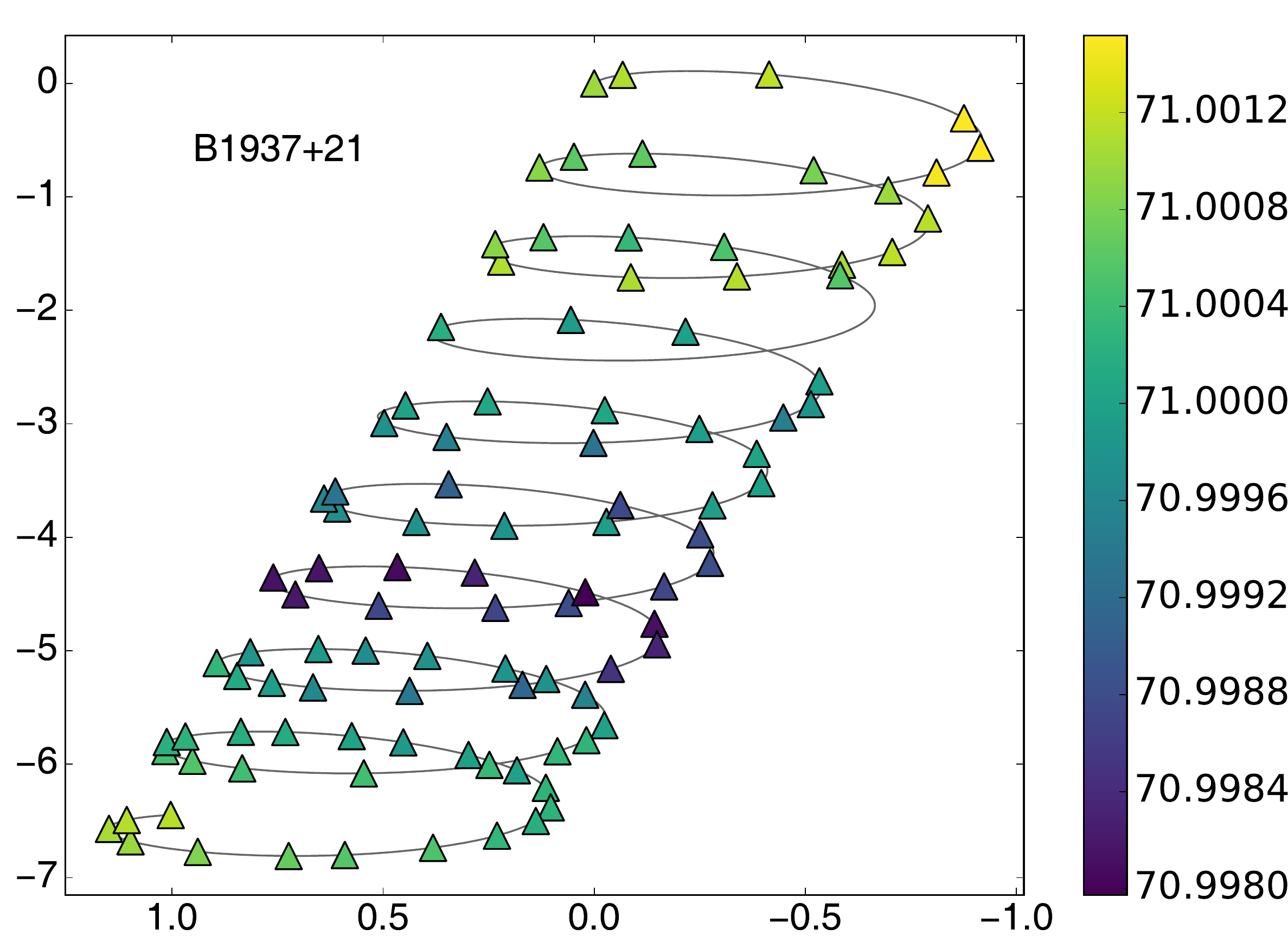}	
		
   \leavevmode\smash{\makebox[0pt]{\hspace{5em}
  \rotatebox[origin=l]{90}{\hspace{30em}
    $\Delta_Y$ (AU)}%
}}\hspace{0pt plus 1filll}\null
\leavevmode\smash{\makebox[0pt]{\hspace{47em}
  \rotatebox[origin=l]{90}{\hspace{30em}
    DM (pc cm$^{-3}$)}%
}}\hspace{0pt plus 1filll}\null

$\Delta_X$ (AU)				
	\caption{MSP trajectories are plotted with color mapping the DM at each epoch. 
The trajectories are calculated assuming that all the free electrons along the LOS are sitting in a phase screen halfway between the Earth and the MSP; the trajectory is then the projected motion of the pulsar on the phase screen. The axes depict the space traversed at the phase screen in AU in the RA and Dec directions.
	The pulsar's motion starts at (0,0). Pulsars closer to the ecliptic will show a tighter sinusoid than those further away. The trajectory plot can be used to show limited localized structure.}
	\label{fig:trajectory3}		
	\end{center}
\end{figure*}

\begin{figure*}
	\begin{center}
		\includegraphics[trim=0cm 0cm 0cm 0cm, clip,width=0.4\textwidth]{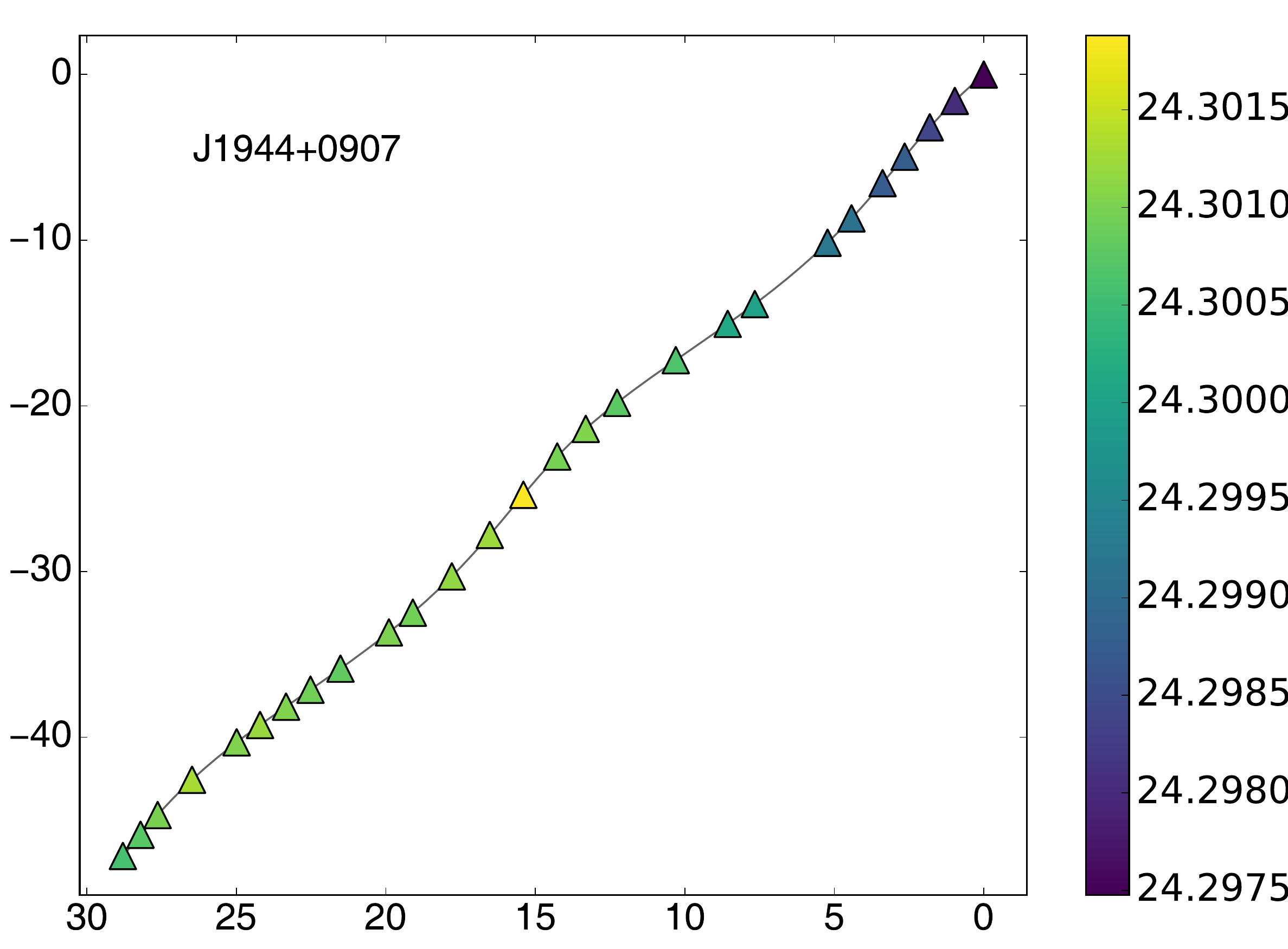}
		\includegraphics[trim=0cm 0cm 0cm 0cm, clip,width=0.4\textwidth]{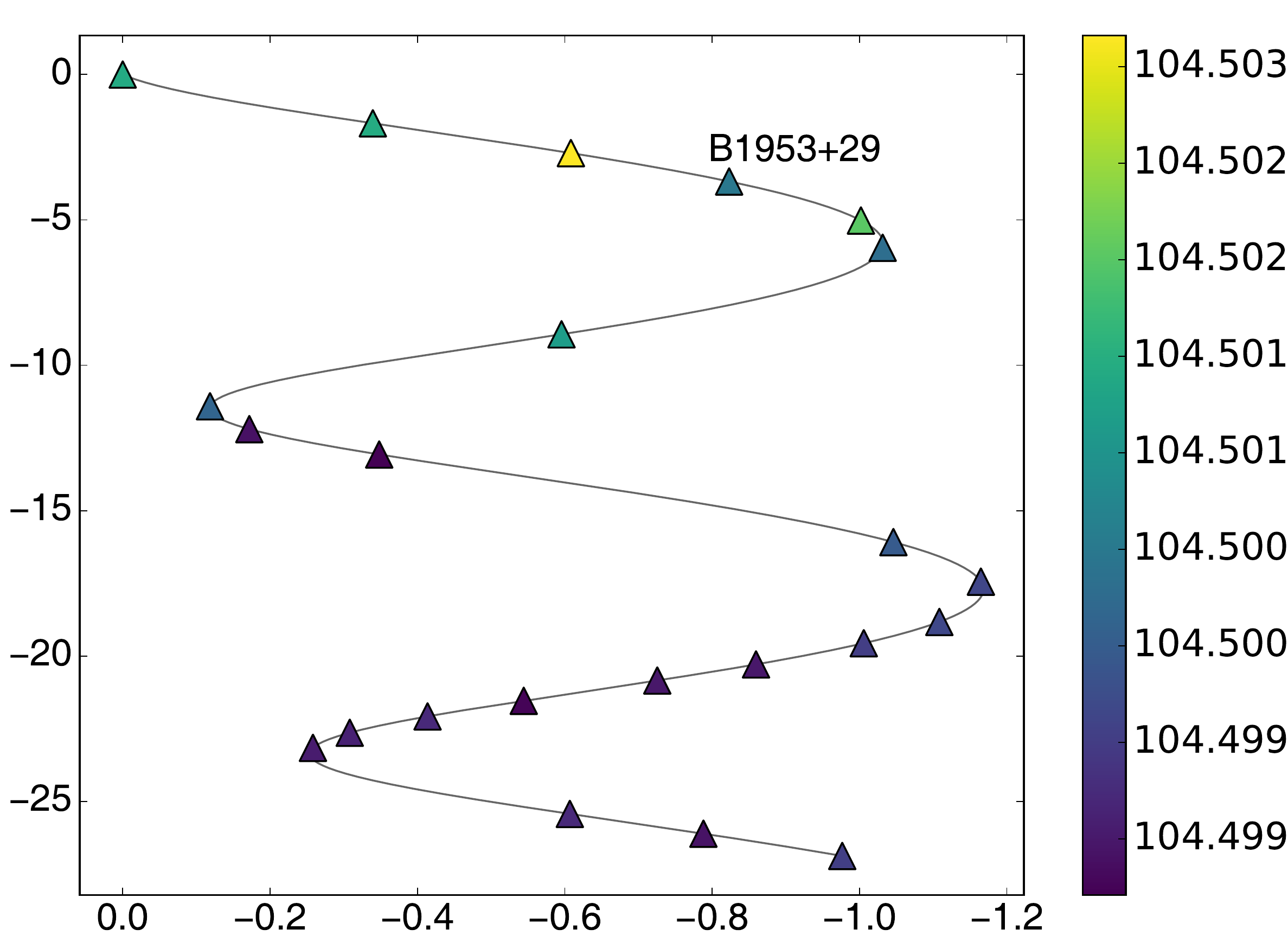}	
		\includegraphics[trim=0cm 0cm 0cm 0cm, clip,width=0.4\textwidth]{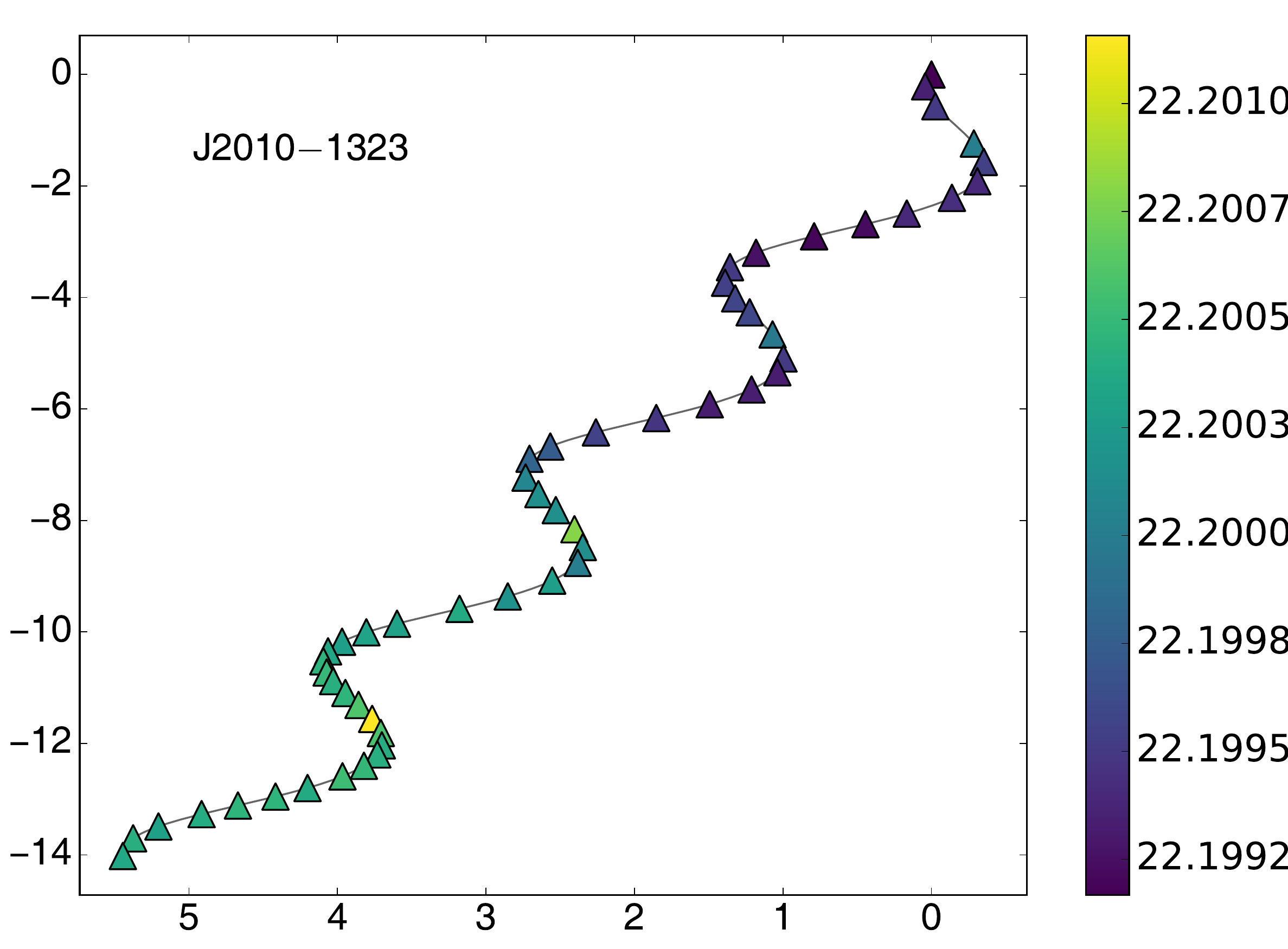}	
		\includegraphics[trim=0cm 0cm 0cm 0cm, clip,width=0.4\textwidth]{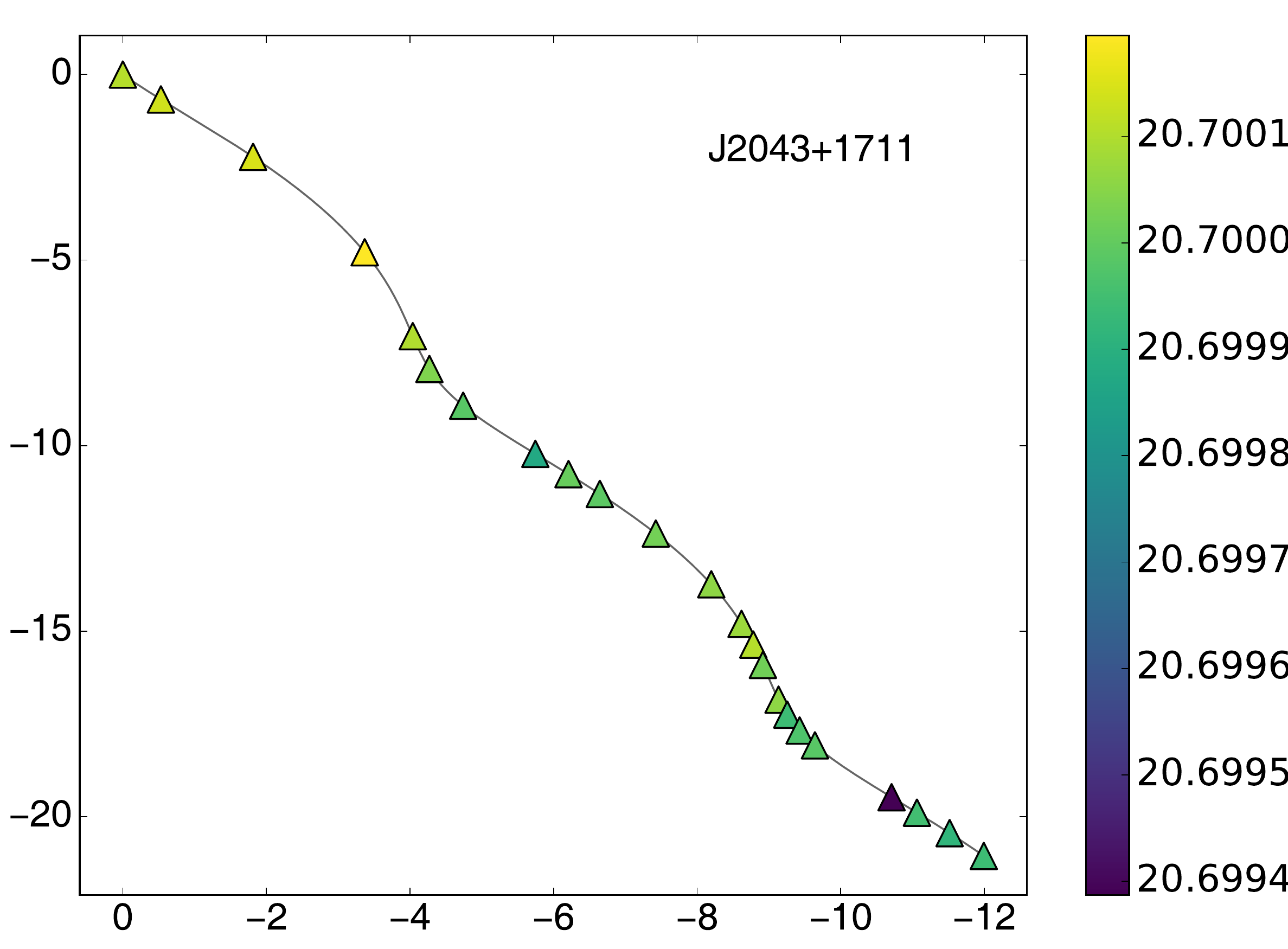}	
		\includegraphics[trim=0cm 0cm 0cm 0cm, clip,width=0.4\textwidth]{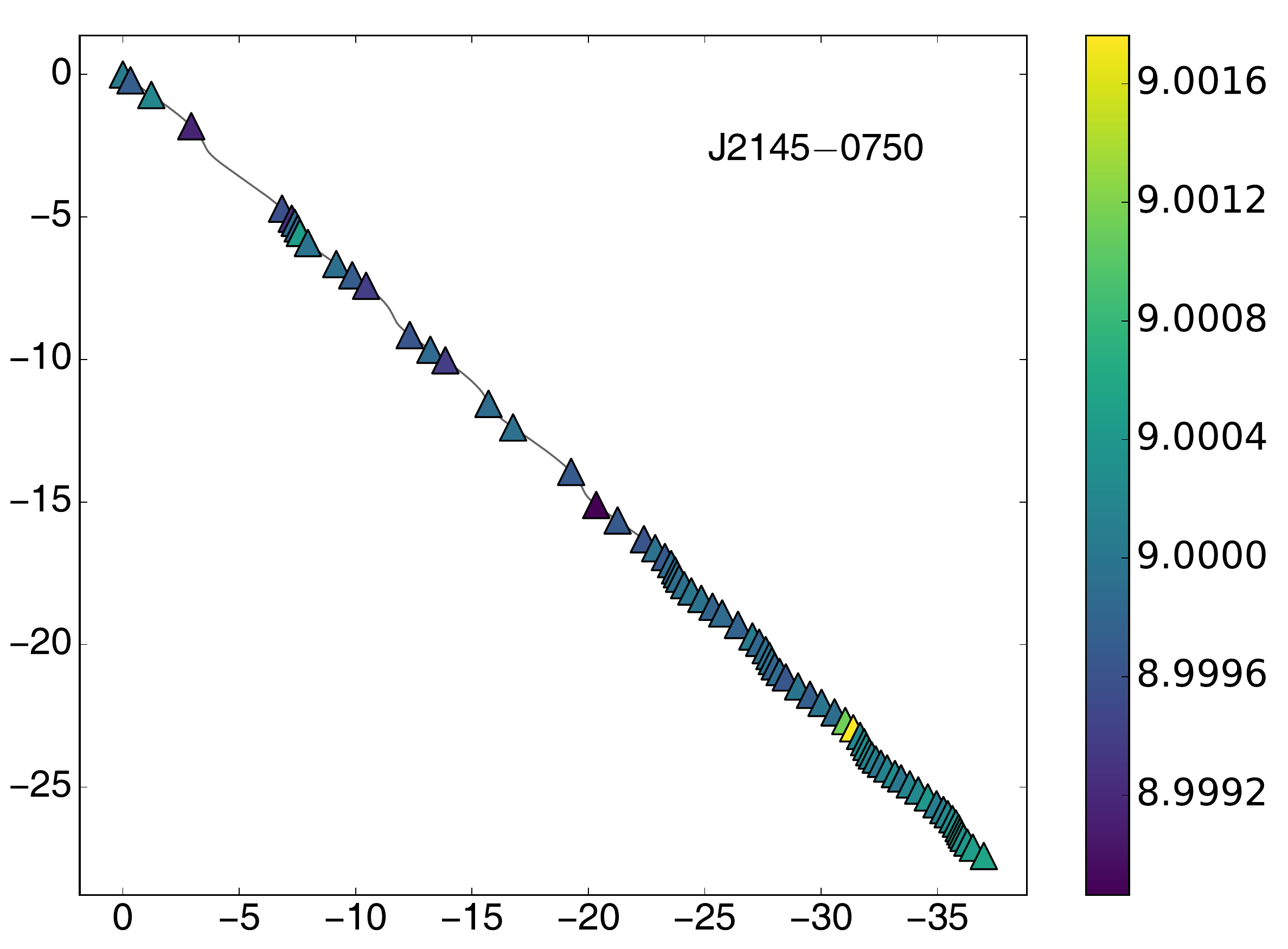}	
  	    \includegraphics[trim=0cm 0cm 0cm 0cm, clip,width=0.4\textwidth]{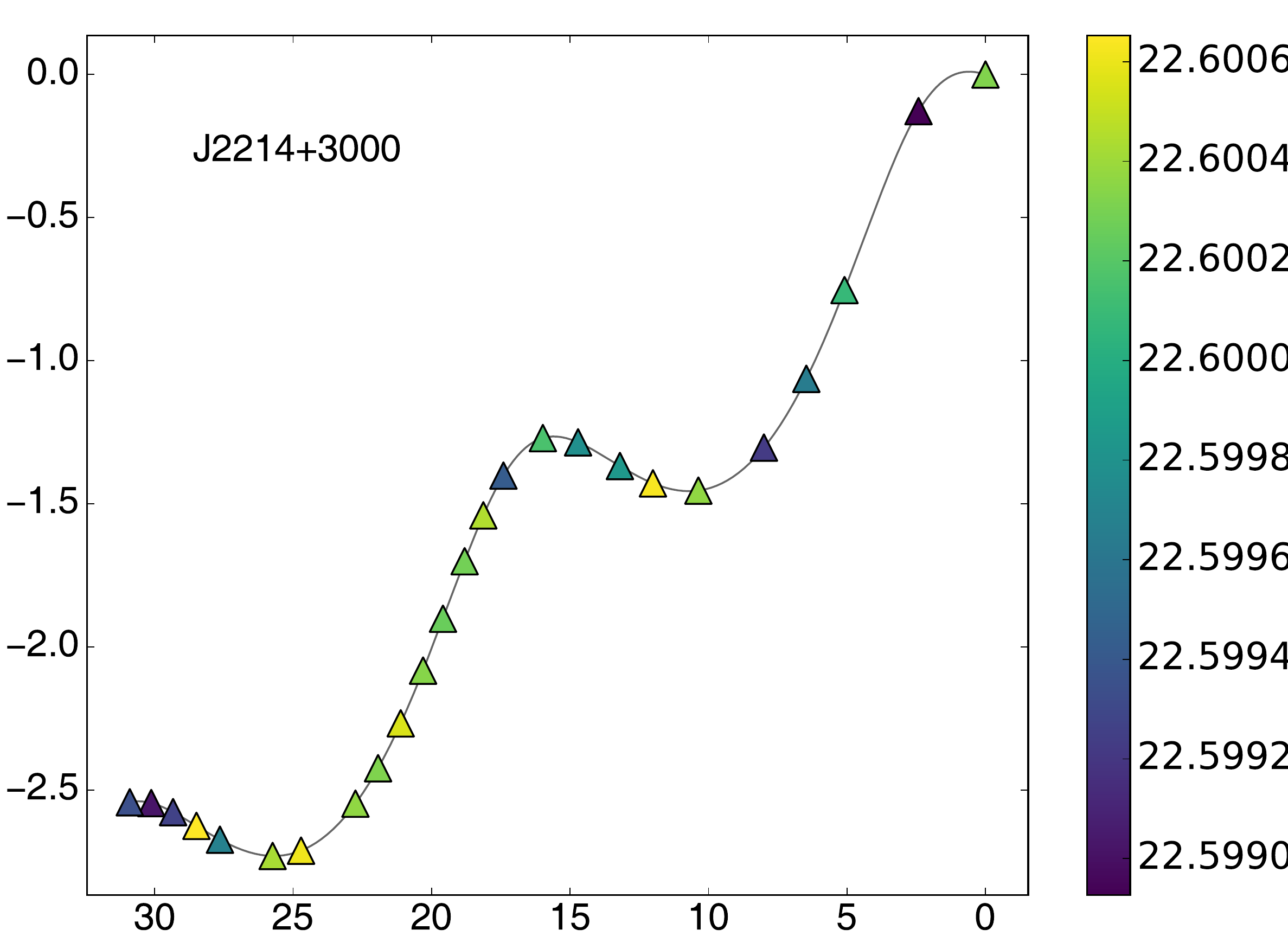}	
	    \includegraphics[trim=0cm 0cm 0cm 0cm, clip,width=0.4\textwidth]{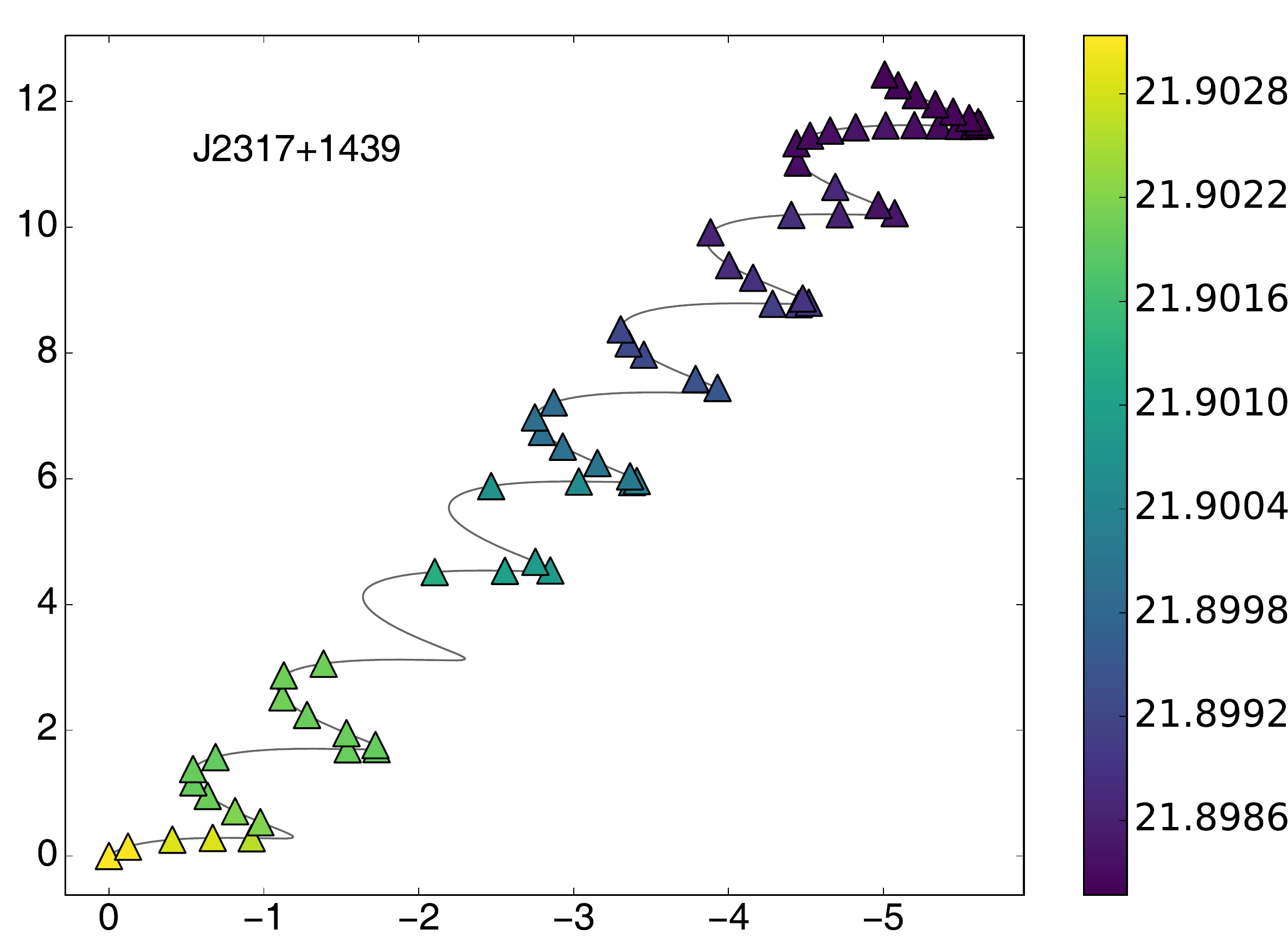}	
	    
  \leavevmode\smash{\makebox[0pt]{\hspace{5em}
  \rotatebox[origin=l]{90}{\hspace{30em}
    $\Delta_Y$ (AU)}%
}}\hspace{0pt plus 1filll}\null
\leavevmode\smash{\makebox[0pt]{\hspace{47em}
  \rotatebox[origin=l]{90}{\hspace{30em}
    DM (pc cm$^{-3}$)}%
}}\hspace{0pt plus 1filll}\null

$\Delta_X$ (AU)		
	\caption{MSP trajectories are plotted with color mapping the DM at each epoch. 
The trajectories are calculated assuming that all the free electrons along the LOS are sitting in a phase screen halfway between the Earth and the MSP; the trajectory is then the projected motion of the pulsar on the phase screen. The axes depict the space traversed at the phase screen in AU in the RA and Dec directions.
	The pulsar's motion starts at (0,0). Pulsars closer to the ecliptic will show a tighter sinusoid than those further away. The trajectory plot can be used to show limited localized structure.}
	\label{fig:trajectory4}		
	\end{center}
\end{figure*}

%% file: Figure12.tex
\begin{figure*}
	\begin{center}	
	\label{fig:structure}
		\includegraphics[trim=1.4cm 2.4cm 0.4cm 0.25cm, clip,width=0.344\textwidth]{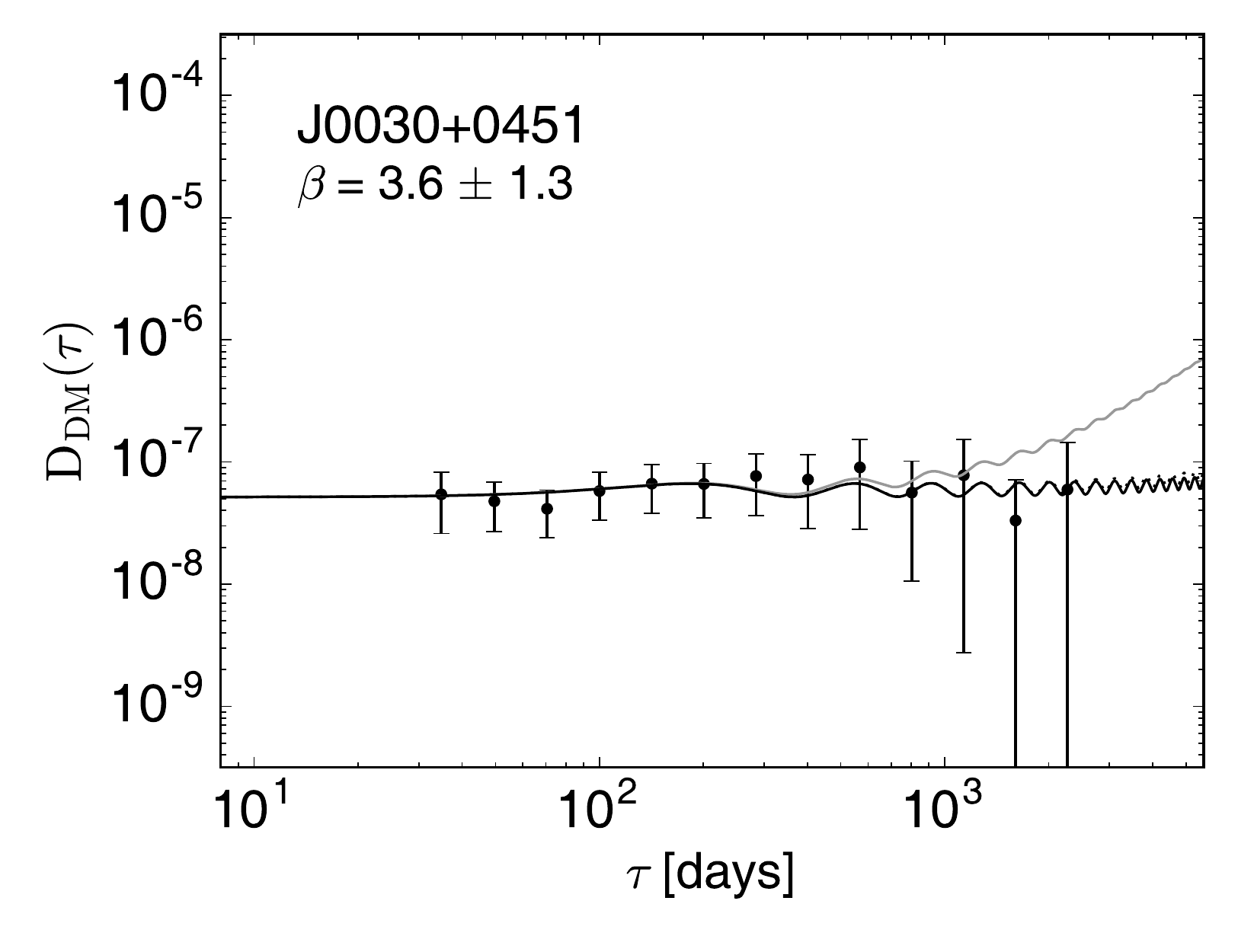}
		\includegraphics[trim=2.8cm 2.4cm 0.4cm 0.25cm, clip,width=0.31\textwidth]{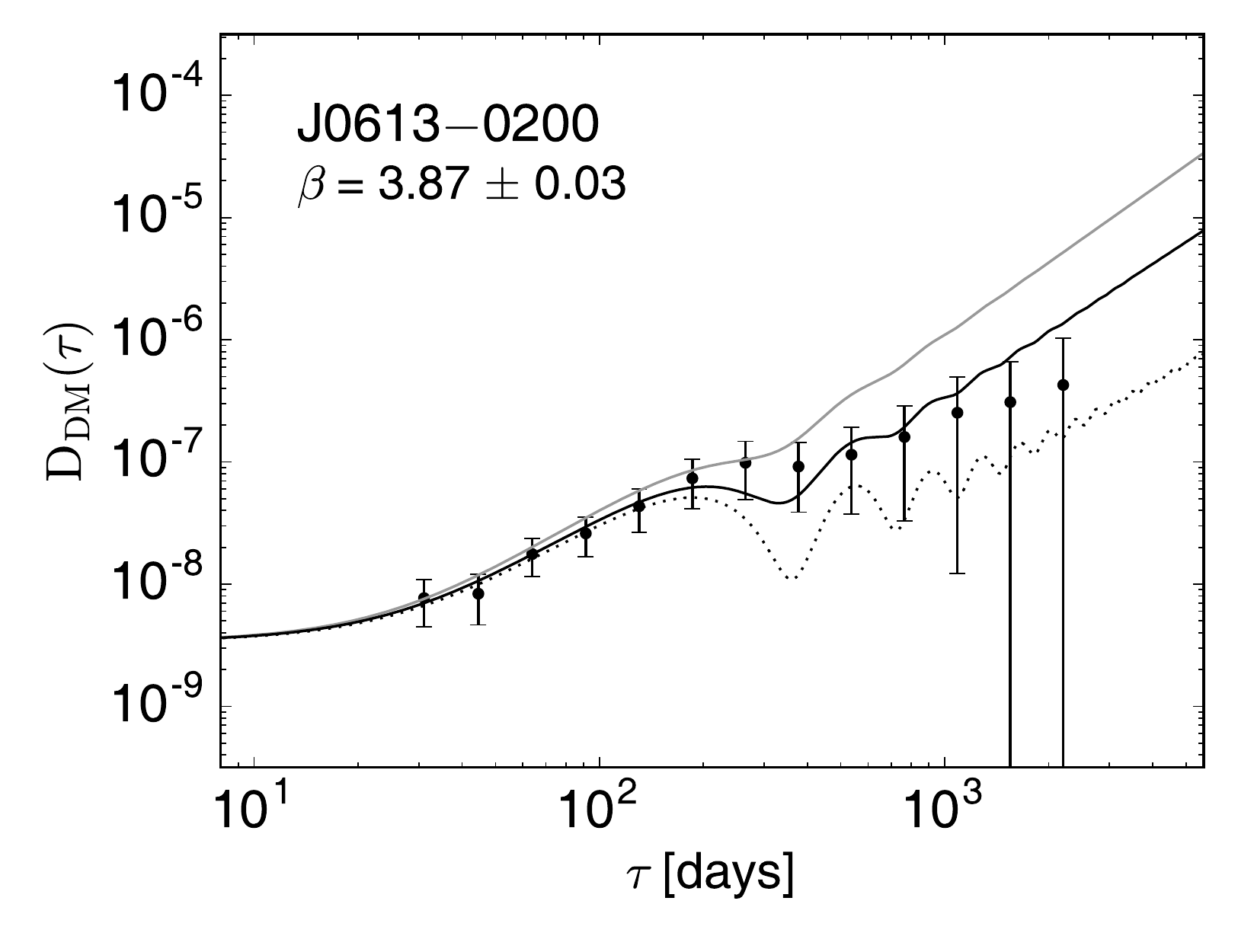}
		\includegraphics[trim=2.8cm 2.4cm 0.4cm 0.25cm, clip,width=0.31\textwidth]{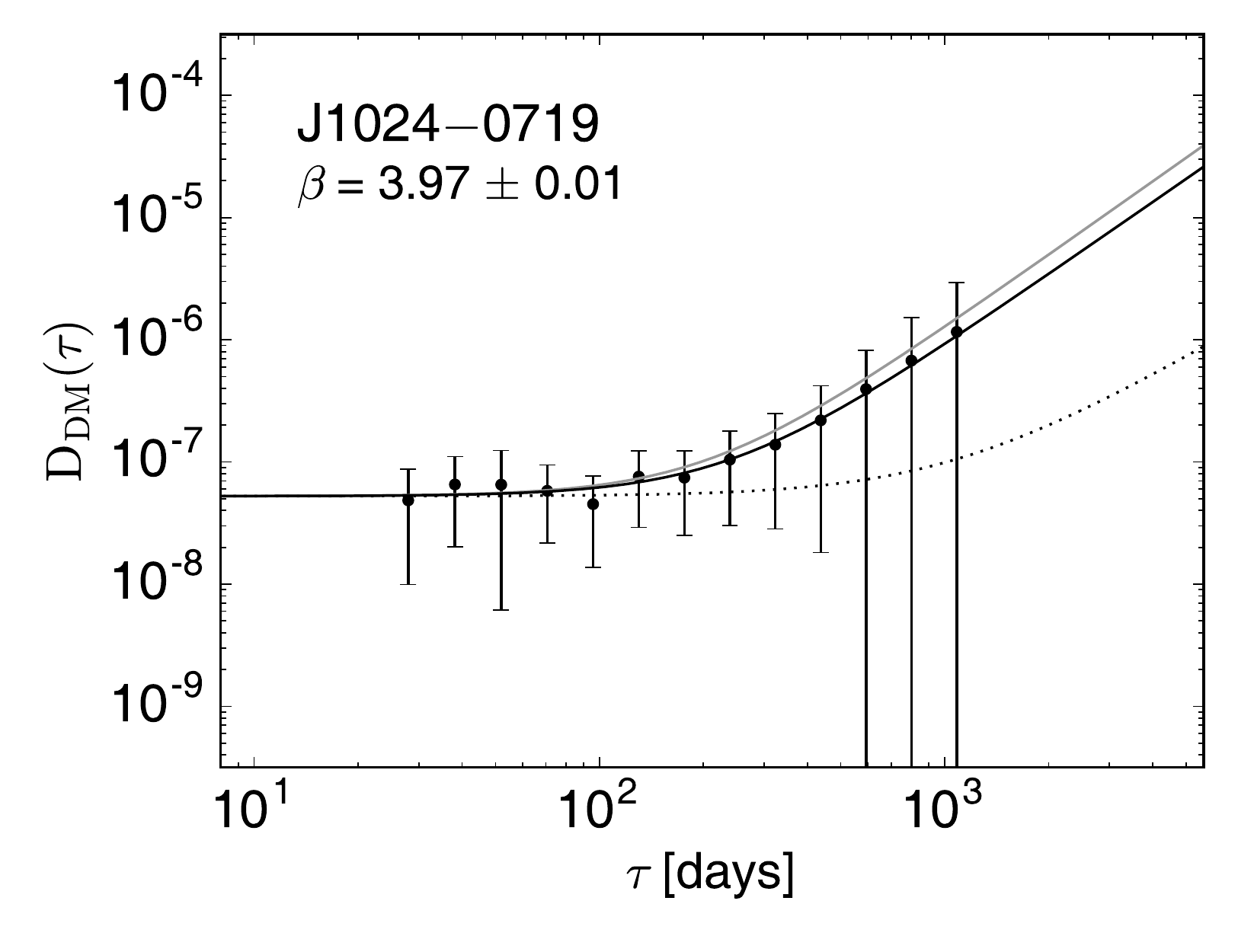}
	
		\includegraphics[trim=1.4cm 2.4cm 0.4cm 0.25cm, clip,width=0.344\textwidth]{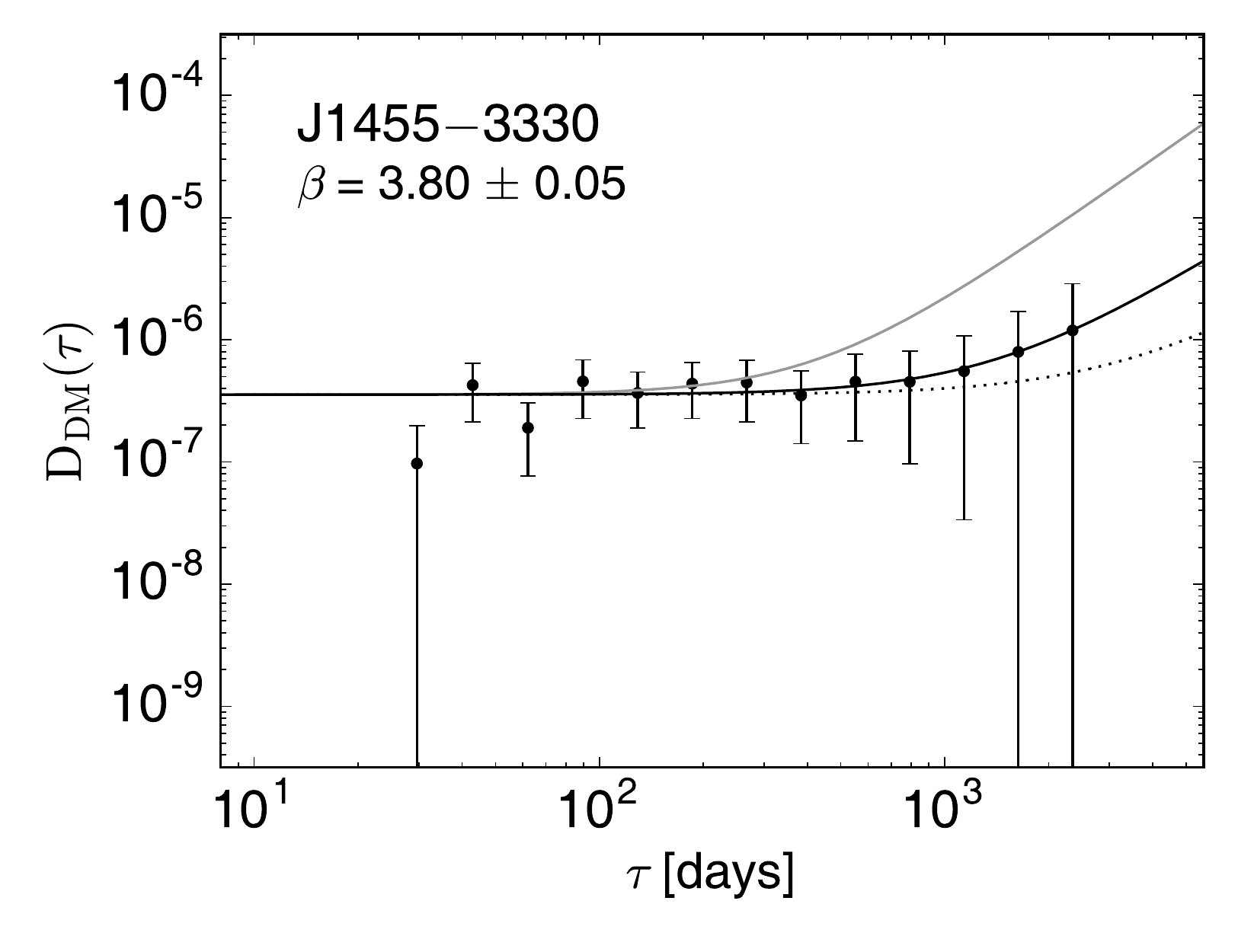}
		\includegraphics[trim=2.8cm 2.4cm 0.4cm 0.25cm, clip,width=0.31\textwidth]{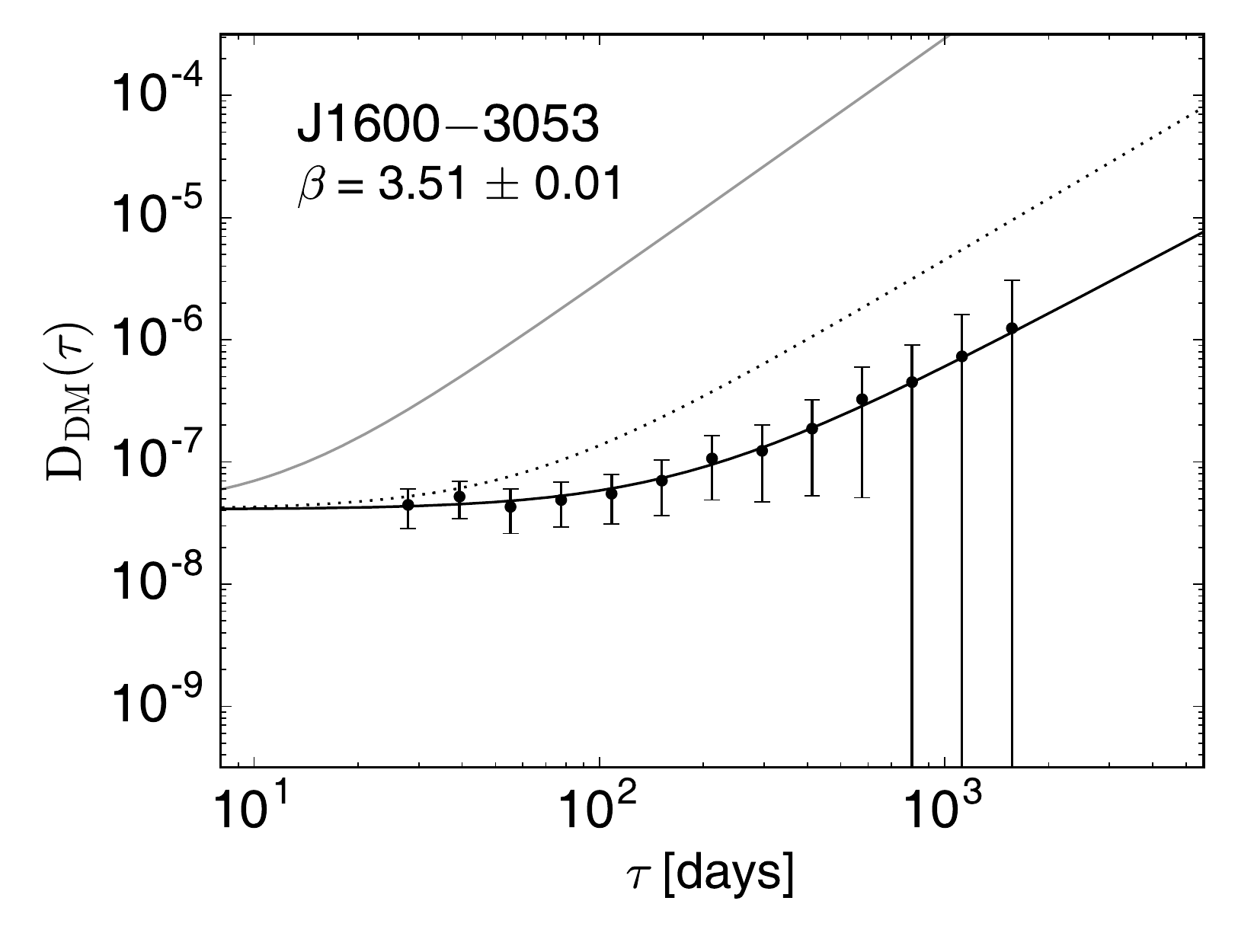}
		\includegraphics[trim=2.8cm 2.4cm 0.4cm 0.25cm, clip,width=0.31\textwidth]{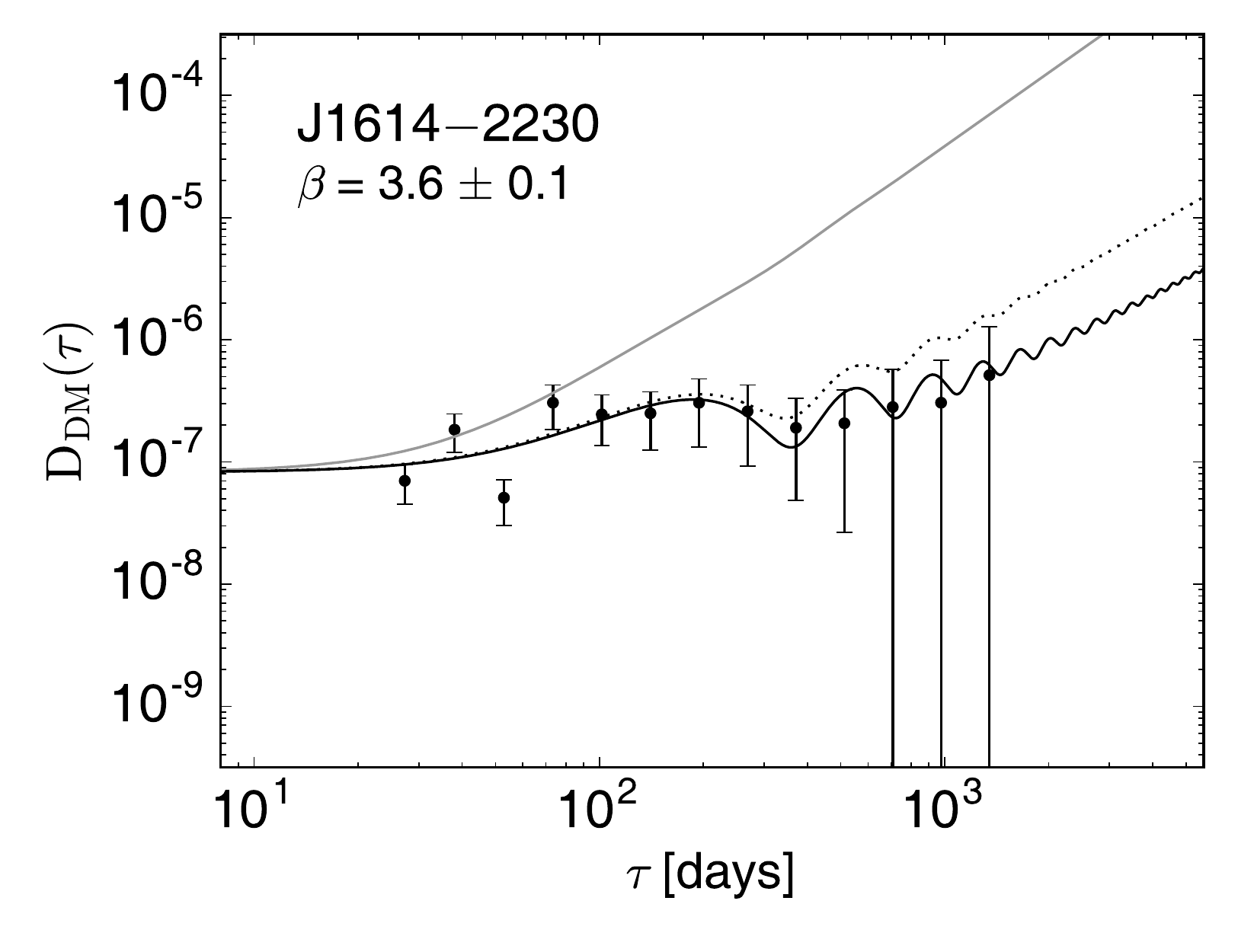}
		
		\includegraphics[trim=1.4cm 1.5cm 0.4cm 0.25cm, clip,width=0.344\textwidth]{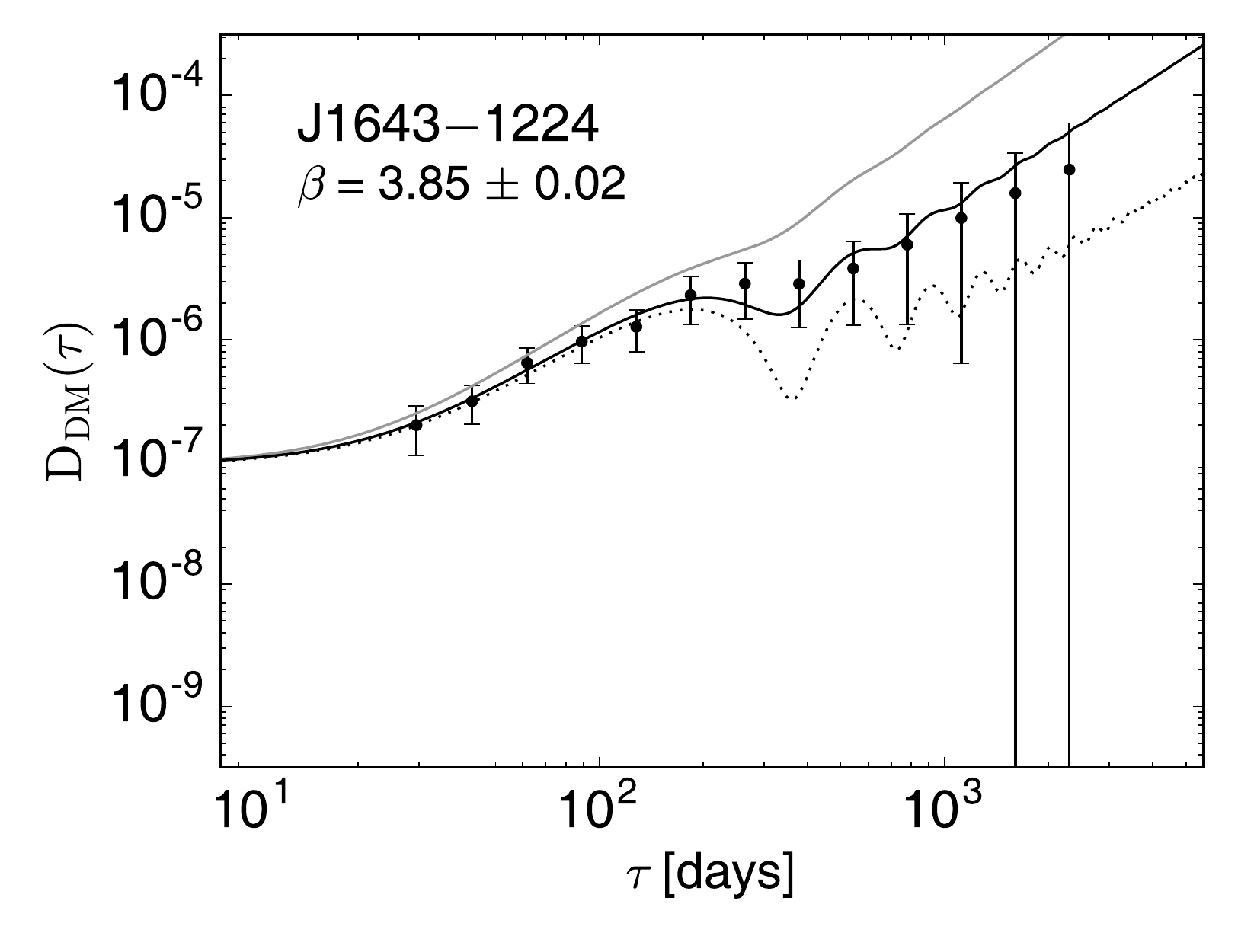}
		\includegraphics[trim=2.8cm 1.5cm 0.4cm 0.25cm, clip,width=0.31\textwidth]{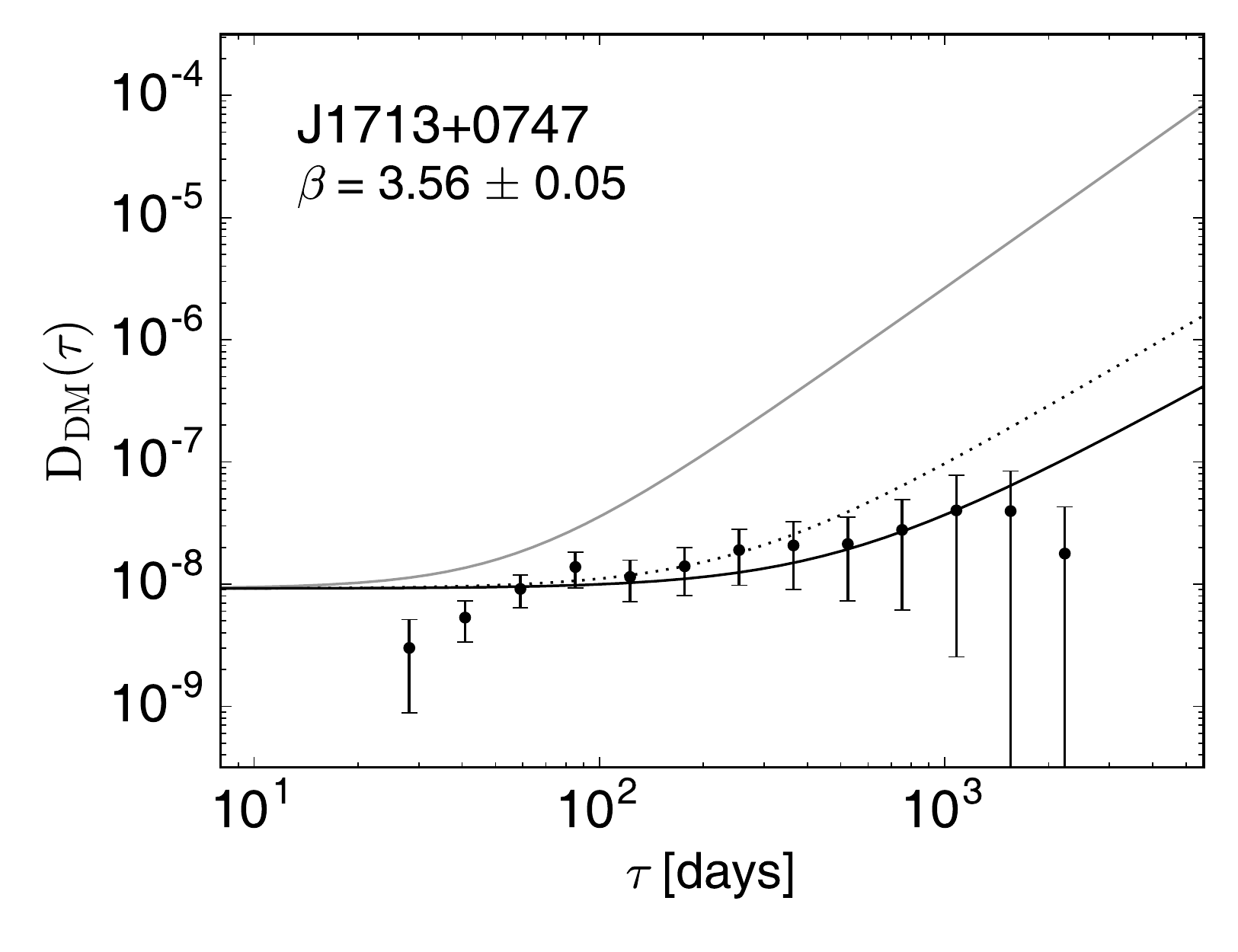}
		\includegraphics[trim=2.8cm 1.5cm 0.4cm 0.25cm, clip,width=0.31\textwidth]{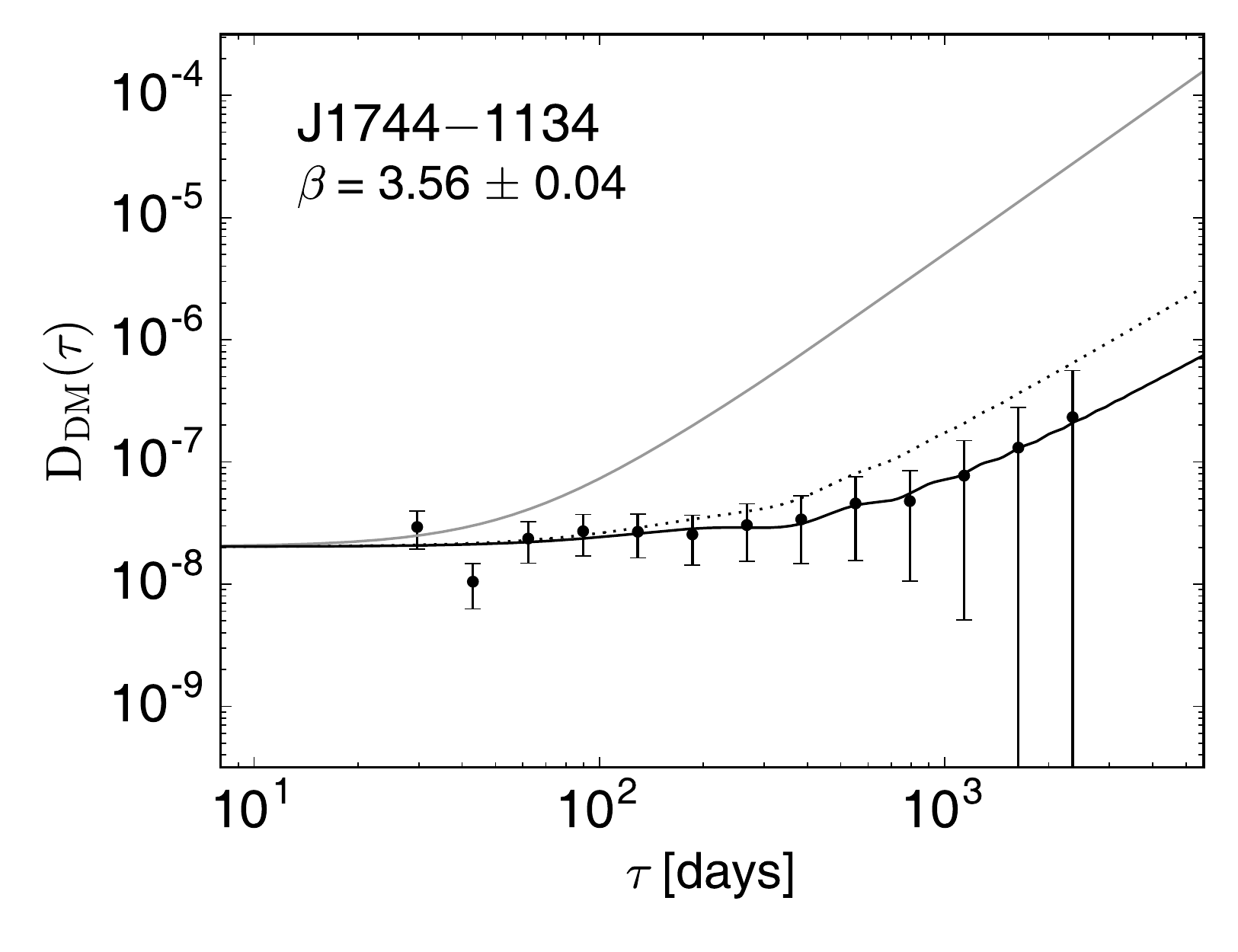}
		
\leavevmode\smash{\makebox[0pt]{\hspace{-3em}
  \rotatebox[origin=l]{90}{\vspace{-10em} \hspace{18em}
    D$_{\textrm{DM}}(\tau)$}%
}}\hspace{0pt plus 1filll}\null

Time Lag [days]\\
    \caption{Structure functions for the DM variations, calculated for the MSPs with measured diffractive timescales. Error bars extending to the bottom of the frame signify an upper limit (in agreement with \cite{you07}). The solid grey lines signify a quadratic power law and the dotted lines signify a Kolmogorov power law, which are anchored to the diffractive timescale, while the solid black lines are the best fits for the model. The errors associated with $\beta$ are $\pm1\sigma$ errors.}
	\end{center} 
\end{figure*}	

%% file: Figure13.tex
\begin{figure*}
	\begin{center}	
	\label{fig:structure2}
		\includegraphics[trim=1.4cm 2.4cm 0.43cm 0.25cm, clip,width=0.344\textwidth]{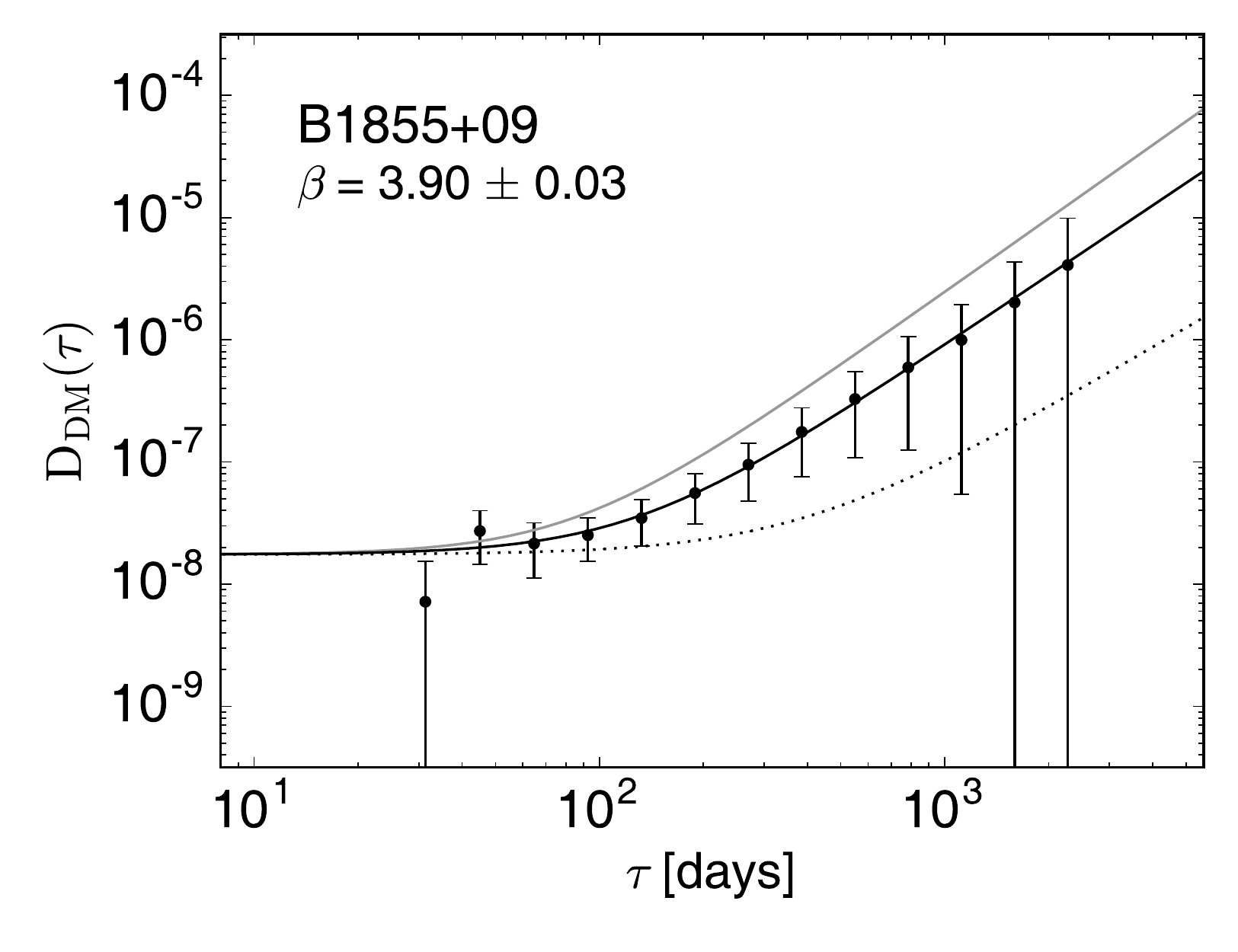}
		\includegraphics[trim=2.8cm 2.4cm 0.43cm 0.25cm, clip,width=0.31\textwidth]{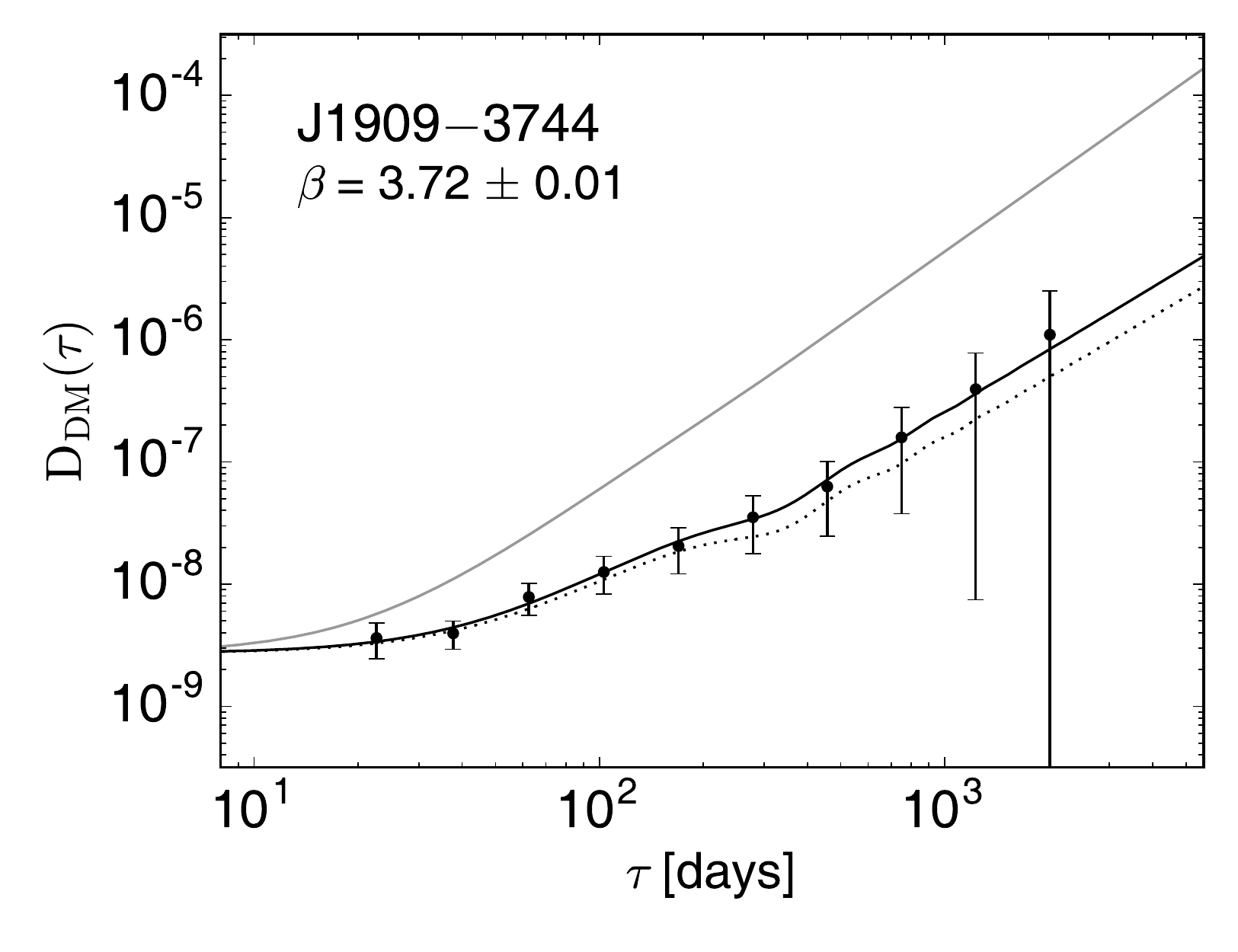}
		\includegraphics[trim=2.8cm 2.4cm 0.43cm 0.25cm, clip,width=0.31\textwidth]{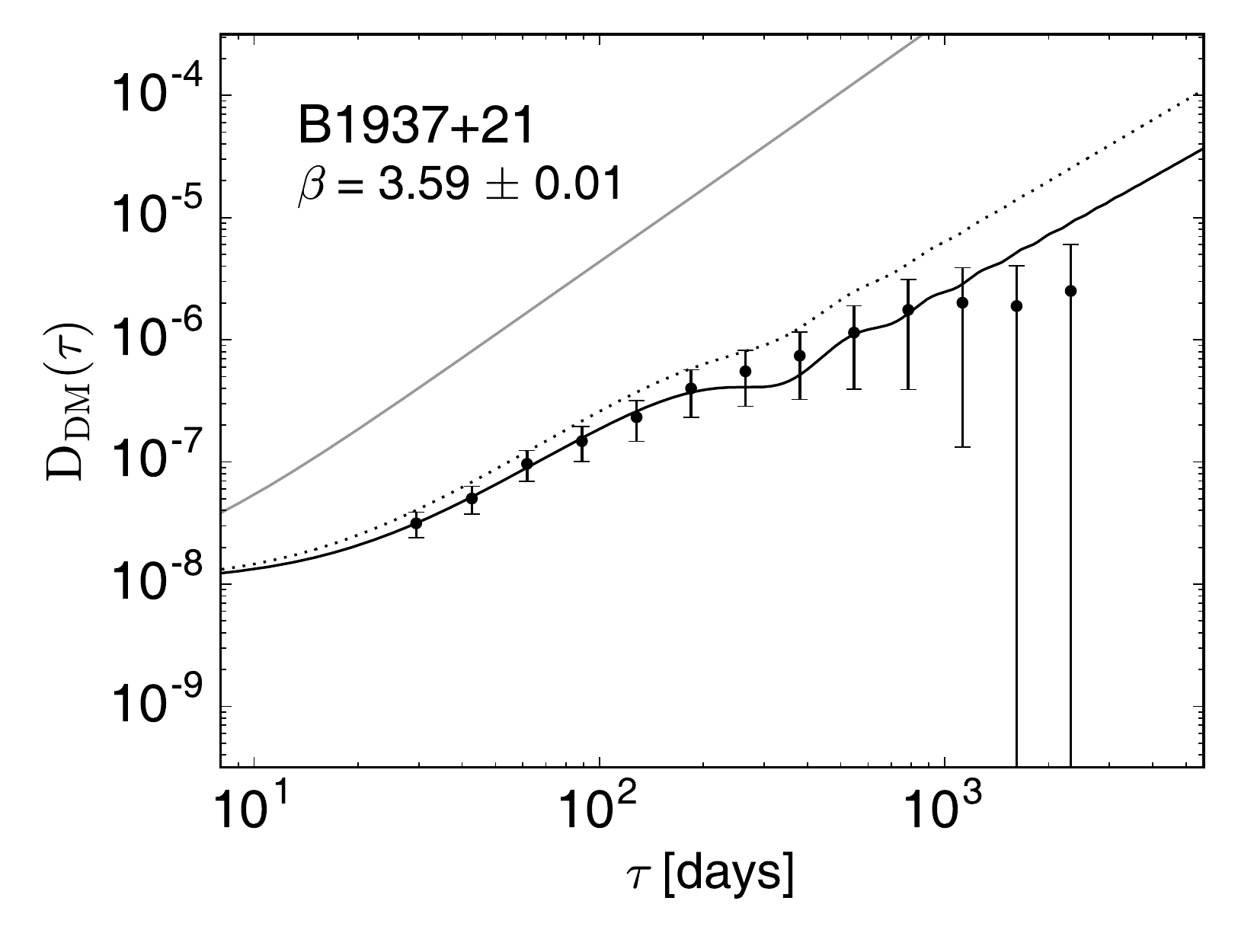}

		\includegraphics[trim=1.4cm 1.5cm 0.4cm 0.25cm, clip,width=0.344\textwidth]{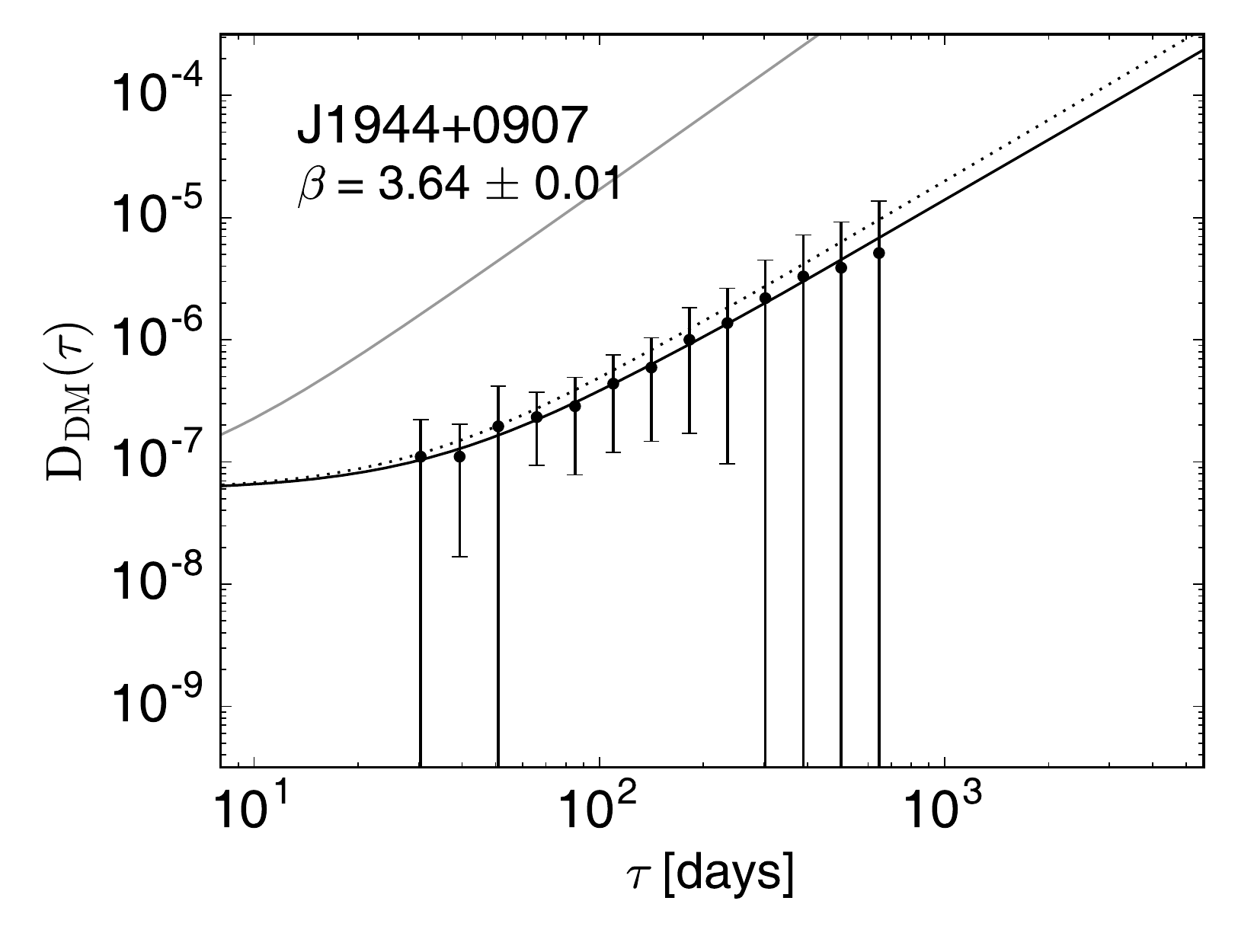} 		
		\includegraphics[trim=2.8cm 1.5cm 0.4cm 0.25cm, clip,width=0.31\textwidth]{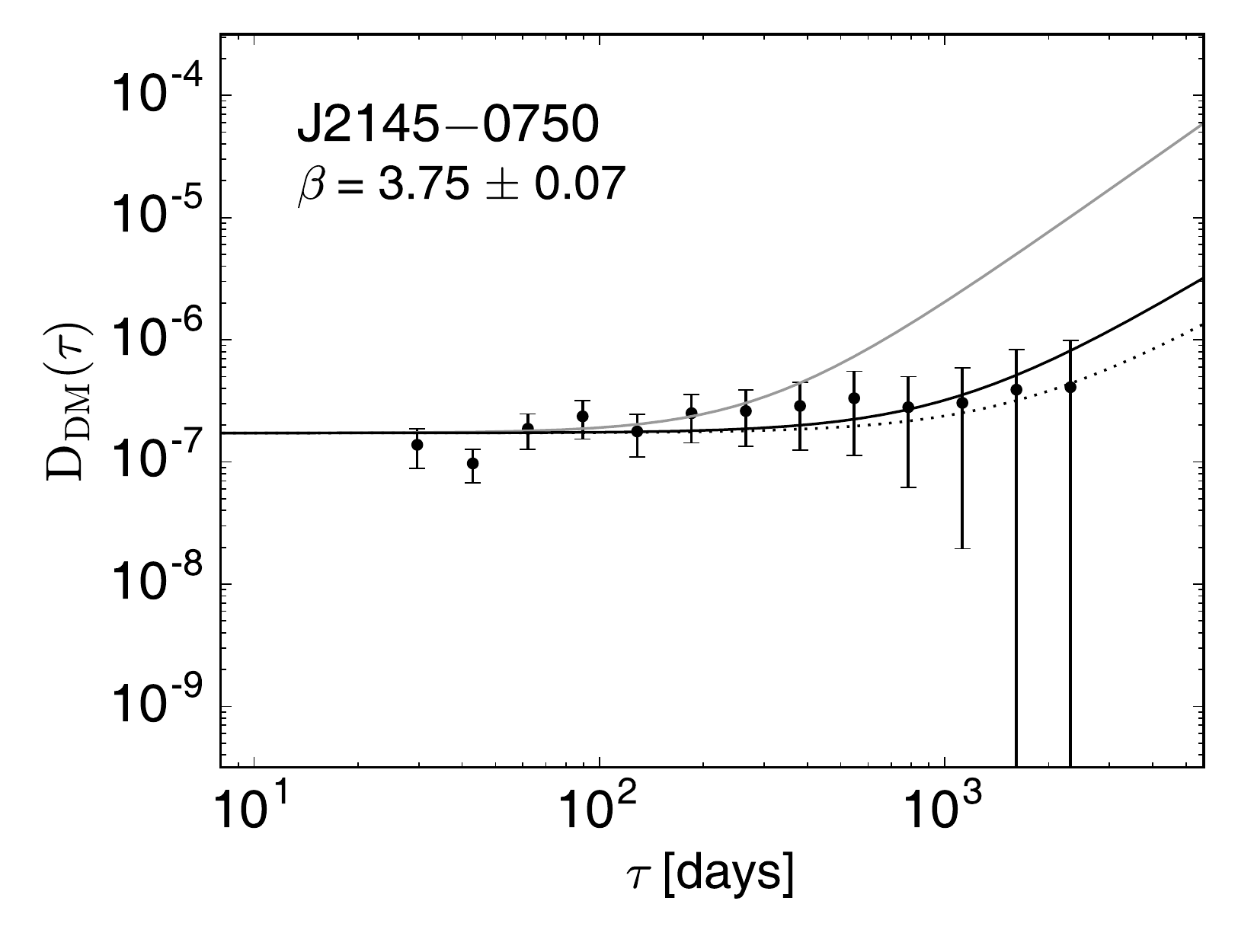}
		\includegraphics[trim=2.8cm 1.5cm 0.4cm 0.25cm, clip,width=0.31\textwidth]{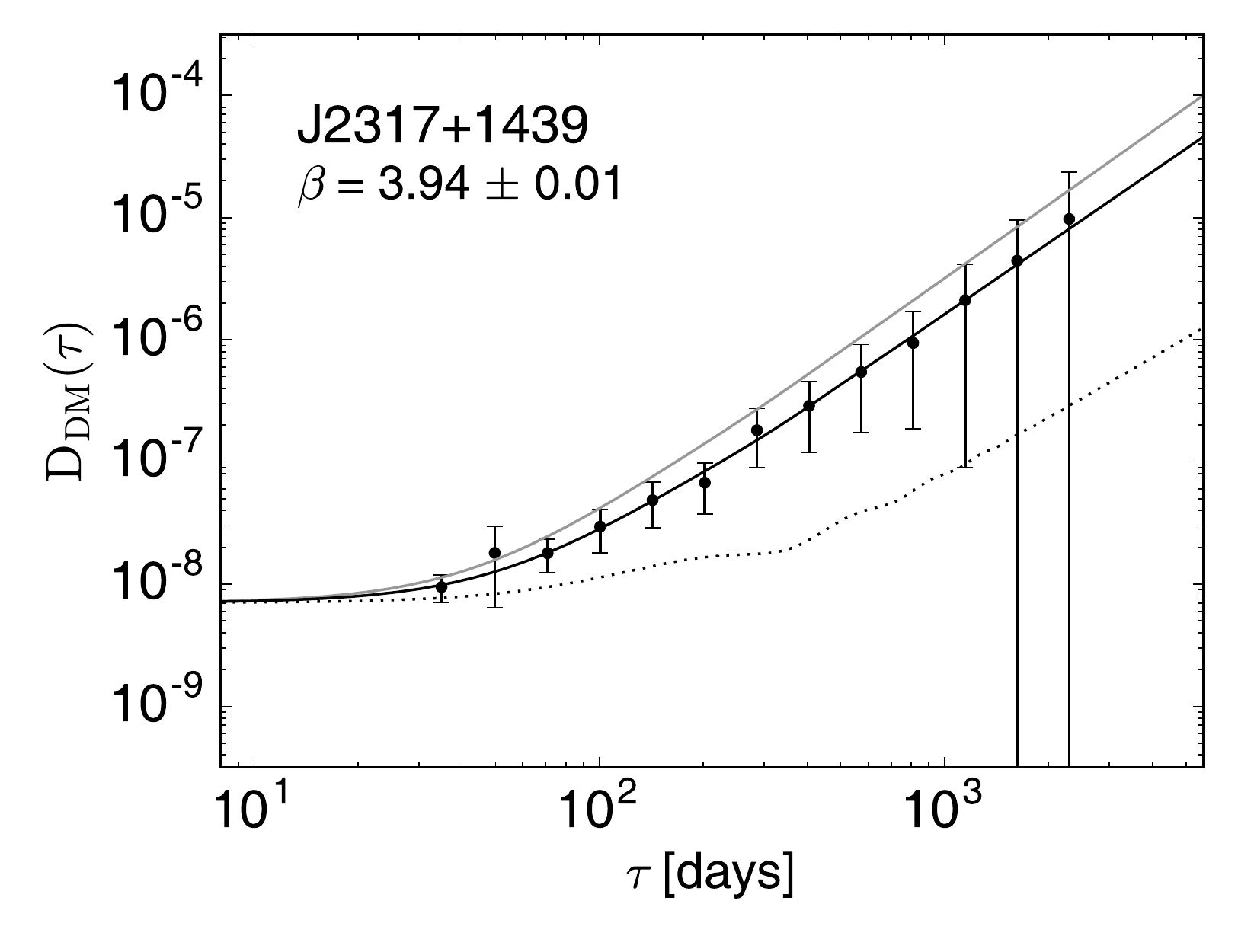}

\leavevmode\smash{\makebox[0pt]{\hspace{-3em}
  \rotatebox[origin=l]{90}{\vspace{-10em} \hspace{12em}
    D$_{\textrm{DM}}(\tau)$}%
}}\hspace{0pt plus 1filll}\null

Time Lag [days]\\
    \caption{Structure functions for the DM variations, calculated for the MSPs with measured diffractive timescales. Error bars extending to the bottom of the frame signify an upper limit (in agreement with \cite{you07}). The solid grey lines signify a quadratic power law and the dotted lines signify a Kolmogorov power law, which are anchored to the diffractive timescale, while the solid black lines are the best fits for the model. The errors associated with $\beta$ are $\pm1\sigma$ errors.}
	\end{center} 
\end{figure*}	

%% file: Table4.tex
\begin{table*}
\begin{center}
\caption{Diffractive timescales for 18 MSPs}
\label{tab:tdiff}
\begin{tabular}{crrrrrrr}
\tableline
PSR &  $\nu_{\rm{obs}}$  &  $\Delta \tau_{\rm{D}}$ & Source & $\beta$ & $\sigma_n$ & A \\
		   & (MHz) & (min) & & & (10$^{-4}$ pc cm$^{-3}$)& (10$^{-4}$ pc cm$^{-3}$)\\
\tableline

J0030+0451 & 436 & 167.7 & 1 & 3.6(1.3) & 1.6(1) & 0.9(4) \\
J0613--0200 & 1400 & 75 & 2 & 3.87(3) & 0.4(1) & 1.5(2) \\ 
J1024--0719	& 1400 & 69.7 & 2 & 3.97(1) & 1.6(5) & ---  \\ 
J1455--3330 & 436 & 17.7 & 3 & 3.80(5) & 3.8(4) & --- \\
J1600--3053 & 1400 & 4.5 & 2 & 3.51(1) & 1.43(3) & --- \\ 
			& 1373 & 4.7 &  4\\ 
J1614--2230 & 1400 & 12.5 &  & 3.6(1) & 2.0(4) & 3.4(8)\\ 
			& 1500 & 8.1 &  \\  
J1643--1224 & 1400 & 9.7 &  2 & 3.85(2) & 2.2(6) & 8.9(7) \\
J1713+0747  & 1400 & 47.6 & 2 & 3.56(5) & 0.68(5) & --- \\ 
			& 430 & 14 &  4\\
			& 436 & 28 &  4\\ 
J1744--1134  & 1400 & 34.5 & 2 & 3.56(4) & 1.01(9) & 0.5(4) \\ 
			& 436 & 21 & 4  \\ 
			& 660 & 20 &  4\\ 
B1855+09    & 1500 & 20.3 &  & 3.90(3) & 0.93(7) & --- \\ 
			& 1500 & 24.4 &  2  \\ 
J1909--3744  & 1300 & 30.6 &  & 3.72(1) & 0.37(2) & 0.63(7)\\ 
			& 1400 & 37.6 & 2  \\ 
B1937+21    & 1500 & 4.0 & & 3.59(1) & 0.7(2) & 3.2(3)\\ 
			& 1500 & 7.4 & \\
			& 320 & 1.1 &  4\\
			& 430 & 1.7 & 4\\
			& 1400 & 7.4 &  4\\
			& 1400 & 5.5 & 2  \\
J1944+0907   & 1500 & 2.0 &  & 3.64(1) & 1.7(2) & --- \\
J2145--0750   & 1400 & 56.6 & 2 & 3.75(7) & 2.9(2) & --- \\
			& 327 & 6.4 & 4 \\
			& 436 & 21-25 &  4\\
J2317+1439 & 436 & 13.5 & 3 & 3.94(1) & 0.59(4) & 0.5(2) \\ 
\tableline
   \end{tabular}
	\end{center}
{\small {\bf Notes.} Diffractive timescales obtained from the PPTA and NANOGrav datasets. Values with no reference were calculated from the 9-year data set. The first value listed for each MSP is the value used in calculating the structure function. The values $\beta$, $\sigma_n$, and A are from fitting the SF and correspond to the stochastic power law exponent, the white noise component, and the periodic amplitude component respectively. Values in parentheses show the uncertainty in the last digit. \\ $^1$\cite{nic01}, $^2$\cite{kei13}, $^3$\cite{joh98}, $^4$\cite{you07}} 
\end{table*}

%% file: Table5.tex
\begin{table}
  \begin{center}
  \caption{Significance of DM peaks for MSPs within 10$^{\circ}$ of the ecliptic}
  \label{tab:sunpos}
  \begin{tabular}{llllll}
	\hline 
	\hline
	PSR &  $\sigma$ &  DM$_{\rm{peak}}$/$\sigma$ & $\theta$\\
		   & (10$^{-3}$ pc cm$^{-3}$) & & (degrees) \\
	\hline	
J0023+0923 & 0.53 & 1.2 & 6.3 \\
J0030+0451 & 0.12 & 44.4 & 1.5 \\
J1614--2230 & 0.22 & 4.3 & 6.8  \\
J2010--1323 & 0.18 & 4.2 & 6.5 \\

	\hline
	\hline
  \end{tabular}
  \end{center}
  {\small {\bf Notes.} Columns are the rms $\sigma$ of DM measurements with a Sun-pulsar angle greater than 30$^{\circ}$, the ratio of the highest DM value in the data set over the rms, and the minimal angle $\theta$ between the Sun and the pulsar. The solar angle corresponds to the minimal angle seen between the pulsar and the Sun in the 9-year data.}
\end{table}

%% file: manuscript.bbl
\begin{thebibliography}{31}
\providecommand{\natexlab}[1]{#1}
\providecommand{\url}[1]{\texttt{#1}}
\expandafter\ifx\csname urlstyle\endcsname\relax
  \providecommand{\doi}[1]{doi: #1}\else
  \providecommand{\doi}{doi: \begingroup \urlstyle{rm}\Url}\fi

\bibitem[{Andrae} et~al.(2010){Andrae}, {Schulze-Hartung}, \&
  {Melchior}]{and10}
{Andrae}, R., {Schulze-Hartung}, T., \& {Melchior}, P., \emph{ArXiv e-prints},
  2010.

\bibitem[{Arzoumanian} et~al.(2015){Arzoumanian}, {Brazier}, {Burke-Spolaor},
  {Chamberlin}, {Chatterjee}, {Christy}, {Cordes}, {Cornish}, {Crowter},
  {Demorest}, {Dolch}, {Ellis}, {Ferdman}, {Fonseca}, {Garver-Daniels},
  {Gonzalez}, {Jenet}, {Jones}, {Jones}, {Kaspi}, {Koop}, {Lam}, {Lazio},
  {Levin}, {Lommen}, {Lorimer}, {Luo}, {Lynch}, {Madison}, {McLaughlin},
  {McWilliams}, {Nice}, {Palliyaguru}, {Pennucci}, {Ransom}, {Siemens},
  {Stairs}, {Stinebring}, {Stovall}, {Swiggum}, {Vallisneri}, {van Haasteren},
  {Wang}, \& {Zhu}]{arz15}
{Arzoumanian}, Z., {Brazier}, A., {Burke-Spolaor}, S., {Chamberlin}, S.,
  {Chatterjee}, S., {Christy}, B., {Cordes}, J.~M., {Cornish}, N., {Crowter},
  K., {Demorest}, P.~B., {Dolch}, T., {Ellis}, J.~A., {Ferdman}, R.~D.,
  {Fonseca}, E., {Garver-Daniels}, N., {Gonzalez}, M.~E., {Jenet}, F.~A.,
  {Jones}, G., {Jones}, M.~L., {Kaspi}, V.~M., {Koop}, M., {Lam}, M.~T.,
  {Lazio}, T.~J.~W., {Levin}, L., {Lommen}, A.~N., {Lorimer}, D.~R., {Luo}, J.,
  {Lynch}, R.~S., {Madison}, D., {McLaughlin}, M.~A., {McWilliams}, S.~T.,
  {Nice}, D.~J., {Palliyaguru}, N., {Pennucci}, T.~T., {Ransom}, S.~M.,
  {Siemens}, X., {Stairs}, I.~H., {Stinebring}, D.~R., {Stovall}, K.,
  {Swiggum}, J.~K., {Vallisneri}, M., {van Haasteren}, R., {Wang}, Y., \&
  {Zhu}, W., \emph{\apj}, 813:\penalty0 65, 2015.

\bibitem[{Cordes} \& {Lazio}(2002)]{cor02}
{Cordes}, J.~M. \& {Lazio}, T.~J.~W., \emph{ArXiv Astrophysics e-prints}, 2002.

\bibitem[{Cordes} \& {Rickett}(1998)]{cor98}
{Cordes}, J.~M. \& {Rickett}, B.~J., \emph{\apj}, 507, \penalty0 846--860,
  1998.

\bibitem[{Cordes} \& {Shannon}(2010)]{cor10}
{Cordes}, J.~M. \& {Shannon}, R.~M., \emph{ArXiv e-prints}, 2010.

\bibitem[{Cordes} et~al.(1990){Cordes}, {Wolszczan}, {Dewey}, {Blaskiewicz}, \&
  {Stinebring}]{cor90}
{Cordes}, J.~M., {Wolszczan}, A., {Dewey}, R.~J., {Blaskiewicz}, M., \&
  {Stinebring}, D.~R., \emph{\apj}, 349, \penalty0 245--261, 1990.

\bibitem[{Cordes} et~al.(2016){Cordes}, {Shannon}, \& {Stinebring}]{cor16}
{Cordes}, J.~M., {Shannon}, R.~M., \& {Stinebring}, D.~R., \emph{\apj},
  817:\penalty0 16, 2016.

\bibitem[{Demorest} et~al.(2013){Demorest}, {Ferdman}, {Gonzalez}, {Nice},
  {Ransom}, {Stairs}, {Arzoumanian}, {Brazier}, {Burke-Spolaor}, {Chamberlin},
  {Cordes}, {Ellis}, {Finn}, {Freire}, {Giampanis}, {Jenet}, {Kaspi}, {Lazio},
  {Lommen}, {McLaughlin}, {Palliyaguru}, {Perrodin}, {Shannon}, {Siemens},
  {Stinebring}, {Swiggum}, \& {Zhu}]{dem13}
{Demorest}, P.~B., {Ferdman}, R.~D., {Gonzalez}, M.~E., {Nice}, D., {Ransom},
  S., {Stairs}, I.~H., {Arzoumanian}, Z., {Brazier}, A., {Burke-Spolaor}, S.,
  {Chamberlin}, S.~J., {Cordes}, J.~M., {Ellis}, J., {Finn}, L.~S., {Freire},
  P., {Giampanis}, S., {Jenet}, F., {Kaspi}, V.~M., {Lazio}, J., {Lommen},
  A.~N., {McLaughlin}, M., {Palliyaguru}, N., {Perrodin}, D., {Shannon}, R.~M.,
  {Siemens}, X., {Stinebring}, D., {Swiggum}, J., \& {Zhu}, W.~W., \emph{\apj},
  762:\penalty0 94, 2013.

\bibitem[{Fonseca} et~al.(2014){Fonseca}, {Stairs}, \& {Thorsett}]{fon14}
{Fonseca}, E., {Stairs}, I.~H., \& {Thorsett}, S.~E., \emph{\apj},
  787:\penalty0 82, 2014.

\bibitem[{Hamilton} et~al.(1985){Hamilton}, {Hall}, \& {Costa}]{ham85}
{Hamilton}, P.~A., {Hall}, P.~J., \& {Costa}, M.~E., \emph{\mnras}, 214,
  \penalty0 5P--8P, 1985.

\bibitem[{Isaacman} \& {Rankin}(1977)]{isa77}
{Isaacman}, R. \& {Rankin}, J.~M., \emph{\apj}, 214, \penalty0 214--232, 1977.

\bibitem[{Jenet} et~al.(2005){Jenet}, {Hobbs}, {Lee}, \& {Manchester}]{jen05}
{Jenet}, F.~A., {Hobbs}, G.~B., {Lee}, K.~J., \& {Manchester}, R.~N.,
  \emph{\apjl}, 625, \penalty0 L123--L126, 2005.

\bibitem[{Johnston} et~al.(1998){Johnston}, {Nicastro}, \& {Koribalski}]{joh98}
{Johnston}, S., {Nicastro}, L., \& {Koribalski}, B., \emph{\mnras}, 297,
  \penalty0 108--116, 1998.

\bibitem[{Kaspi} et~al.(1994){Kaspi}, {Taylor}, \& {Ryba}]{kas94}
{Kaspi}, V.~M., {Taylor}, J.~H., \& {Ryba}, M.~F., \emph{\apj}, 428, \penalty0
  713--728, 1994.

\bibitem[{Keith} et~al.(2013){Keith}, {Coles}, {Shannon}, {Hobbs},
  {Manchester}, {Bailes}, {Bhat}, {Burke-Spolaor}, {Champion}, {Chaudhary},
  {Hotan}, {Khoo}, {Kocz}, {Os{\l}owski}, {Ravi}, {Reynolds}, {Sarkissian},
  {van Straten}, \& {Yardley}]{kei13}
{Keith}, M.~J., {Coles}, W., {Shannon}, R.~M., {Hobbs}, G.~B., {Manchester},
  R.~N., {Bailes}, M., {Bhat}, N.~D.~R., {Burke-Spolaor}, S., {Champion},
  D.~J., {Chaudhary}, A., {Hotan}, A.~W., {Khoo}, J., {Kocz}, J.,
  {Os{\l}owski}, S., {Ravi}, V., {Reynolds}, J.~E., {Sarkissian}, J., {van
  Straten}, W., \& {Yardley}, D.~R.~B., \emph{\mnras}, 429, \penalty0
  2161--2174, 2013.

\bibitem[{Lam} et~al.(2015){Lam}, {Cordes}, {Chatterjee}, \& {Dolch}]{lam15}
{Lam}, M.~T., {Cordes}, J.~M., {Chatterjee}, S., \& {Dolch}, T., \emph{\apj},
  801:\penalty0 130, 2015.

\bibitem[{Lam} et~al.(2016){Lam}, {Cordes}, {Chatterjee}, {Jones},
  {McLaughlin}, \& {Armstrong}]{lam16}
{Lam}, M.~T., {Cordes}, J.~M., {Chatterjee}, S., {Jones}, M.~L., {McLaughlin},
  M.~A., \& {Armstrong}, J.~W., \emph{\apj}, 821:\penalty0 66, 2016.

\bibitem[{Levin} et~al.(2016){Levin}, {McLaughlin}, {Jones}, {Cordes},
  {Stinebring}, {Chatterjee}, {Dolch}, {Lam}, {Lazio}, {Palliyaguru},
  {Arzoumanian}, {Crowter}, {Demorest}, {Ellis}, {Ferdman}, {Fonseca},
  {Gonzalez}, {Jones}, {Nice}, {Pennucci}, {Ransom}, {Stairs}, {Stovall},
  {Swiggum}, \& {Zhu}]{lev16}
{Levin}, L., {McLaughlin}, M.~A., {Jones}, G., {Cordes}, J.~M., {Stinebring},
  D.~R., {Chatterjee}, S., {Dolch}, T., {Lam}, M.~T., {Lazio}, T.~J.~W.,
  {Palliyaguru}, N., {Arzoumanian}, Z., {Crowter}, K., {Demorest}, P.~B.,
  {Ellis}, J.~A., {Ferdman}, R.~D., {Fonseca}, E., {Gonzalez}, M.~E., {Jones},
  M.~L., {Nice}, D.~J., {Pennucci}, T.~T., {Ransom}, S.~M., {Stairs}, I.~H.,
  {Stovall}, K., {Swiggum}, J.~K., \& {Zhu}, W., \emph{\apj}, 818:\penalty0
  166, 2016.

\bibitem[{Lorimer} \& {Kramer}(2012)]{lor12}
{Lorimer}, D.~R. \& {Kramer}, M.
\newblock \emph{{Handbook of Pulsar Astronomy}}.
\newblock 2012.

\bibitem[{Matthews} et~al.(2016){Matthews}, {Nice}, {Fonseca}, {Arzoumanian},
  {Crowter}, {Demorest}, {Dolch}, {Ellis}, {Ferdman}, {Gonzalez}, {Jones},
  {Jones}, {Lam}, {Levin}, {McLaughlin}, {Pennucci}, {Ransom}, {Stairs},
  {Stovall}, {Swiggum}, \& {Zhu}]{mat16}
{Matthews}, A.~M., {Nice}, D.~J., {Fonseca}, E., {Arzoumanian}, Z., {Crowter},
  K., {Demorest}, P.~B., {Dolch}, T., {Ellis}, J.~A., {Ferdman}, R.~D.,
  {Gonzalez}, M.~E., {Jones}, G., {Jones}, M.~L., {Lam}, M.~T., {Levin}, L.,
  {McLaughlin}, M.~A., {Pennucci}, T.~T., {Ransom}, S.~M., {Stairs}, I.~H.,
  {Stovall}, K., {Swiggum}, J.~K., \& {Zhu}, W., \emph{\apj}, 818:\penalty0 92,
  2016.

\bibitem[{McLaughlin}(2013)]{mcl13}
{McLaughlin}, M.~A., \emph{Classical and Quantum Gravity}, 30\penalty0
  (22):\penalty0 224008, 2013.

\bibitem[{Nicastro} et~al.(2001){Nicastro}, {Nigro}, {D'Amico}, {Lumiella}, \&
  {Johnston}]{nic01}
{Nicastro}, L., {Nigro}, F., {D'Amico}, N., {Lumiella}, V., \& {Johnston}, S.,
  \emph{\aap}, 368, \penalty0 1055--1062, 2001.

\bibitem[{Petroff} et~al.(2013){Petroff}, {Keith}, {Johnston}, {van Straten},
  \& {Shannon}]{pet13}
{Petroff}, E., {Keith}, M.~J., {Johnston}, S., {van Straten}, W., \& {Shannon},
  R.~M., \emph{\mnras}, 435, \penalty0 1610--1617, 2013.

\bibitem[{Ramachandran} et~al.(2006){Ramachandran}, {Demorest}, {Backer},
  {Cognard}, \& {Lommen}]{ram06}
{Ramachandran}, R., {Demorest}, P., {Backer}, D.~C., {Cognard}, I., \&
  {Lommen}, A., \emph{\apj}, 645, \penalty0 303--313, 2006.

\bibitem[{Rankin} \& {Roberts}(1971)]{ran71}
{Rankin}, J.~M. \& {Roberts}, J.~A.
\newblock \emph{{Time Variability of the Dispersion of the Crab Nebula
  Pulsar}}.
\newblock In {Davies}, R.~D. \& {Graham-Smith}, F., editors, \emph{The Crab
  Nebula}, volume~46 of \emph{IAU Symposium}, page 114, 1971.

\bibitem[{Reardon} et~al.(2016){Reardon}, {Hobbs}, {Coles}, {Levin}, {Keith},
  {Bailes}, {Bhat}, {Burke-Spolaor}, {Dai}, {Kerr}, {Lasky}, {Manchester},
  {Os{\l}owski}, {Ravi}, {Shannon}, {van Straten}, {Toomey}, {Wang}, {Wen},
  {You}, \& {Zhu}]{rea16}
{Reardon}, D.~J., {Hobbs}, G., {Coles}, W., {Levin}, Y., {Keith}, M.~J.,
  {Bailes}, M., {Bhat}, N.~D.~R., {Burke-Spolaor}, S., {Dai}, S., {Kerr}, M.,
  {Lasky}, P.~D., {Manchester}, R.~N., {Os{\l}owski}, S., {Ravi}, V.,
  {Shannon}, R.~M., {van Straten}, W., {Toomey}, L., {Wang}, J., {Wen}, L.,
  {You}, X.~P., \& {Zhu}, X.-J., \emph{\mnras}, 455, \penalty0 1751--1769,
  2016.

\bibitem[{Rickett}(1990)]{ric90}
{Rickett}, B.~J., \emph{\araa}, 28, \penalty0 561--605, 1990.

\bibitem[{Scargle}(1982)]{sca82}
{Scargle}, J.~D., \emph{\apj}, 263, \penalty0 835--853, 1982.

\bibitem[{Siemens} et~al.(2013){Siemens}, {Ellis}, {Jenet}, \& {Romano}]{sie13}
{Siemens}, X., {Ellis}, J., {Jenet}, F., \& {Romano}, J.~D., \emph{Classical
  and Quantum Gravity}, 30\penalty0 (22):\penalty0 224015, 2013.

\bibitem[{Vigeland} \& {Siemens}(2016)]{vig16}
{Vigeland}, S.~J. \& {Siemens}, X., \emph{ArXiv e-prints}, 2016.

\bibitem[{You} et~al.(2007){You}, {Hobbs}, {Coles}, {Manchester}, {Edwards},
  {Bailes}, {Sarkissian}, {Verbiest}, {van Straten}, {Hotan}, {Ord}, {Jenet},
  {Bhat}, \& {Teoh}]{you07}
{You}, X.~P., {Hobbs}, G.~B., {Coles}, W.~A., {Manchester}, R.~N., {Edwards},
  R., {Bailes}, M., {Sarkissian}, J.~M., {Verbiest}, J.~P.~W., {van Straten},
  W., {Hotan}, A.~W., {Ord}, S., {Jenet}, F.~A., {Bhat}, N.~D.~R., \& {Teoh},
  A., \emph{\mnras}, 378, \penalty0 493--506, 2007.

\end{thebibliography}
